\documentclass[fleqn,usenatbib]{mnras}
\usepackage{newtxtext,newtxmath}
\usepackage[T1]{fontenc}

\usepackage{graphicx}	% Including figure files
\usepackage{amsmath}	% Advanced maths commands
\usepackage{longtable}
\usepackage{subfig}
\usepackage{comment}

\title[Radio variability of OH Megamasers]{Radio Variability and Broad-Band Spectra of Infrared Galaxies\\ with and without OH Megamaser Emission}

\author[Yu. Sotnikova et al.]{
Yu. V. Sotnikova,$^{1}$\thanks{E-mail: lacerta999@gmail.com}
T. V. Mufakharov,$^{1,2}$
A. G. Mikhailov,$^{1}$
V. A. Stolyarov,$^{1,2,3}$
Z. Z. Wu,$^{4}$
\newauthor
M. G. Mingaliev,$^{1,2}$
T. A. Semenova,$^{1}$
A. K. Erkenov,$^{1}$
N. N. Bursov,$^{1}$
and R. Y. Udovitskiy$^{1}$
\\
$^{1}$Special Astrophysical Observatory of RAS, Nizhny Arkhyz 369167, Russia\\
$^{2}$Kazan Federal University, 18 Kremlyovskaya St, Kazan 420008, Russia\\
$^{3}$Astrophysics group, Cavendish laboratory, University of Cambridge, J J Thomson Ave, Cambridge, CB3 0HE, UK\\
$^{4}$College of Physics, Guizhou University, 550025 Guiyang, China
}

\date{\bf Published in Astrophysical Bulletin, Vol. 37, No. 3, pp. 246-263 (2022)\\ DOI: 10.1134/S1990341322030117} 
\pubyear{2022}
\begin{document}
\label{firstpage}
\pagerange{\pageref{firstpage}--\pageref{lastpage}}
\maketitle

\begin{abstract}
We study the radio variability of galaxies with and without sources of hydroxyl (OH) megamaser radiation based on the continuum radio measurements conducted in 2019--2022 with the radio telescope RATAN-600 at frequencies of 2.3, 4.7, 8.2, and 11.2 GHz. Presumably, radio continuum emission significantly affects the megamaser radiation brightness, therefore, such a characteristic as the variability of radio emission is important for determining the OHM galaxies parameters. With additional data from the literature, the parameters of radio variability on a time scale up to 30 years were estimated. The median values of the  variability index for 48 OHM galaxies are in the range $V_{S}=0.08$--$0.17$, and for 30 galaxies without OH emission they are $V_{S}=0.08$--$0.28$. For some individual galaxies in both samples, flux density variations reach 30--50\%. These sources either are commonly associated with AGNs or reveal active star formation. Generally, the variability of luminous infrared galaxies with and without OH megamaser emission is moderate and of the same order of magnitude on long time scales. From estimating the spectral energy distribution parameters in a broad frequency range (from MHz to THz), we determined the spectral index below 50~GHz and the color temperatures of dust components for megamaser and control sample galaxies. 
At a level of $\rho<0.05$, there are no statistically significant differences in the distribution of these parameters for the two samples, as well there are no statistically significant correlations between the dust color temperatures and the variability index or luminosity in the OH line.
\end{abstract}


\section{INTRODUCTION}

The galaxies hosting OH megamaser emission (OH megamaser galaxies -- OHMs) are radio sources with the enhancement of spectral lines corresponding to the quantum transitions between the sub-levels of hydroxyl molecule (OH) hyperfine structure at 
frequencies of 1665 and 1667 MHz \citep{1992AJ....103..728B}. Galaxies are classified as OHMs in the case of $L_{\rm OH}>10L_{\odot}$ \citep{2014A&A...570A.110Z}. Almost all known megamasers are located in luminous and ultra-luminous infrared galaxies (U/LIRGs) with luminosity in the far infrared range $L_{\rm FIR}$ exceeding $10^{11} L_{\odot}$. It is believed that their emission is generated due to strong tidal interaction during the merging of rich molecular gas galaxies~\cite{1996MNRAS.279..477C}. There is an opinion 
that the OH emission is characteristic of the evolutionary stage of formation of
quasars and massive elliptical galaxies~\citep{2011AJ....141..100H}.

The dominant energy source in the central parts of OHMs can be extensive star formation or radiation from the accreting supermassive black hole in an AGN, or a combination of these two processes \citep{2015A&A...574A...4V}. One of the main conditions of megamaser radiation is the presence of seed radio emission. By measuring the radio spectra parameters and variability for as many OHM galaxies as possible, we can determine the statistical radio properties of this class of objects and their possible difference from the properties of infrared galaxies without OHM emission sources. To date there are few systematic OHM  radio measurements due to their weak flux densities (units or tens of mJy) \citep{2005Ap.....48..237K,2005Ap.....48...99K}.  

In 2019--2021 with the radio telescope RATAN-600 we carried out one of the largest OHM radio continuum surveys. A sample of 74~OHM galaxies was observed at 
frequencies 1.2--22.3 GHz \citep{2022MNRAS.510.2495S}. 
The aim of the study was to determine the parameters of the sample of OHM galaxies in the radio domain and compare them with the same parameters obtained for a control sample of 128 infrared galaxies without megamaser emission~\citep{2014A&A...570A.110Z}.
As a result, statistical differences of radio properties in the two samples were obtained. We found significant correlations between the OH line and  radio continuum luminosities ($L_{\rm OH}$--$P_{1.4}$), between the OH line luminosities and the radio spectral index ($L_{\rm OH}$--$\alpha_{4.7}$), between the FIR and radio continuum luminosities ($L_{\rm FIR}-P_{1.4}$). Despite the fact that both samples were dominated by objects with steep radio spectra (53\% and 61\%), it was found that the sample of OHM galaxies contains 2 times more objects with flat spectra (32\%) than the control one (17\%).   

The results may indicate a significant impact of radio continuum emission on the intensity of megamaser radiation. Obviously, a significant proportion of AGNs in both samples (47\% and 30\% for OHM and non-OHM respectively) influenced the obtained properties. The purpose of this work is to estimate the radio flux density variability of infrared galaxies in 
both samples. The variability of radio emission was estimated using both the  multifrequency RATAN-600 measurements in 2019--2022 and the literature data. Light curves were analyzed for individual objects on time scales up to 34 years.

The objects with high frequency measurements (THz) were selected to assess the contribution of the synchrotron and dust components to the continuum radio spectra of galaxies. The spectral index of the low-frequency component was determined and estimates of dust color temperature were made. The estimates of the correlation of dust  color temperature with some characteristic parameters of megamaser radiation were obtained.

\section{Sample and observations}

Our target and control samples, containing 74~OHM galaxies and 128 galaxies without megamaser emission greater than 5~mJy at the 1.4~GHz frequency, are borrowed from \cite{2014A&A...570A.110Z}. This list of OHM galaxies had been the most complete at the time of their study, also the authors collected a control sample that included all Arecibo survey galaxies without detected megamaser emission \citep{2002AJ....124..100D}.  
The observations were conducted in 2019--2022 with the radio telescope RATAN-600 at 1.2, 2.3, 4.7, 8.2, 11.2, and 22.3 GHz. Flux densities measured in 2019--2021 and 
the radio properties of the target and control samples are published in  \cite{2022MNRAS.510.2495S}. 
In total, with RATAN-600 we managed to detect 63 OHMs and 35~galaxies of 
the control sample, noticing  the low detection level (4--75\% at different frequencies).

\section{Variability}

To estimate variability, we used both the RATAN-600 measurements obtained in 2019--2022 and the literature data from CATS\footnote{https://www.sao.ru/cats}~\citep{1997BaltA...6..275V,2005BSAO...58..118V},  NED,\!\footnote{https://ned.ipac.caltech.edu} and  
VizieR\!\footnote{https://vizier.u-strasbg.fr/viz-bin/VizieR}. For 61 OHMs and 56 galaxies of the control sample, there are radio continuum measurements in the literature
for at least one observing epoch and at frequencies within $\pm$20\% of the \mbox{RATAN-600} frequencies. 

As a quantitative estimate of variability, we used the variability index $V_{S}$ from \cite{1992ApJ...399...16A}:
\begin{equation}
\label{eq:Var}
V_{S}=\frac{(S_{\rm max}-\sigma_{S_{\rm max}})-(S_{\rm min}+\sigma_{S_{\rm min}})}
{(S_{\rm max}-\sigma_{S_{\rm max}})+(S_{\rm min}+\sigma_{S_{\rm min}})} ,
\end{equation}
where $S_{\rm max}$ and $S_{\rm min}$ are the maximum and minimum values of flux density
for all epochs of observations; $\sigma_{S_{\rm max}}$ and $\sigma_{S_{\rm min}}$
are their errors. Thus we can partially exclude the influence of flux density measurement uncertainties that could cause overestimation of the variability index. The negative value of $V_{S}$ corresponds to the case when the errors of measured flux density are greater than their scatter, which makes impossible to detect object's variability.

Uncertainties of variability indices were estimated as:
\begin{equation}
\label{eq:deltaVar}
\Delta V_{S}=\frac{2S_{\rm min}(\sigma_{S_{\rm min}}+\sigma_{S_{\rm max}})}
{(S_{\rm min}+S_{\rm max})^{2}} .
\end{equation}

The modulation index, defined as the standard deviation of flux
density divided by the mean flux density, was calculated as in \cite{2003A&A...401..161K}:
\begin{equation}
\label{eq:mod}
M=\frac{\sqrt{\cfrac{1}{N}\displaystyle\sum_{i=1}^N\left(S_i-\cfrac{1}{N}\displaystyle\sum_{i=1}^N{S_i}\right)^2}}{\cfrac{1}{N}\displaystyle\sum_{i=1}^N{S_i}} ,
\end{equation}
where $S_{\rm i}$ is the $i$-th flux density, $N$ is the number of observations. 

The calculated variability indices at frequencies of 2.3, 4.7, and 8.2 GHz for the OHM and control sample galaxies are presented in Table \ref{tab:OHM_var} and Table \ref{tab:control_var}.

For 48 OHM galaxies and 30 control sample galaxies, there are at least two measurements at least at one frequency. The statistics of variability indices for them are presented in
Table \ref{tab:stat}. Only positive values of $V_{S}$ were used to calculate the mean and median values of the variability indices. On average, at a frequency of 4.7 GHz, which represents the largest number of sources with available measurements,
variability of 15--20\% for the sample of megamasers and 10--15\% for the control sample is obtained. When considering galaxies identified as having AGNs in both samples, it is obvious that their variability is higher, but it also lies within 15--20\% for 
the OHM galaxies, and for a few galaxies in the control sample the spread 
is slightly wider: 12--22\% (Table \ref{tab:stat}).

The most variable source in the control sample, J0854+2006, is identified with 
a blazar in~\citet{2010ApJ...715..429A} ($V_{S_{2.3}}=0.86\,(0.02)$, $V_{S_{4.7}}=0.81\,(0.03)$, $V_{S_{8.2}}=0.82\,(0.02)$), it also has the biggest number of measurements ($N=776$--$2830$). The most variable galaxies among OHMs are: the AGN \citep{2011ApJ...743..171A} NGC 253 (J0047$-$2517 with $V_{S_{2.3}}=0.34\,(0.04)$, $V_{S_{8.2}}=0.35\,(0.05)$) and 
the Seyfert 2 galaxy \citep{2006A&A...455..773V} IRAS 15065$-$1107 (J1509$-$1119 with $V_{S_{4.7}}=0.49\,(0.06)$). 

\begin{figure*}
\centering
\begin{tabular}{cc}
\begin{minipage}{0.5\linewidth}
\center{\includegraphics[width=\textwidth]{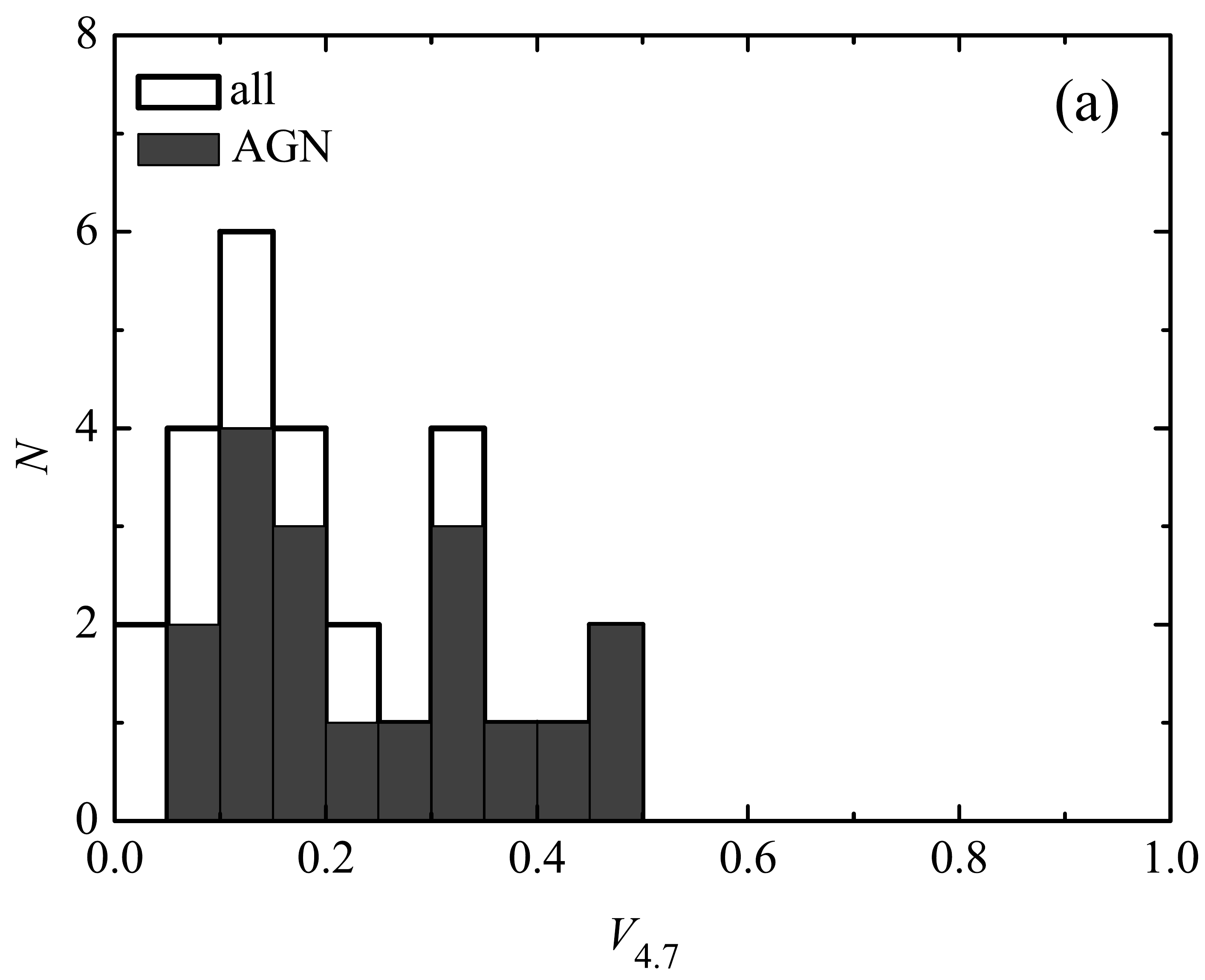}} 
\end{minipage}
\hfill
\begin{minipage}{0.5\linewidth}
\center{\includegraphics[width=\textwidth]{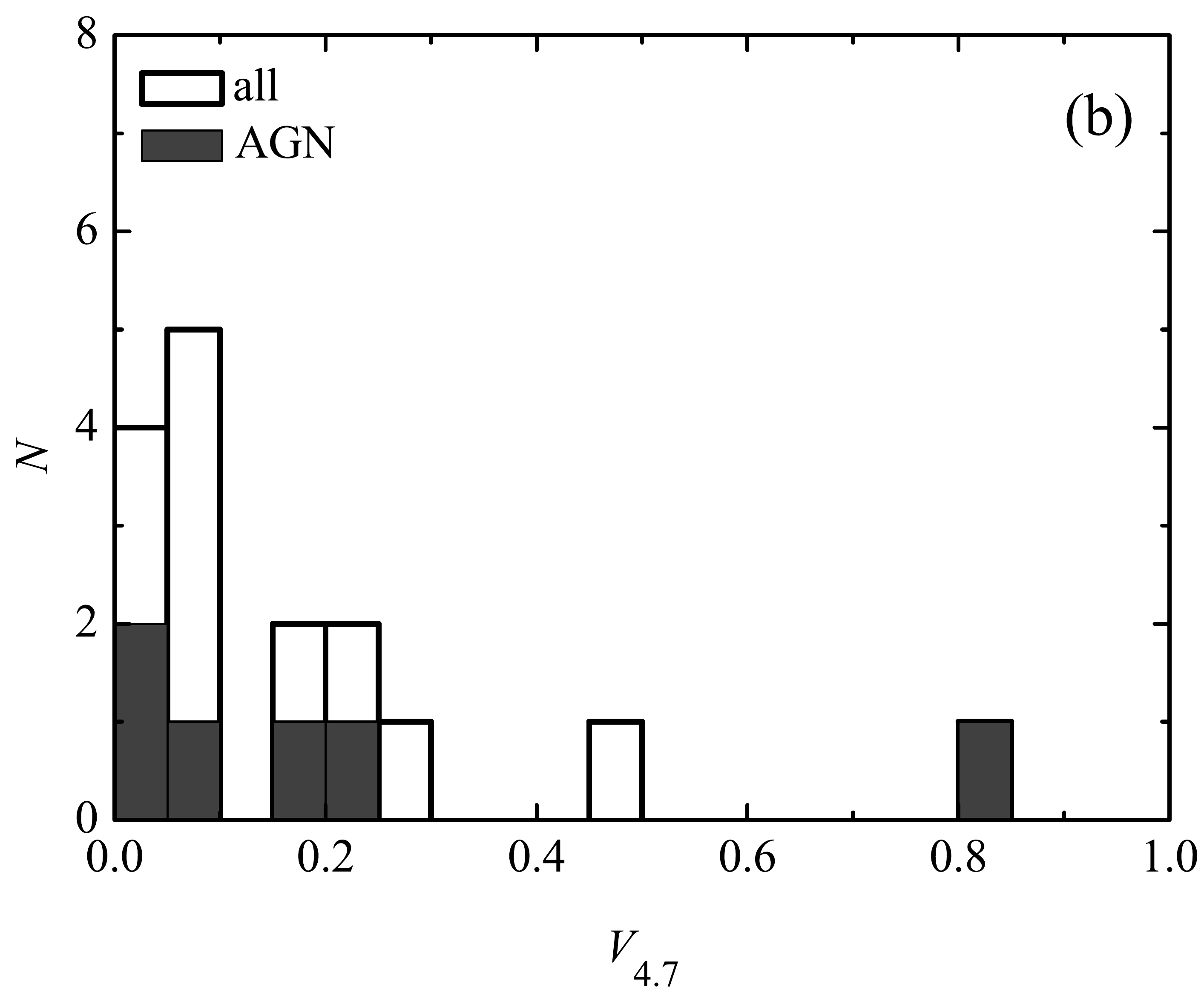}}  
\end{minipage}
\end{tabular}
\caption{The distribution of variability indices at 4.7 GHz for the (a) OHM and (b) control samples. AGNs are shown with the  dark grey color.}
\label{fig:var.distr}
\end{figure*}

\begin{figure*}
\centering
\begin{tabular}{lr}
\begin{minipage}{0.5\linewidth}
\center{\includegraphics[width=\textwidth]{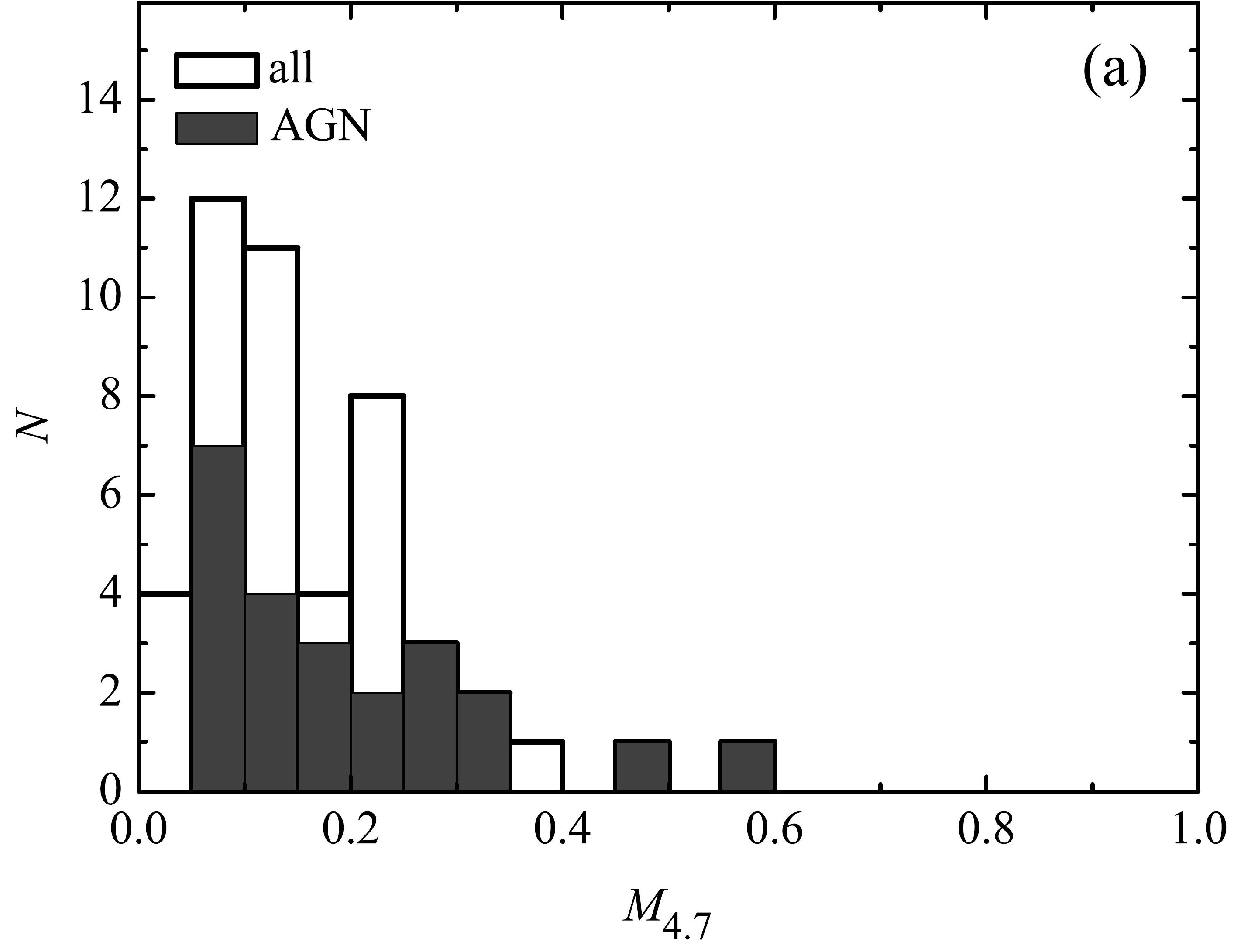}} 
\end{minipage}
\hfill
\begin{minipage}{0.5\linewidth}
\center{\includegraphics[width=\textwidth]{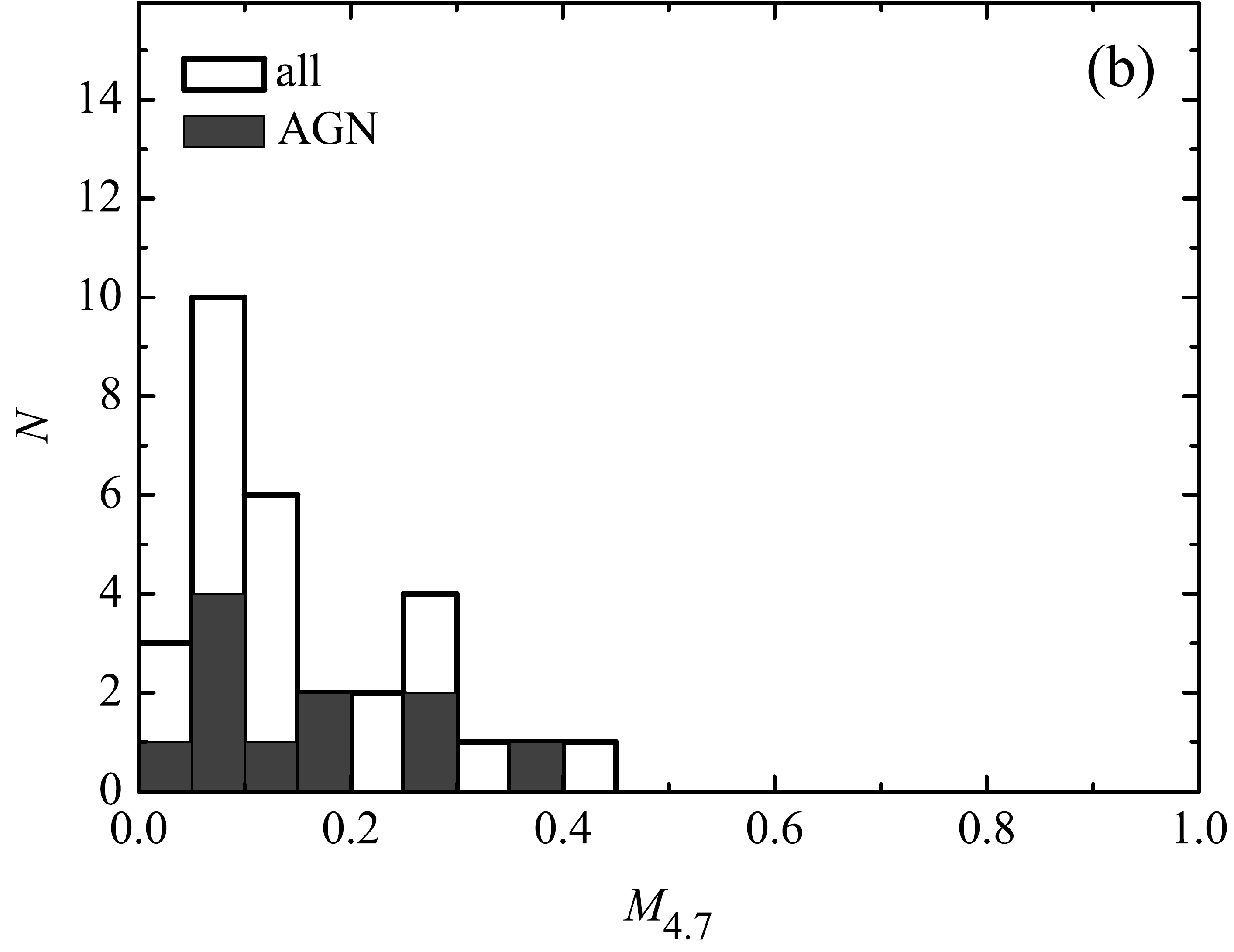}}  
\end{minipage}
\end{tabular}
\caption{The distribution of modulation indices at 4.7 GHz for the (a) OHM and (b) control samples. AGNs are shown with the dark grey color.}
\label{fig:mod.distr}
\end{figure*}

The distributions of variability and modulation indices at 4.7~GHz for megamaser and control sample galaxies are shown in Fig.\ref{fig:var.distr} and Fig.\ref{fig:mod.distr}.

We have not found any statistically significant (with p-value\,$<\!0.05$) 
correlations between the modulation and variability indices at frequencies of 4.7 and 8.2~GHz  and other parameters such as the luminosity in the $L_{\rm OH}$ hydroxyl line, luminosity in the far infrared region $L_{\rm FIR}$, spectral index in the FIR, radio loudness, $q$~parameter, and radio luminosity at 1.4 GHz for both the OHM and control samples.

The distribution of variability and modulation indices at 4.7 and 8.2 GHz in both samples does not differ statistically according to the Kolmogorov--Smirnov and Mann--Whitney tests ($\rho<0.05$). Also we found that variability and modulation indices belong to the same distribution according to the Kruskal--Wallis test ($\rho<0.05$).

It is obvious that to assess the relationship between the activity of the host galaxy nucleus and the megamaser emission (or its variability), simultaneous measurements are essential. In this work we analyze individual measurements of 
integrated radio flux density, which does not allow us to estimate the possible correlation of such events. 

\begin{table}
\caption{\label{tab:stat} 
The median and mean values of the variability indices at different frequencies for the OHM and control samples. The left part of the table is for the full sample, and the right part is for the AGN subsample.}
\centering
\begin{tabular}{|l|c|c|c|c|c|c|}
\hline
     & $N$ & median  & mean  & $N$ & median  & mean  \\
\hline 
\multicolumn{4}{|c|}{OHM} & \multicolumn{3}{|c|}{OHM-AGN}\\
\hline
$M_{2.3}$      & 9  & 0.08  & 0.12 (0.08) & 8  & 0.12  & 0.12 (0.08) \\
$V_{S_{2.3}}$  & 6  & 0.17  & 0.19 (0.11) & 6  & 0.17  & 0.19 (0.11) \\
\hline
$M_{4.7}$      & 47  & 0.13  & 0.16 (0.12) & 23  & 0.17  & 0.20 (0.13)\\
$V_{S_{4.7}}$  & 27  & 0.17  & 0.20 (0.14) & 18  & 0.20  & 0.25 (0.13)\\
\hline
$M_{8.2}$      & 29  & 0.15  & 0.17 (0.13) & 18  & 0.15  & 0.20 (0.14)\\
$V_{S_{8.2}}$  & 13  & 0.08  & 0.13 (0.13) & 9   & 0.16  & 0.17 (0.14)\\
\hline
\multicolumn{4}{|c|}{Control} & \multicolumn{3}{|c|}{Control-AGN}\\
\hline
$M_{2.3}$      & 5  & 0.08  & 0.11 (0.12) & 2  & 0.16  & 0.16 (0.23)\\
$V_{S_{2.3}}$  & 5  & 0.22  & 0.24 (0.26) & 2  & 0.34  & 0.34 (0.48)\\
\hline
$M_{4.7}$      & 30  & 0.11  & 0.15 (0.11) & 11  & 0.13  & 0.16 (0.11)\\
$V_{S_{4.7}}$  & 16  & 0.08  & 0.17 (0.21) & 6  & 0.12  & 0.22 (0.31)\\
\hline
$M_{8.2}$      & 11  & 0.14  & 0.18 (0.13) & 6  & 0.12  & 0.20 (0.16)\\
$V_{S_{8.2}}$  & 5   & 0.28  & 0.38 (0.32) & 2  & 0.53  & 0.53 (0.42)\\
\hline
\end{tabular}
\end{table}


\subsection{Light curves}

\begin{figure*}
\centering
\includegraphics[width=1\textwidth]{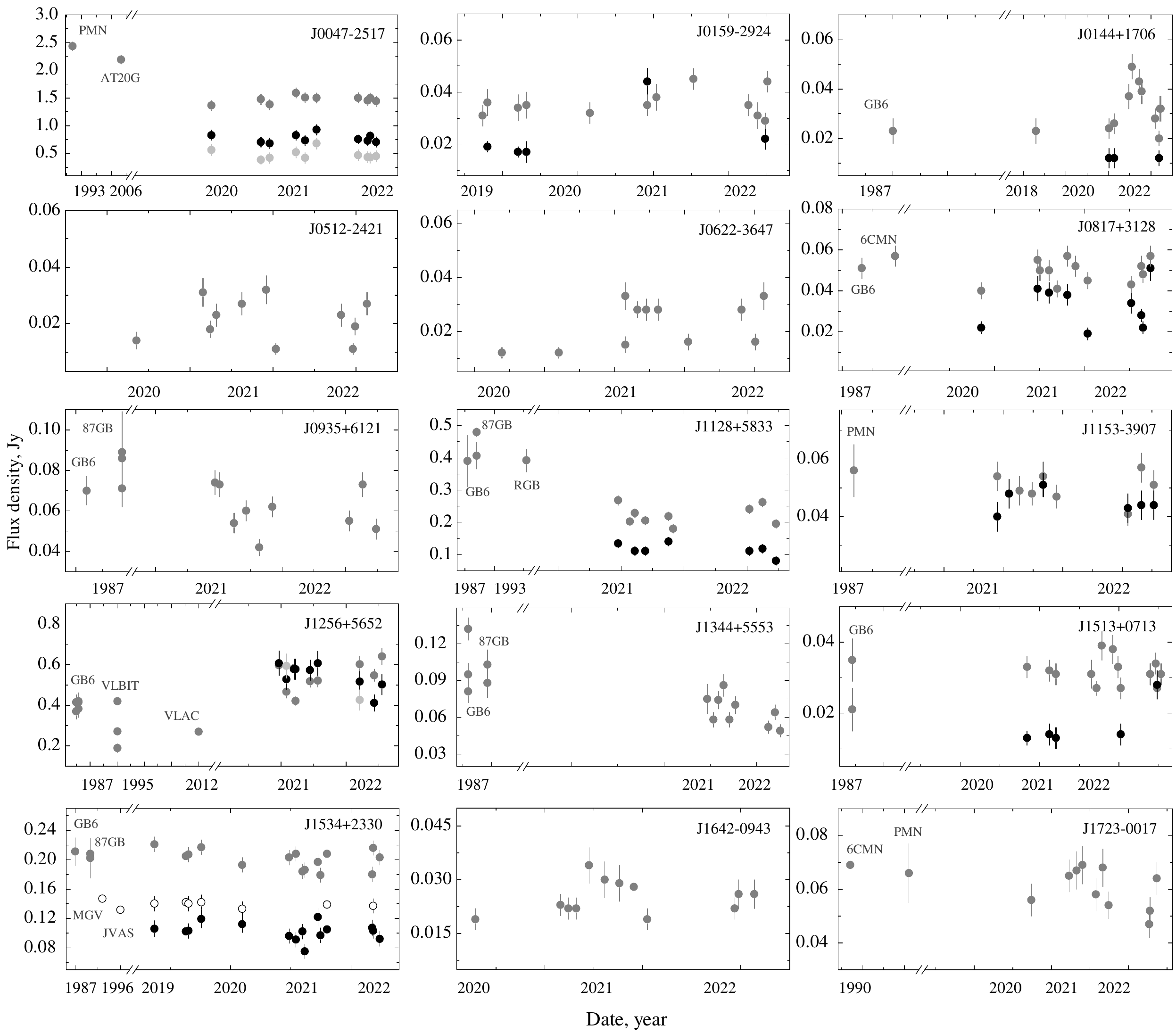}
\caption{Light curves of the galaxies at frequencies of 22.3 (light gray), 11.2 (black), and 4.7 GHz (dark grey) obtained with RATAN-600 in 2019--2022 and from the literature.}
\label{var1}
\end{figure*}

For 14 OHM galaxies we estimated radio variability near a frequency of 5 GHz  
on a time scale of 2--34 yrs, and near a frequency of 11.2 GHz on a scale of 1--9 yrs. We used available literature data provided by the GB6 \citep{1996ApJS..103..427G}, 87GB \citep{1991ApJS...75.1011G}, PMN \citep{1993AJ....105.1666G}, RGB \citep{1997A&AS..122..235L}, VLBIT \citep{1996ApJS..107...37T}, VLAC (Taylor Very Large Array Calibration Source List), JVAS \citep{1998MNRAS.300..790W}, AT20G \citep{2010MNRAS.402.2403M}, and MGV (Identifications of Radio Sources in the MG-VLA Survey) catalogs. The light curves are presented in Fig.\ref{var1}. The classification of the galaxies, number of observing epochs $N$, time scale $t$, radio morphology, and references are presented in Table~\ref{tab:var}. For eight galaxies the variability index reaches values of 0.20--0.49 at 5 GHz. Seven of them are identified as AGNs according to \cite{2006A&A...455..773V}. Information about angular structure is available for a few objects.

For four galaxies, J0159$-$2924, J0512$-$2421, J0622$-$3647, and J1642$-$0943, 
the variability of integrated flux density was estimated based only on the RATAN measurements at 5 GHz in 2019--2022 with $V_{S_{4.7}}$ equal to 0.12, 0.35, 0.33, and 0.14 respectively. For the galaxies J0817$+$3128, J1153$+$3905, J1534$+$2330, and J1723$-$001, the variability was calculated on
a long time scale with a rare number of observations, and $V_{S_{4.7}}$ does not exceed 0.04--0.09. And one galaxy, J1513$+$0713, with the ``core'' type morphology turned out to be low-variable, $V_{S_{4.7}}=0.13$ (15~measurements over 34~years).

One of the brightest objects, J0047-2517 (Table~\ref{tab:var}), according to the VLA \citep{2006AJ....132.1333L} has a complex radio morphology with several dozen compact components which can be identified with supernovae, supernova remnants, or HII regions~\citep{1991AJ....102..875U}. In one of the components a flux density variability of 6.5\%($\pm2.5\%$) was found at 4.7 GHz over a time interval of 18 months in 1989--1990. According to the VLBI measurements at 2.3~GHz, the central region of the galaxy has a core$+$jet type morphology (\textit{astrogeo}\footnote{https://http://astrogeo.org/vlbi\_images/}, an angular resolution of about 5 mas). We estimated (Fig.~\ref{var1}) the variability $V_{S_{4.7}}=0.25\pm0.02$ on a time scale of 28~years 
(12~observing epochs) and an insignificant variability $V_{S_{11.2}}=0.03\pm0.004$ over 2 years (10~epochs).

The galaxy Arp 299 (J1128$+$5833) with an active nucleus is one of the most variable in our sample, $V_{S_{4.7}}=0.42\,(0.02)$ on a 34-year scale (Fig.~\ref{var1}) and $V_{S_{11.2}}=0.18\,(0.03)$ on a 2-year scale. The VLA measurements \citep{2011MNRAS.415.2688R} revealed variable radio emission from one of the compact components at 8.4 GHz with a variability index of about 0.22 in 1990--2006 on a time scale of more than 10 years.

The galaxy J0144$+$1706 (III Zw 35) has a similar angular structure with many compact components. The radio continuum VLA and MERLIN measurements \citep{1990MNRAS.244..281C,2001A&A...377..413P} at 18 cm  in 1986--1998 revealed no change in spectral flux density. We found variations in radio emission $V_{S_{4.7}}=0.31\pm0.06$ in the period of 1987--2022. The most variable galaxy with $V_{S_{4.7}}=0.49\pm0.06$ is J1256$+$5652 with a resolved core and a 2.3~GHz two-side jet.

We estimated the variability of individual galaxies using low-cadence measurements. Due to this reason the obtained results most likely correspond to the lower bound of variability. The integral flux density variability of OHM galaxies results from the contributions of an active nucleus and the variability of compact regions (supernova remnants or regions of ionized hydrogen HII), which are potential megamaser emission sources \citep{2001A&A...377..413P}. VLBI measurements of compact components show that radio emission variations in them can reach a few dozen per cent on scales of several years. Spectral flux density in such regions can be an order of magnitude less than the integral flux density of the host galaxy. Obviously, in the case of RATAN-600 measurements we observe the dominant contribution of a variable radio nucleus.

\begin{table*}
\centering
\caption{\label{tab:var} The variability index $V_{S}$ of the OHM galaxies calculated using the RATAN-600 and literature data.}
\begin{tabular}{|c|c|c|c|c|c|c|c|c|c|}
\hline
 NVSS name & t$_{5}$, & N$_{4.7}$, & $V_{S_{4.7}}$ & t$_{11.2}$, & N$_{11.2}$, & $V_{S_{11.2}}$ & type & morph & reference\\
      & years    &  epochs &     &  years   & epochs &  &  & &\\
\hline
004733$-$251717 &  28 & 12 & 0.25 (0.02) & 2 & 10 & 0.03 (0.004) & AGN & core+jet & \cite{1991AJ....102..875U}\\
 &   &  &  &  &  &  &  &  & \cite{2006AJ....132.1333L}\\
014439$+$170608 & 29 & 12 & 0.31 (0.06) & -- & -- & -- & AGN & diffused & \text{\cite{2001A&A...377..413P}} \\
015913$-$292435 &  3 & 12 & 0.12 (0.02) & -- & -- & -- & AGN & -- & \cite{2019ApJS..241...19I} \\
051209$-$242156 &  2 & 11 & 0.35 (0.08) & -- & -- & -- & AGN & -- & \cite{1996Ap.....39..237K} \\
062222$-$364742 &  2 & 11 & 0.33 (0.07) & -- & -- & -- &  & -- & \cite{1996Ap.....39..237K}\\
 &   &  &  &  &  &  &  &  & \text{\cite{2011A&A...525A..91T}}\\
081755$+$312827 &  34 & 14 & 0.08 (0.01) & 2 & 9 & 0.34 (0.07) & AGN & -- & -- \\
093551$+$612112 &  34 & 13 & 0.20 (0.05) & -- & -- & -- & AGN & core & \textit{astrogeo} \\
112832$+$583346 &  34 & 13 & 0.42 (0.02) & 1 & 7 & 0.18 (0.03) & AGN & core+jet & \text{\cite{2010A&A...519L...5P}} \\
 &   &  &  &  &  &  &  &  & \cite{2011MNRAS.415.2688R}\\
 &   &  &  &  &  &  &  &  & \text{\cite{2011A&A...525A..91T}}\\
115311$+$390748 & 28 & 10 & 0.07 (0.01) & 1 & 9 & 0.02 (0.002) & & & \\
125614$+$565223 & 34 & 18 & 0.49 (0.06) & 1 & 9 & 0.10 (0.01) & AGN & core+double jet & \textit{astrogeo} \\
134442$+$555313 & 34 & 14 & 0.39 (0.05) & -- & -- & --  & AGN & multi-components & \cite{2005MNRAS.361..748B} \\
151313$+$071331 & 34 & 15 & 0.13 (0.04) & 2 & 5 & 0.20 (0.05) & & core & \textit{astrogeo} \\
153457$+$233011 & 34 & 18 & 0.04 (0.003) & 9 & 30 & -0.04 & AGN & core & \textit{astrogeo} \\
164240$-$094315 & 2 & 12 & 0.14 (0.03) & -- & -- & -- & AGN & resolved & EVN archive \\
172321$-$001702 & 32 & 12 & 0.09 (0.01) & 2 & 4 & 0.31 (0.07) & AGN & diffused & \cite{2006ApJ...653.1172M} \\
\hline
\end{tabular}
\end{table*}

\section{Modeling the Spectral Energy Distribution of OHM Galaxies}

\subsection{Problem formulation---an overview}

Continuum radiation from galaxies containing hydroxyl megamasers is formed
due to several mechanisms that dominate in certain frequency ranges. At low frequencies (up to 30--50~GHz), the main contribution is made by the non-thermal mechanisms, in particular by
synchrotron radiation. An additional contribution in this frequency range is provided by the free--free ({\it Bremsstrahlung}) emission of electrons in the field of positive ions, which is an example of the thermal mechanism.

At higher frequencies above 100 GHz and up to about 10~THz, the main contribution is made by the thermal radiation of various dust fractions. \cite{2008MNRAS.388.1595D} suggested a model that included two dust components. One of them has temperature in the range of 15--60~K
and gives the main contribution to the far infrared emission, while the other one with a temperature of 130--250~K makes the main contribution to the mid-IR range. Since this simplified model allows us to obtain good approximations of the observed IR spectral energy distributions, we decided to use it in our study.

Finally, at mid-IR frequencies strong emission details between 3.3 and $12.7$~$\mu$m (from 20 to 90~THz) are observed in the spectral energy distribution (SED). It is assumed that this radiation refers to the molecules of the polycyclic aromatic hydrocarbons (PAH~\citep{2009ARA&A..47..427H}), the levels of which are excited by the absorption of individual UV photons of the interstellar radiation field into which PAH molecules are immersed~\citep{2008MNRAS.388.1595D}.

It should be noted that the main contribution to the galaxy radio continuum and IR emission is provided by parts of the galaxy radically different in terms of their location, which makes it possible to search for parameters of the non-thermal and thermal radiation independently. In the case of observations with a wide (in comparison with the angular size of an object) beam pattern (BP), we can talk only about
spectral indices and dust color temperatures averaged over the BP. At the same time, a good result is shown by fairly simple multicomponent models that do not take into account the effects of radiation transfer.  

In general, it is of interest to model the contribution of all these components to the obtained spectra of objects using both low-frequency observations on RATAN-600 and the literature data from different catalogs. However, in this paper the simulation was limited only to 
the synchrotron and free--free components at frequencies below 50~GHz and the emission of ``cold'' and ``warm'' dust fractions at higher frequencies up to tens of THz. To model the contribution of PAH molecules, as a rule more measurements are required at frequencies above 10~THz (see, e.g.,~\citealt{2011ApJS..193...18W}).

At frequencies below 50~GHz, the total contribution of the synchrotron and free--free components was approximated by a power law
\begin{equation}
    S_{\nu} = A\cdot\nu^{\beta} ,
\end{equation}
where $S_{\nu}$ is the flux density, $\nu$ is the frequency, $\beta$ is the power law index, which is mainly determined by the synchrotron component emission, and $A$ is a scale coefficient.

The emission of dust at frequencies of 100~GHz--10~THz was approximated using a model where two components of dust emit: ``cold'' with a temperature of up to 40~K and ``warm'' with a temperature of up to 200~K \citep{1999ApJ...524..867F},
\begin{equation}
    S_{\nu} = w_1 C \nu^\alpha B(\nu, T_1) + w_2 C \nu^\alpha B(\nu, T_2) ,
\end{equation}
where $\alpha$ is the emissivity of dust, $T_1$ and $T_2$ are the physical temperature of the two dust components in Kelvin, $B(\nu,T)$ is
the Planck's blackbody radiation law, $C$ is the scaling factor, $w_1$ and $w_2 = 1 - w_1$ are the component weights normalized by one. As a rule, the emissivity of dust varies in the range of $\alpha=1.5$--$2.5$; however, to reduce the number of free parameters, this coefficient was fixed with the value of $\alpha=2$. It should be noted that this approach is a very simplified description of the radiation model of an ensemble of dust particles having a continuous temperature distribution in a certain range, which, however, gives acceptable results in the SED modeling.

The main task of the continuum SED modeling was to obtain the key model parameters ($\beta$, $T_1$, and $T_2$) for the objects, evaluate their statistical properties, and compare them in the megamaser and control samples.
It is worth noting that the studied samples are quite small, and these estimates may be biased.

\subsection{Data used for the SED modelling}

In addition to the proprietary flux density data obtained at the radio telescope \mbox{RATAN-600}, the catalogs and databases listed in 
Table~\ref{tab:Catalogs} were used to construct spectra in a broad
range from low frequencies to the mid-IR wavelengths.

\begin{table*}
\caption{\label{tab:Catalogs}
 A list of the databases and catalogs with flux density data used to construct the OHM spectra in a broad frequency range from megahertz to IR. The procedure from \protect\cite{2012wise.rept....1C} was used to convert the photometric stellar magnitudes of the ALLWISE catalog into the flux units (Jy).}
\begin{center}
\begin{tabular}{ |l|l|l|l| } 
 \hline
 Name of the catalog  & Frequency       & URL                       & References            \\ 
 or database               & bands, GHz      &                           &                             \\
 \hline
 CATS                & < 30            & \url{http://cats.sao.ru}  &  \cite{1997BaltA...6..275V}, \cite{2005BSAO...58..118V} \\ 
 IRSA                & > 30            & \url{https://irsa.ipac.caltech.edu/frontpage/}  & \cite{2008SPIE.7016E..18B} \\
 Planck PCCS2 LFI    & 30, 44, 70      & \url{http://pla.esac.esa.int/pla/#catalogues}   & \cite{2016AA...594A..26P} \\ 
 Planck PCCS2 HFI    & 100, 143, 217,  & \url{http://pla.esac.esa.int/pla/#catalogues}   & \cite{2016AA...594A..26P} \\
                     & 545, 857        &                                                 & \\
 Herschel HPPSC      & 1874, 2998, 4283 & \url{https://www.cosmos.esa.int/web/herschel/} & \cite{2017arXiv170505693M} \\ 
                     &                  & \url{pacs-point-source-catalogue}              & \\
 Herschel SPSC       & 600, 857, 1199   & \url{https://www.cosmos.esa.int/web/herschel/} & \cite{2017arXiv170600448S} \\ 
                     &                  & \url{spire-point-source-catalogue}             &  \\
 Akari               & 16655, 33310     & \url{https://irsa.ipac.caltech.edu/Missions/}  & \cite{2010AA...514A...1I} \\ 
                     &                  & \url{akari.html}                               & \\
 AllWISE             & 13636, 25000,    & \url{https://irsa.ipac.caltech.edu/data/download/} & \cite{2010AJ....140.1868W}, \cite{2012wise.rept....1C} \\ 
                     & 65217, 88235     & \url{wise-allwise/}                            & \\
 \hline
\end{tabular}
\end{center}
\end{table*}

Although there are some flux data in the near and middle IR range 
(mainly from the WISE catalog~\citep{2010AJ....140.1868W})
for all 74 megamasers from the list under study, in the sub-millimeter and far infrared range the data were found for only 28 objects (38\% of the list).
In the control sample, the contribution of the sub-mm and IR components turned out to be even smaller, and the spectral energy distribution
in this range was approximated only for 10 objects out of 128 (8\%).

\subsection{Main results and parameter statistics}

Modeling of the SED was carried out for 28 megamasers and 10 objects from the control sample where maser radiation was not detected. With a few exceptions, 
the low-frequency spectral indices $\beta$ are in the range from $-1.0$ to $-0.2$, which illustrates the variants of the superposition of the synchrotron and free-free components for different cases. The color temperatures 
for the dust component lie in the intervals $T_1 = 5$--$33$~K and $T_2 = 40$--$160$~K. Notice that the contribution of ``warm'' dust, despite 
an insignificant weight $w$, has a significant effect on the shape of the spectrum. The parameters for all the sources for which the simulation was performed are given in Table~\ref{tab:SED_models}. The weighted average color temperature $T_{\rm aver} = w_1\cdot T_1 + w_2\cdot T_2$, where $w_1+w_2 = 1$, is also included in this table.

\begin{figure}
   \centering
       \subfloat[]{\includegraphics[width=1.0\columnwidth]{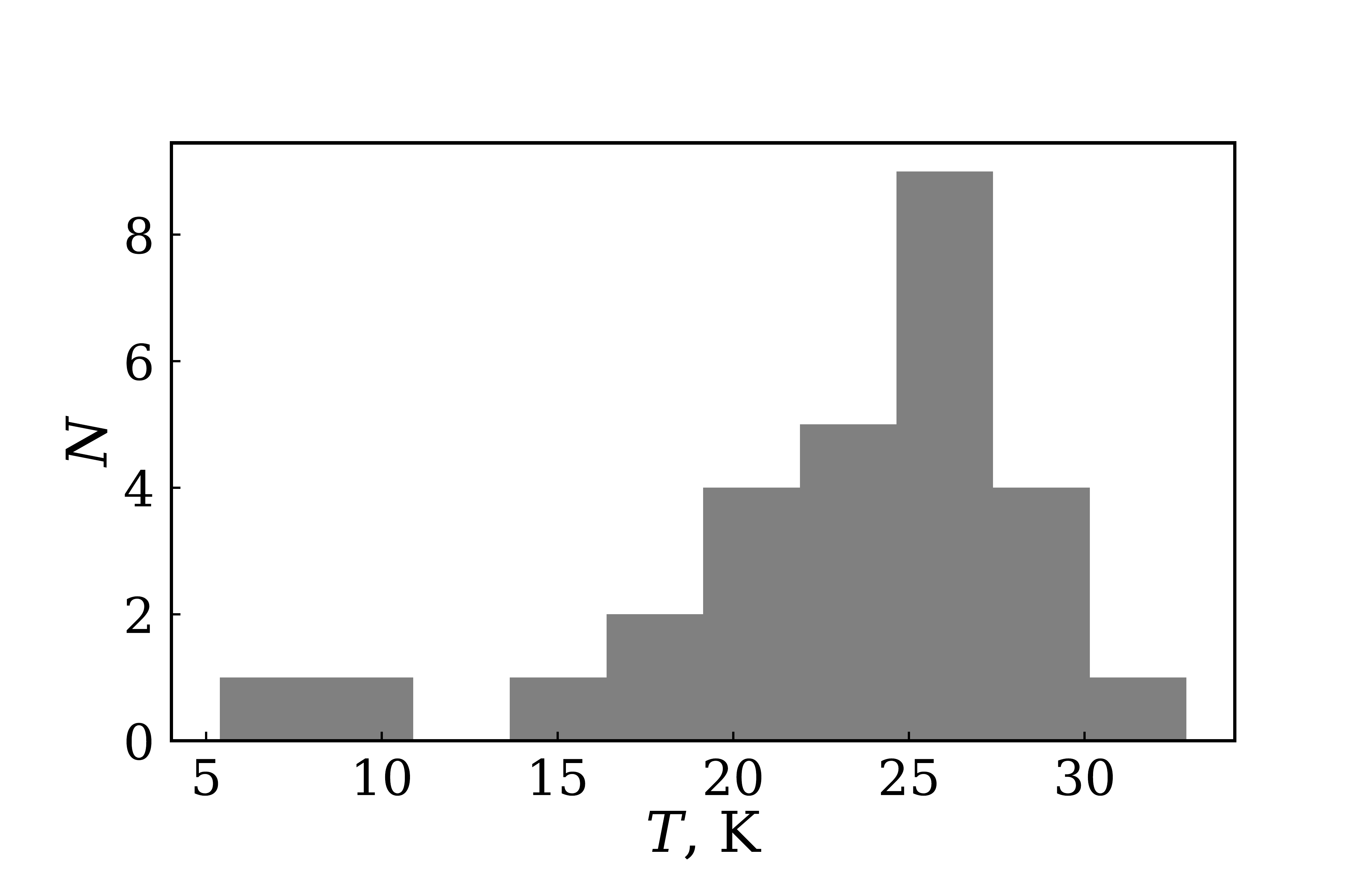}} \\ [-3ex]
        \subfloat[]{\includegraphics[width=1.0\columnwidth]{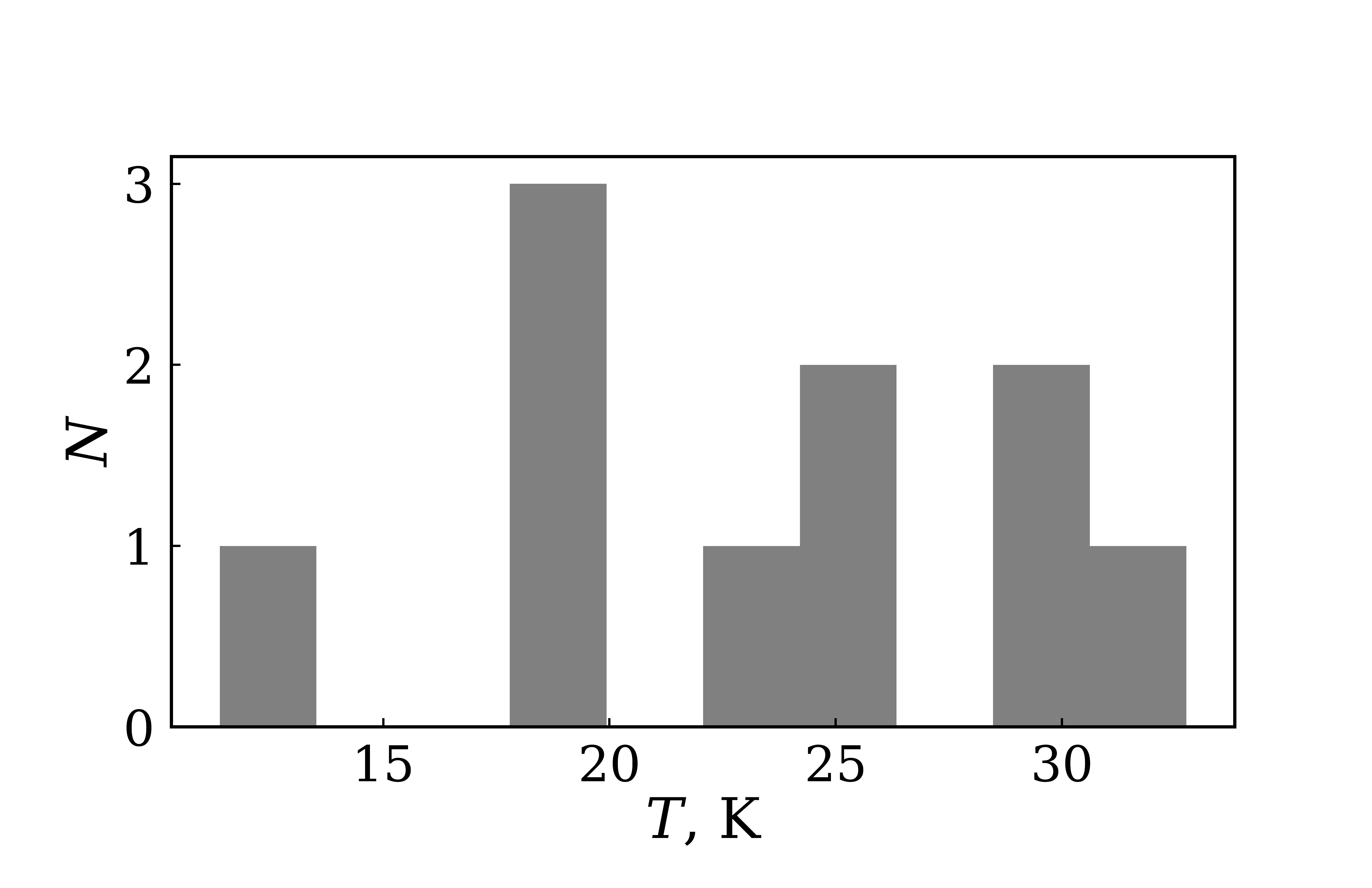}} 
\caption{Distribution of dust temperature values $T_1$: a) the sample of megamasers; b) the control sample.}
\label{ris:SED_T1}
\end{figure}

The distribution of the ``cold'' dust color temperature $T_1$ for the OHM and control samples is shown in Fig.~\ref{ris:SED_T1}. The average values for the two dust components are $\langle T_1 \rangle = 23.2\pm 5.9$~K, $\langle T_2\rangle = 60.9\pm 24.6$~K for megamasers and $\langle T_1\rangle = 23.1\pm 6.2$~K, $\langle T_2\rangle = 73.1 \pm 45.6$~K for the control sample.

The calculations performed do not show significant correlation between the dust color temperature $T_1$ in the sample of megamasers and such parameters as the
variability index $V_{S_{4.7}}$ or the luminosity in the OH line $L_{\rm OH}$. Also, no significant correlation was found between the synchrotron spectral index $\beta$ and $T_1$ (the Pearson correlation coefficient $r=0.22$, p-value~$=0.27$).

The Pearson correlation coefficient for the sample of megamasers between the parameters $T_1$ and $T_2$ themselves is 0.4 (p-value~$ = 0.03$), which is not surprising, since with an increase in the average temperature of the ensemble of dust particles the color temperature should increase for all components of the model.

The Kolmogorov--Smirnov test for the two samples of $T_1$, megamasers and control, gives a value of $ks=0.23$ (p-value~$=0.75$), which makes it impossible to talk about statistical differences between these two distributions.

\begin{figure*}
    \centering
        \subfloat[]{\includegraphics[width=0.8\columnwidth]{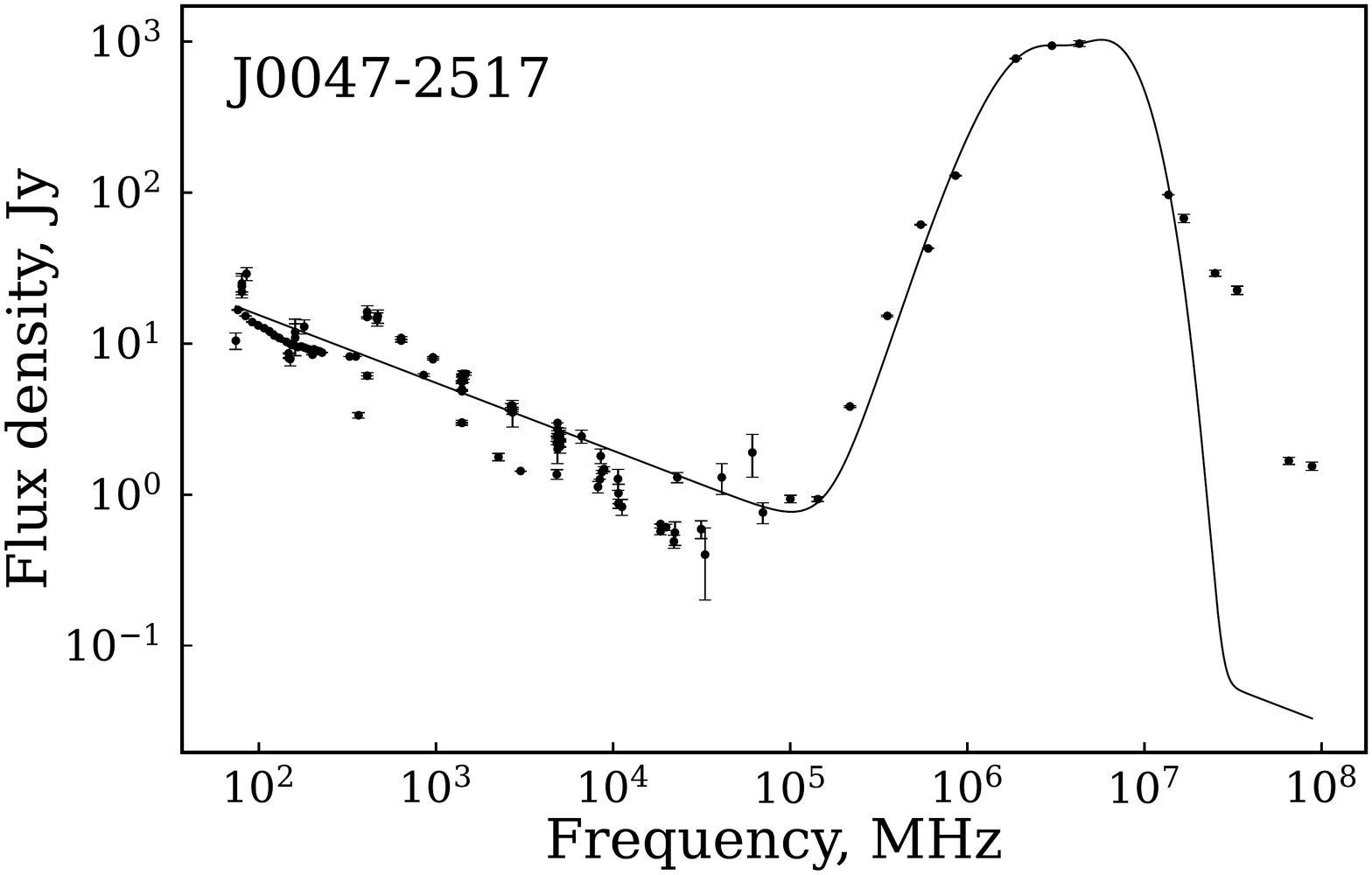}}
        \subfloat[]{\includegraphics[width=0.8\columnwidth]{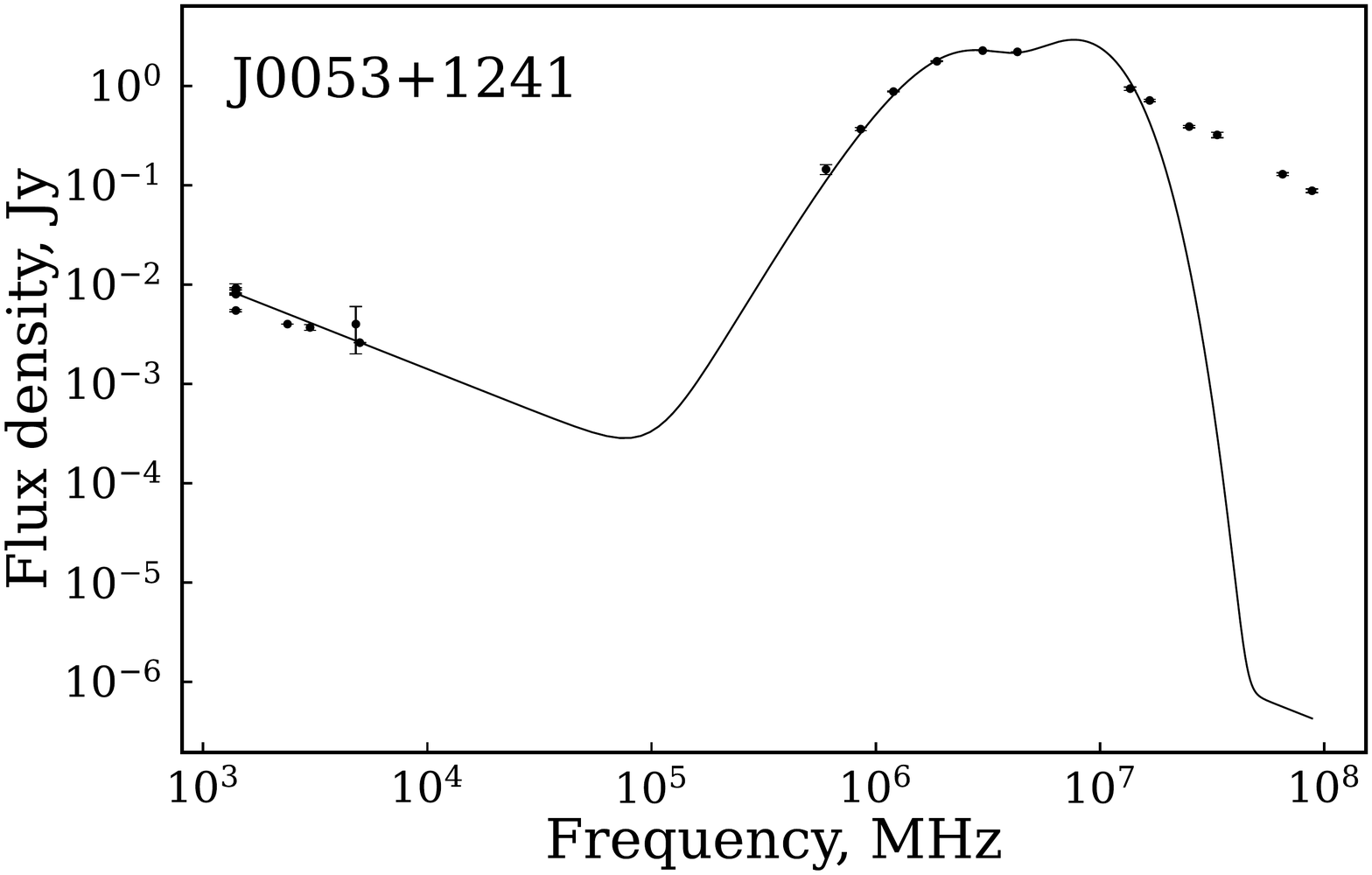}} \\ [-3ex]
        \subfloat[]{\includegraphics[width=0.8\columnwidth]{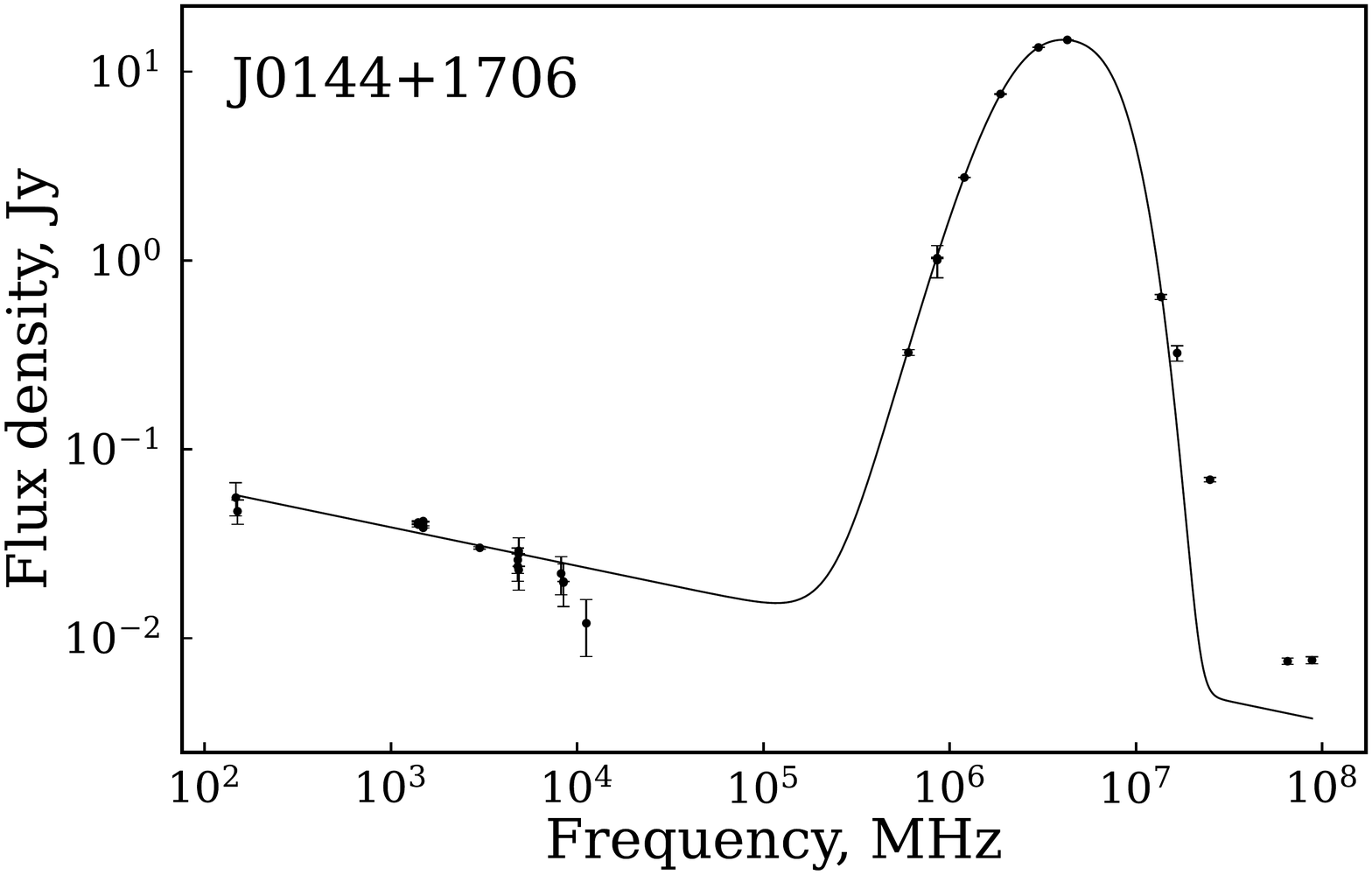}}
        \subfloat[]{\includegraphics[width=0.8\columnwidth]{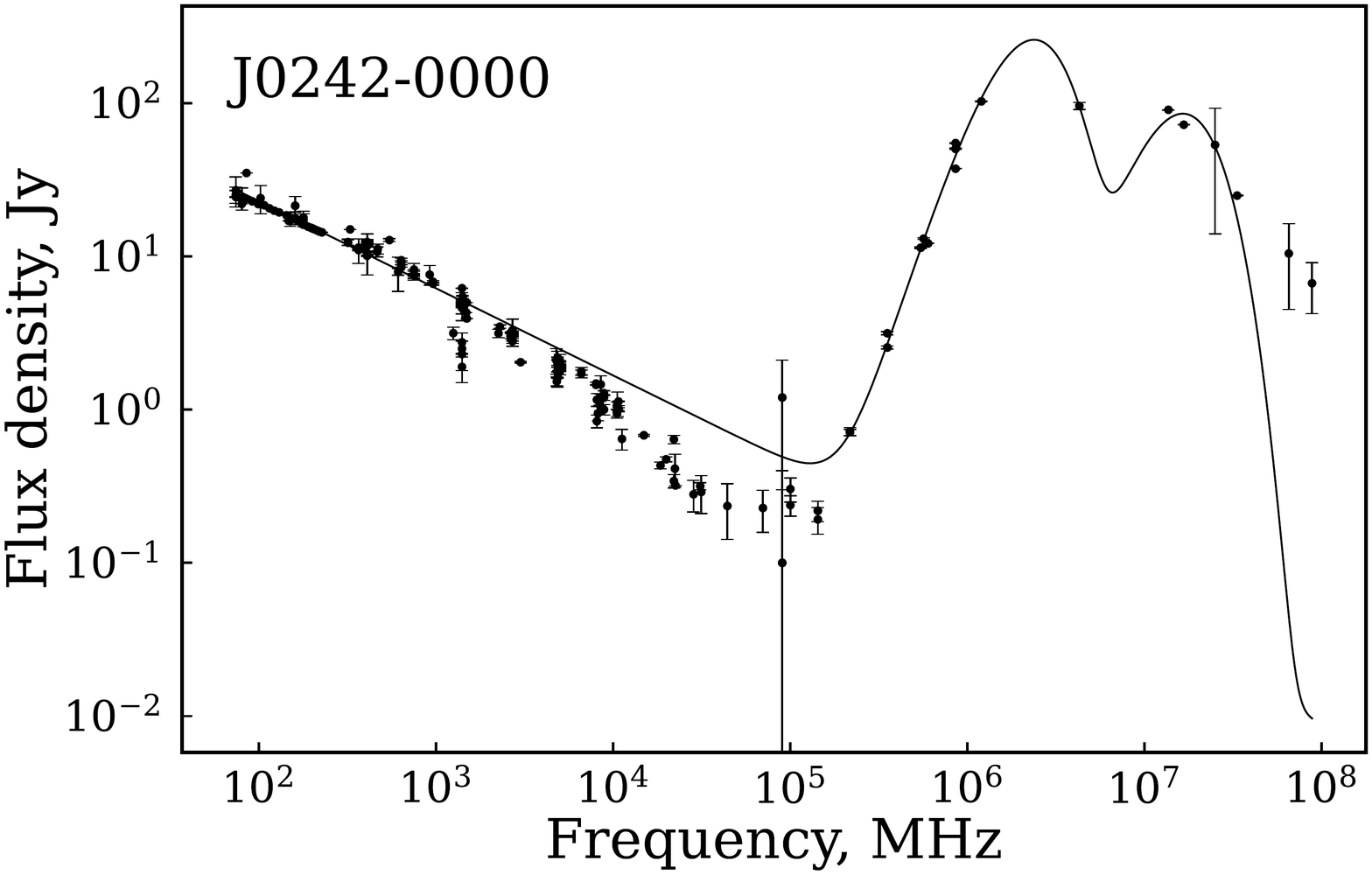}} \\ [-3ex]
        \subfloat[]{\includegraphics[width=0.8\columnwidth]{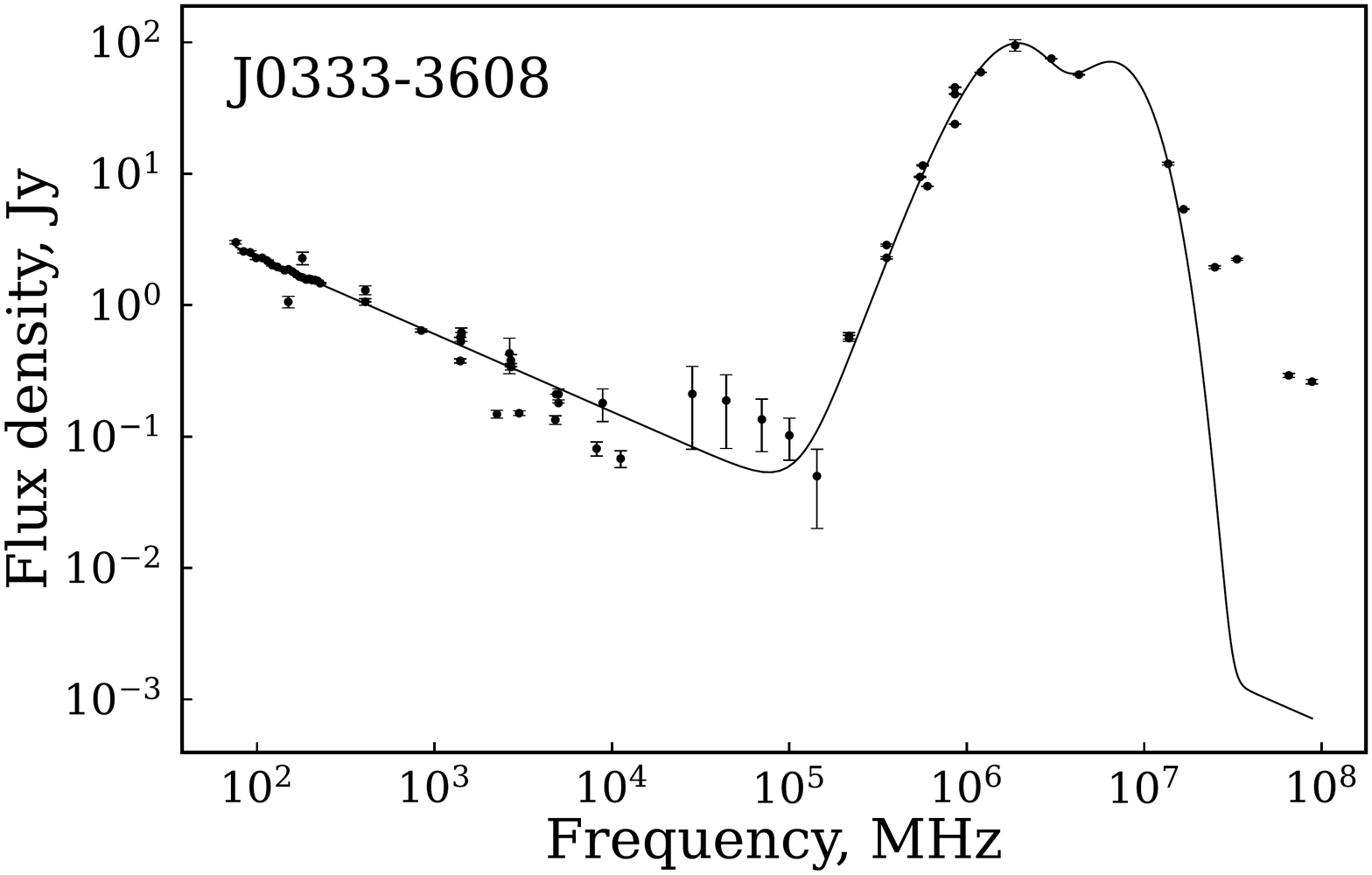}}
        \subfloat[]{\includegraphics[width=0.8\columnwidth]{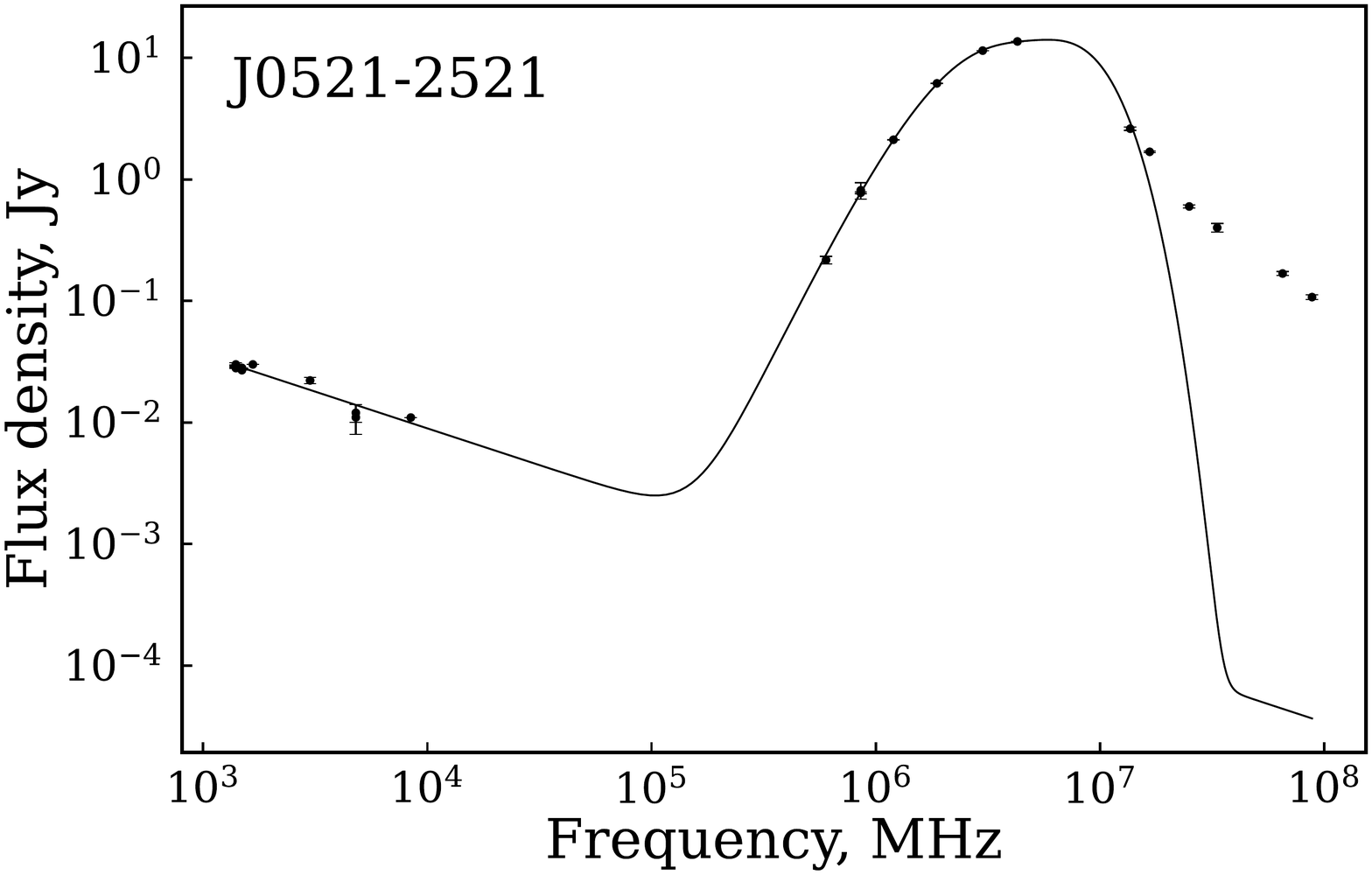}} \\ [-3ex]
        \subfloat[]{\includegraphics[width=0.8\columnwidth]{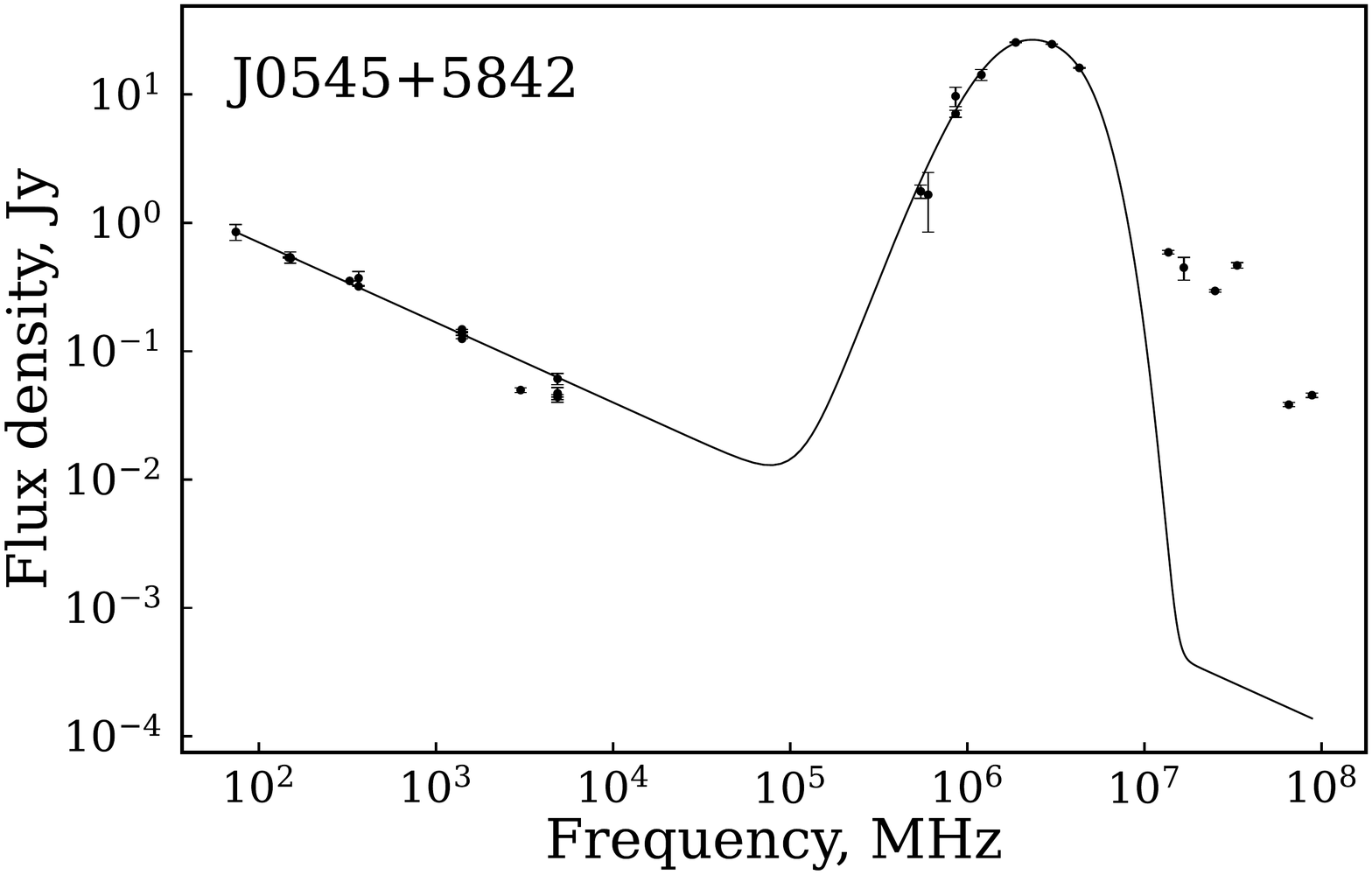}}
        \subfloat[]{\includegraphics[width=0.8\columnwidth]{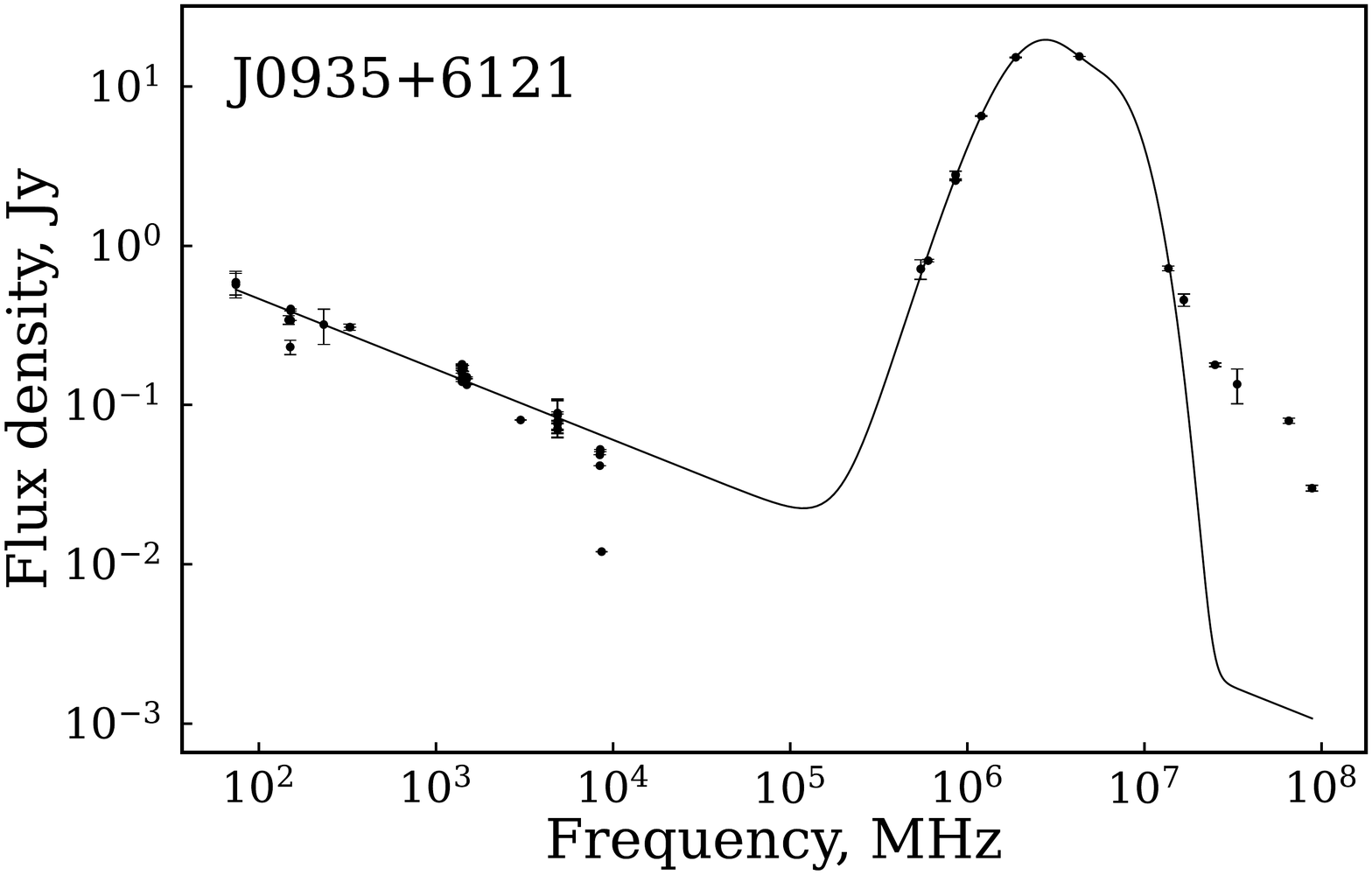}} \\
\caption{SED approximation for the OHMs having flux data in the sub-mm and IR regions. The fluxes in the mid-IR region at frequencies greater 10~THz are formed by the PAH molecules radiation and have not been modelled in this paper.}
\label{ris:SED_models1}
\end{figure*}

\begin{figure*}
    \centering
        \subfloat[]{\includegraphics[width=0.8\columnwidth]{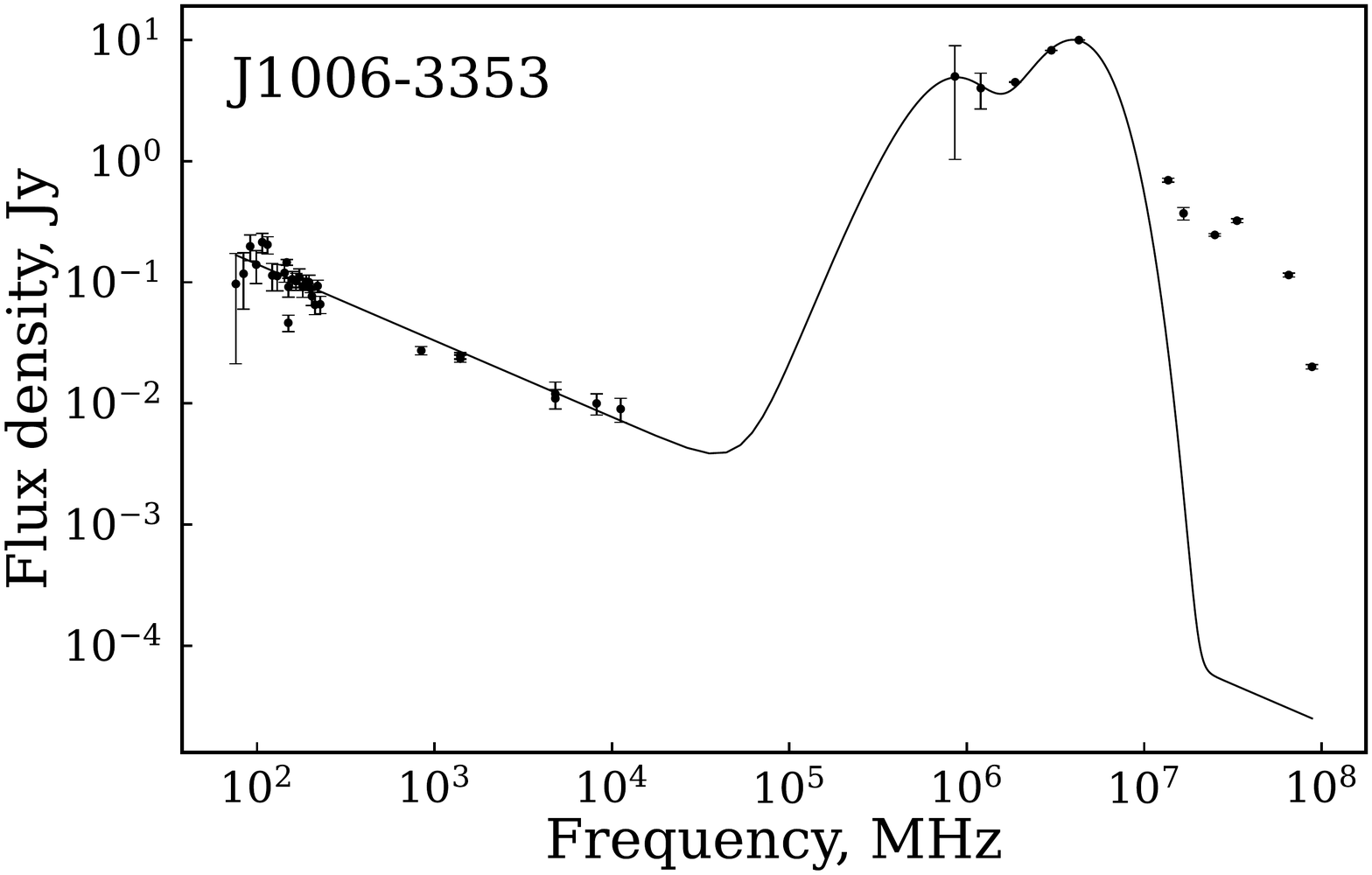}}
        \subfloat[]{\includegraphics[width=0.8\columnwidth]{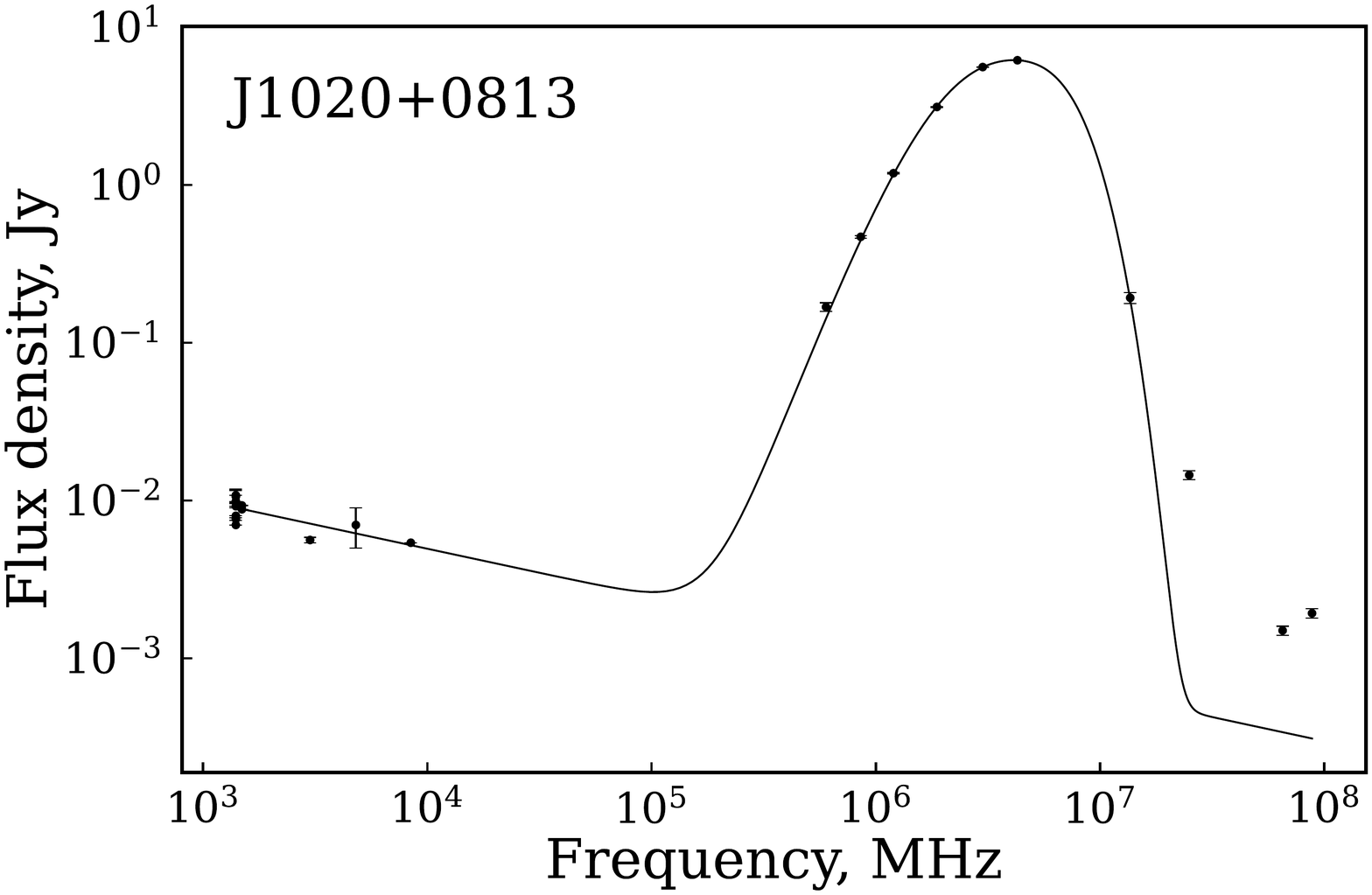}} \\ [-3ex]
        \subfloat[]{\includegraphics[width=0.8\columnwidth]{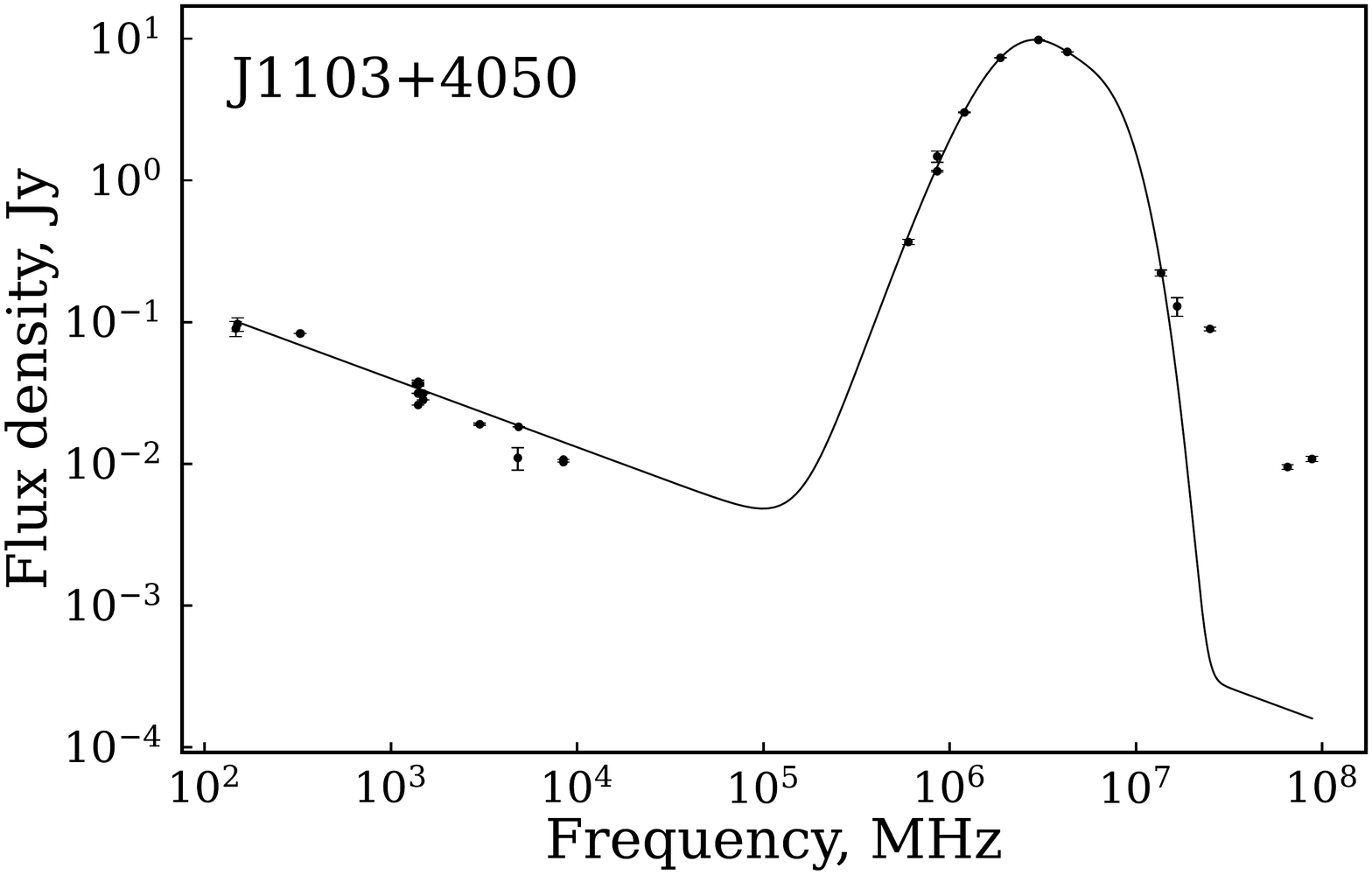}}
        \subfloat[]{\includegraphics[width=0.8\columnwidth]{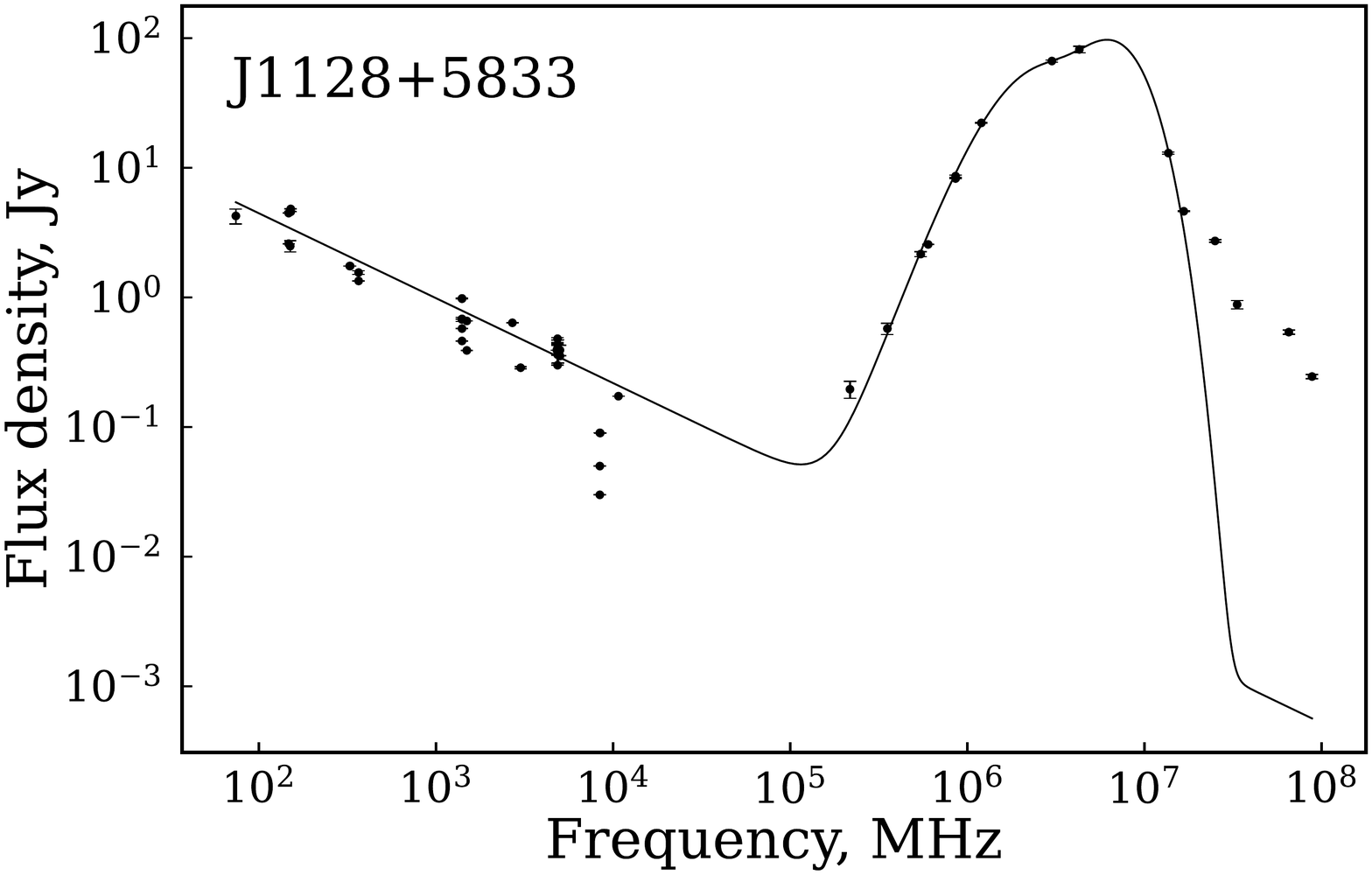}} \\ [-3ex]
        \subfloat[]{\includegraphics[width=0.8\columnwidth]{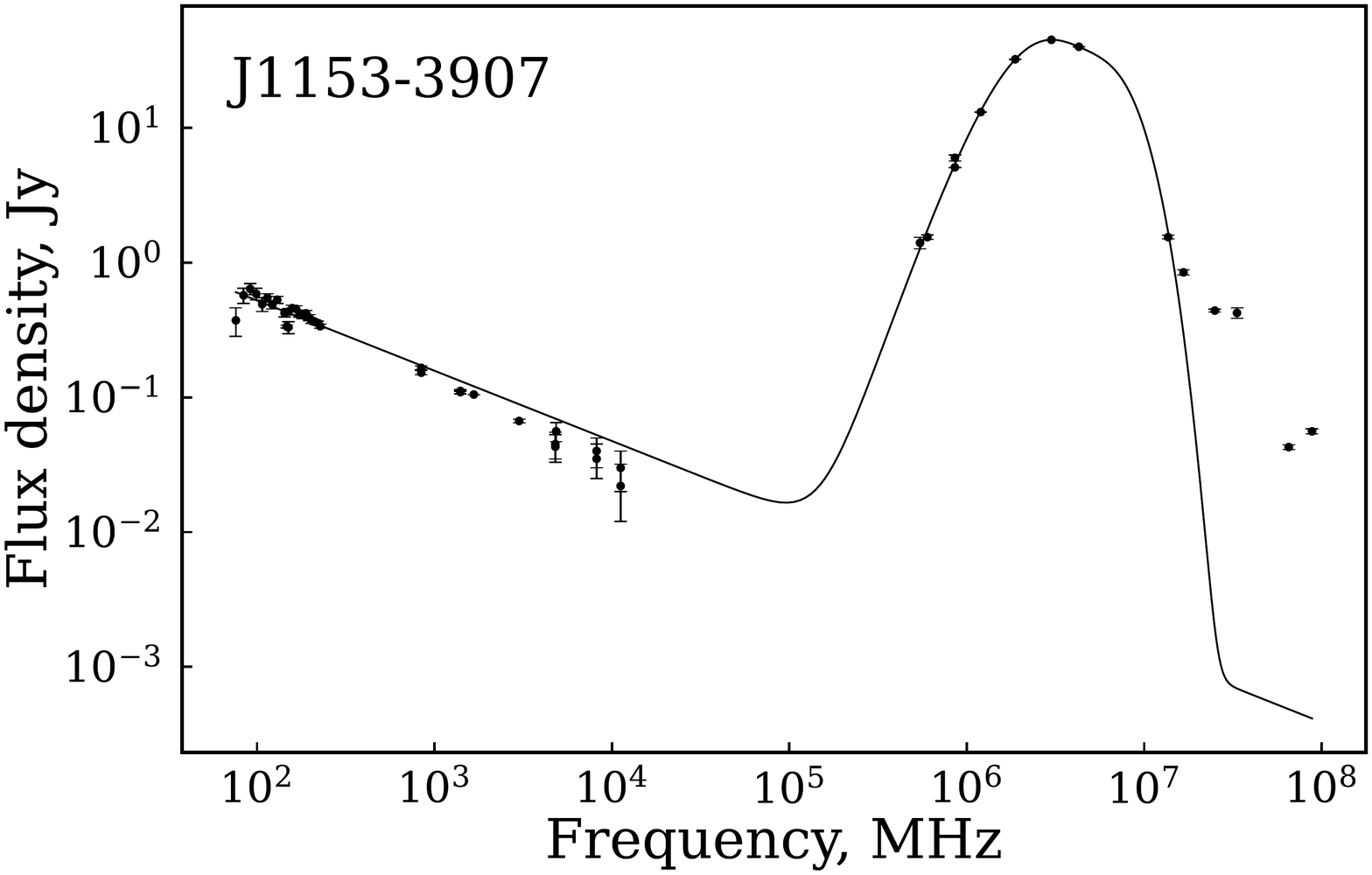}}
        \subfloat[]{\includegraphics[width=0.8\columnwidth]{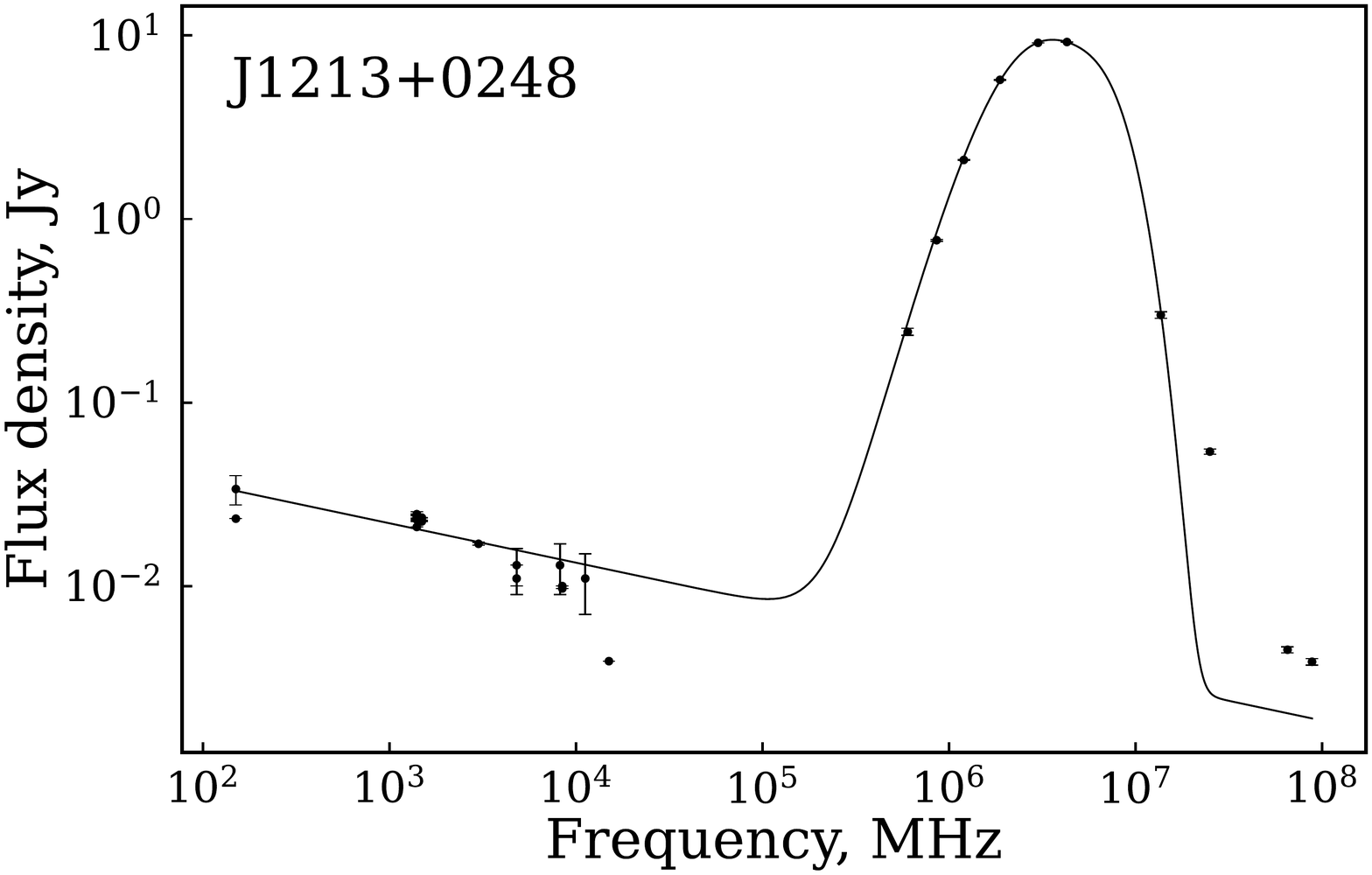}} \\ [-3ex]
        \subfloat[]{\includegraphics[width=0.8\columnwidth]{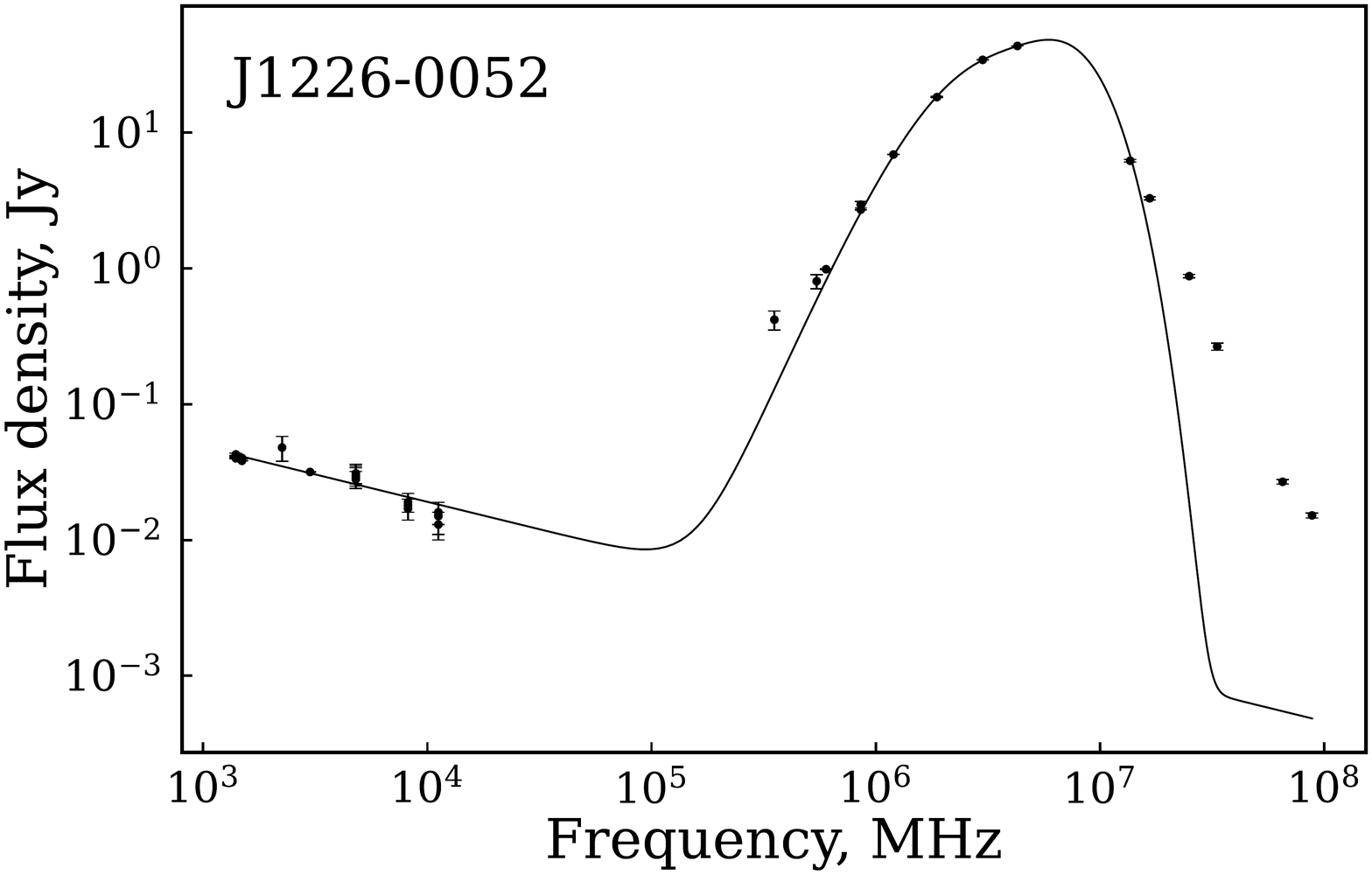}}
        \subfloat[]{\includegraphics[width=0.8\columnwidth]{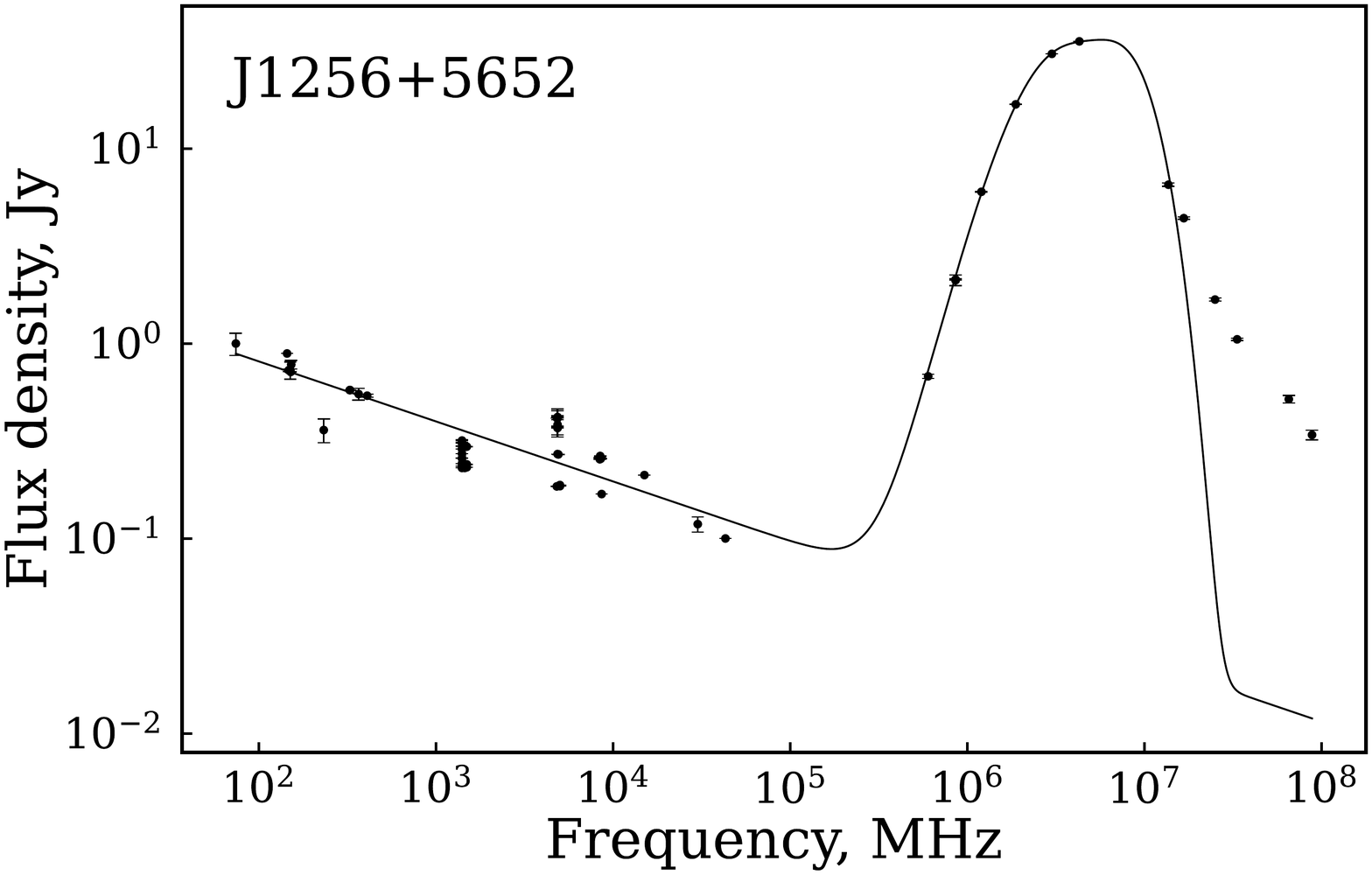}} \\
\caption{SED approximation for the OHMs having flux data in the sub-mm and IR regions (continued).}
\label{ris:SED_models2}
\end{figure*}

\begin{figure*}
    \centering
        \subfloat[]{\includegraphics[width=0.8\columnwidth]{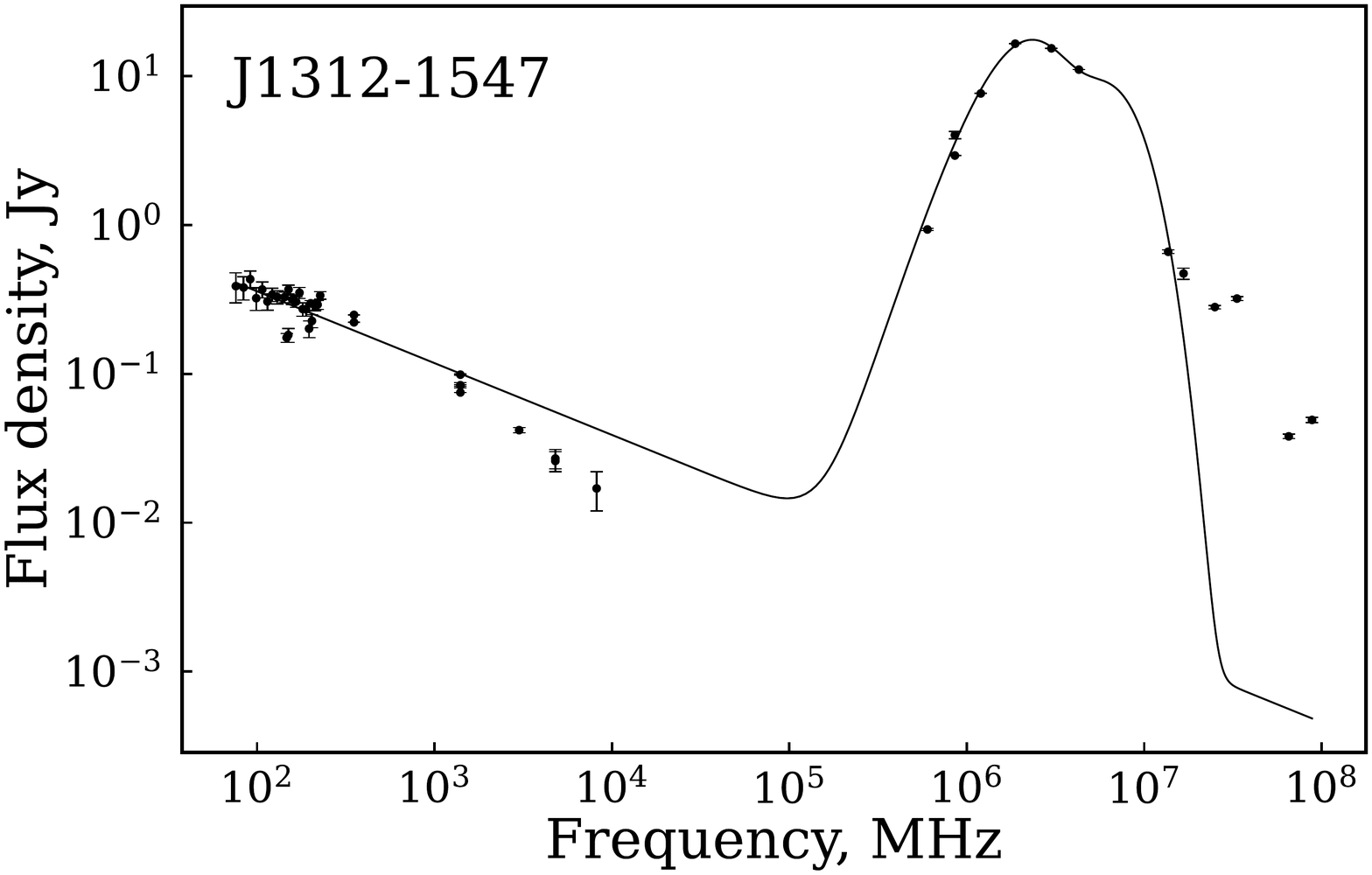}}
        \subfloat[]{\includegraphics[width=0.8\columnwidth]{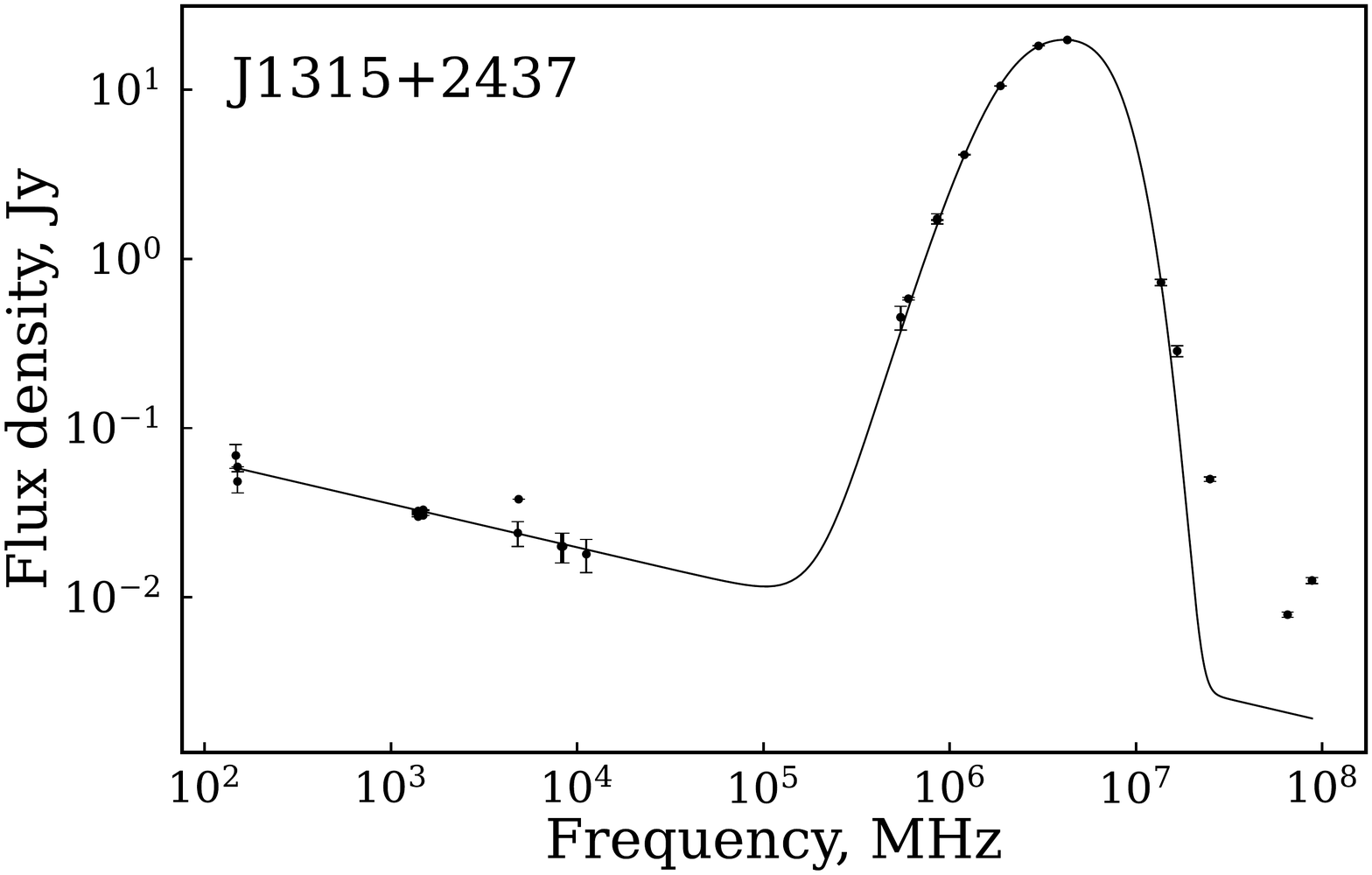}} \\ [-3ex]
        \subfloat[]{\includegraphics[width=0.8\columnwidth]{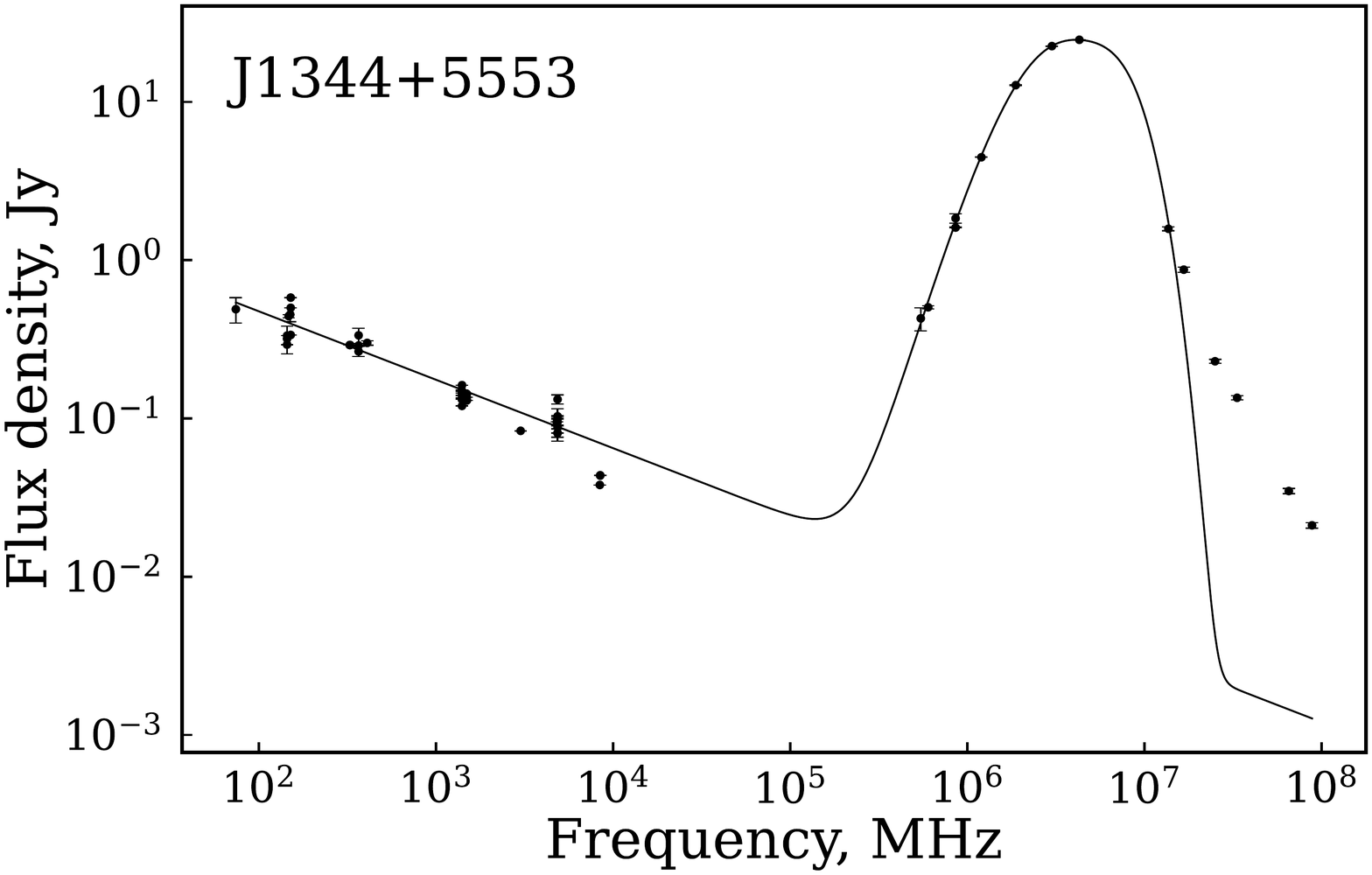}}
        \subfloat[]{\includegraphics[width=0.8\columnwidth]{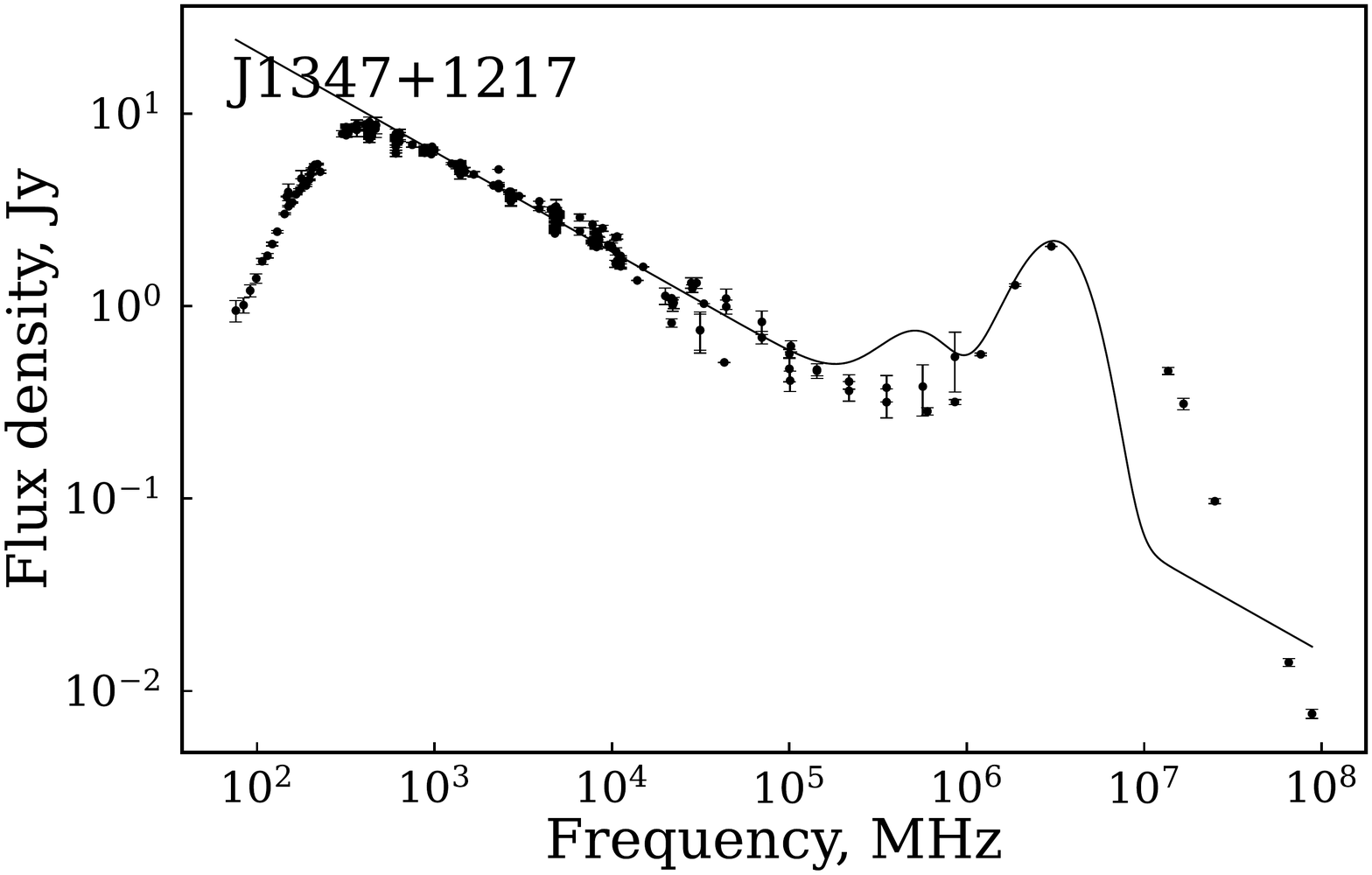}} \\ [-3ex]
        \subfloat[]{\includegraphics[width=0.8\columnwidth]{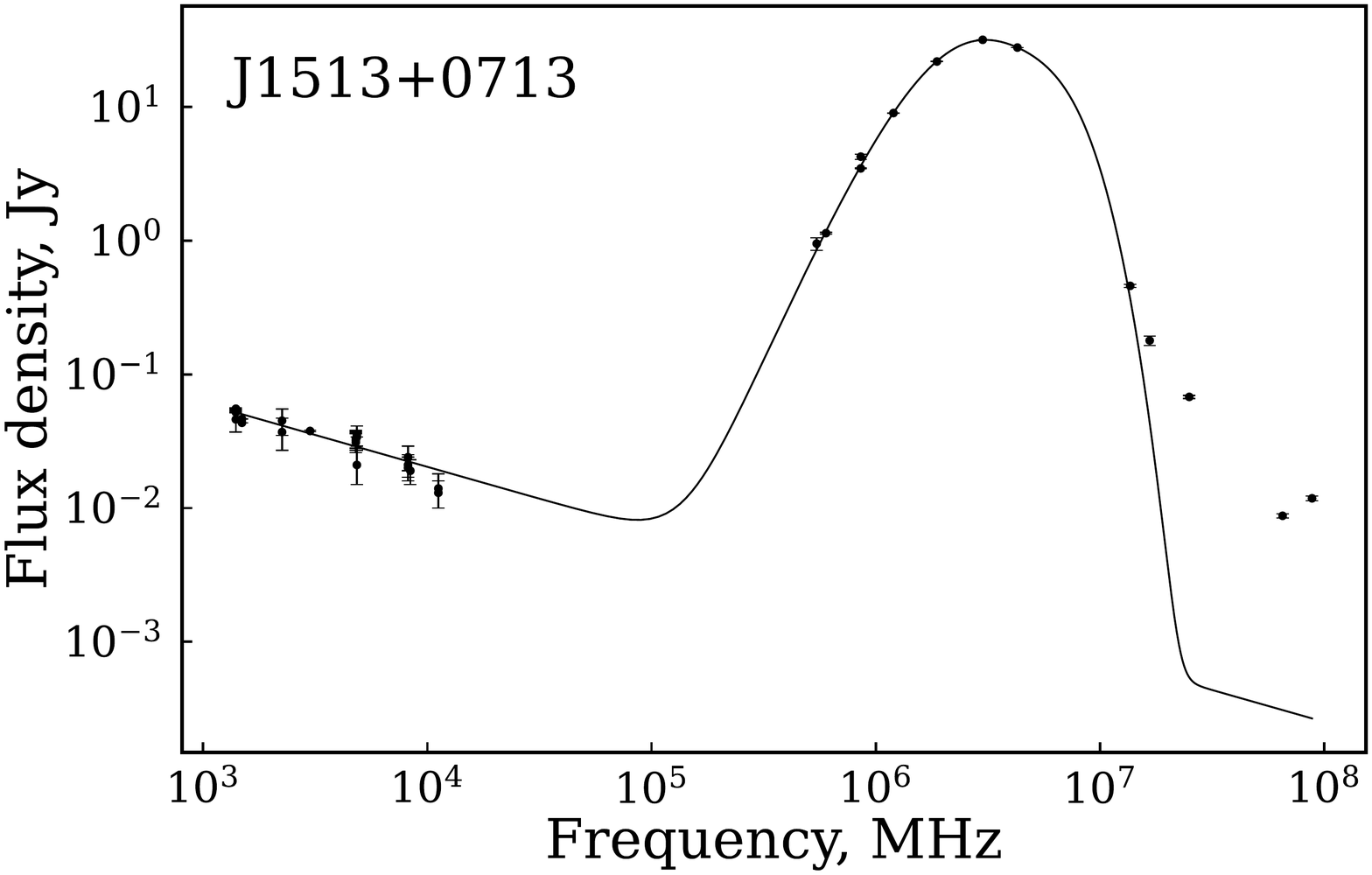}}
        \subfloat[]{\includegraphics[width=0.8\columnwidth]{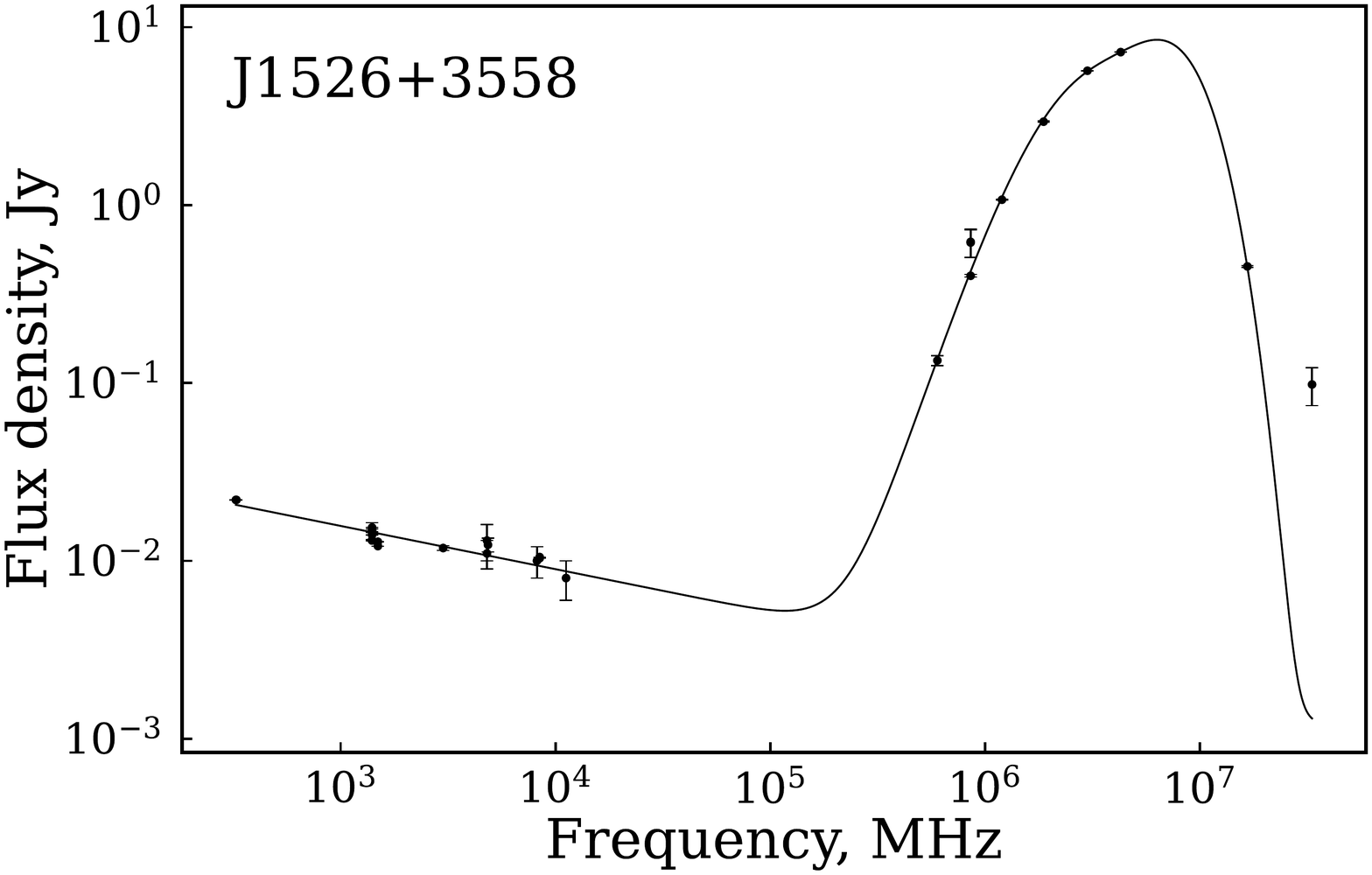}} \\ [-3ex]
        \subfloat[]{\includegraphics[width=0.8\columnwidth]{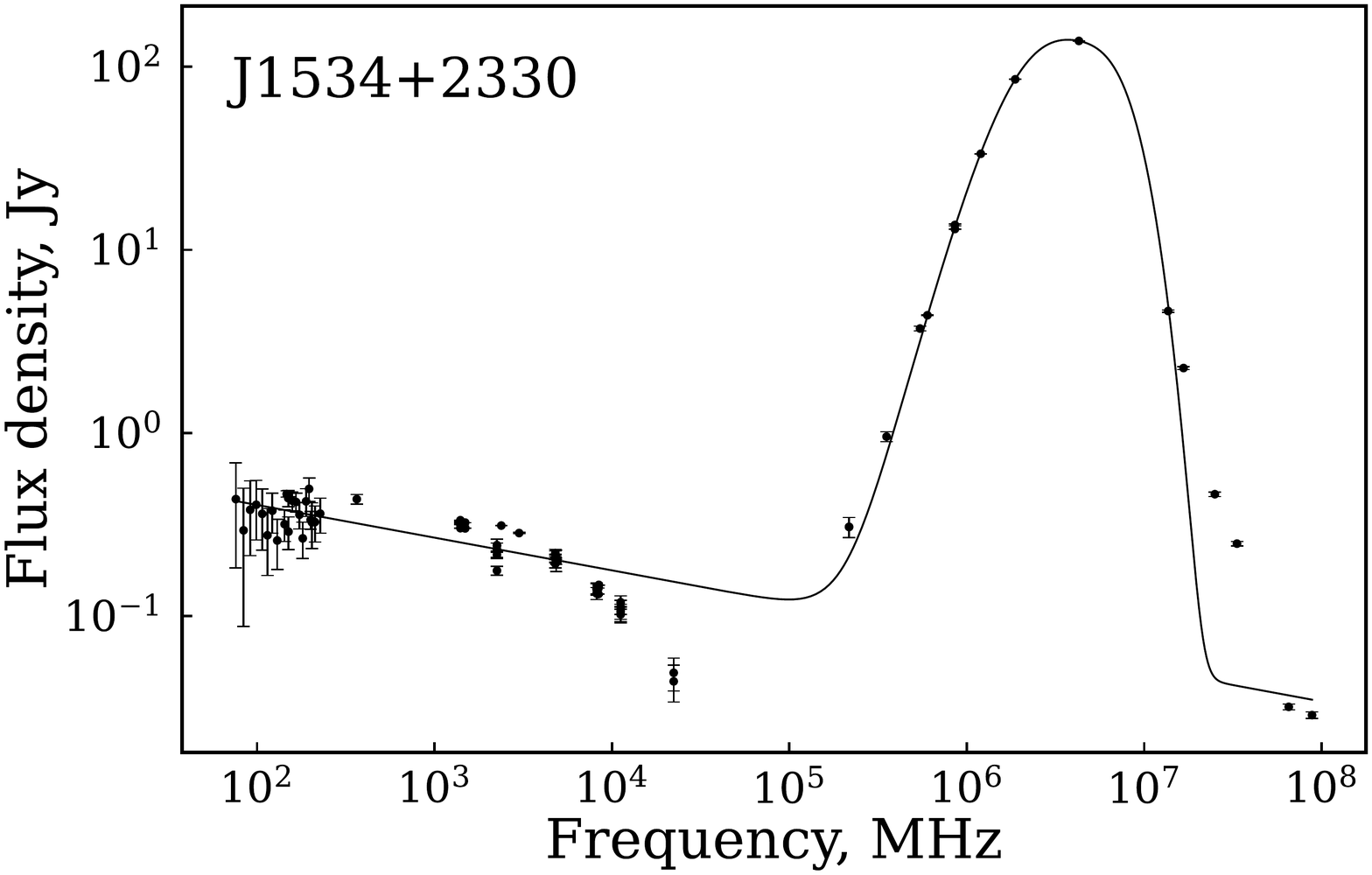}}
        \subfloat[]{\includegraphics[width=0.8\columnwidth]{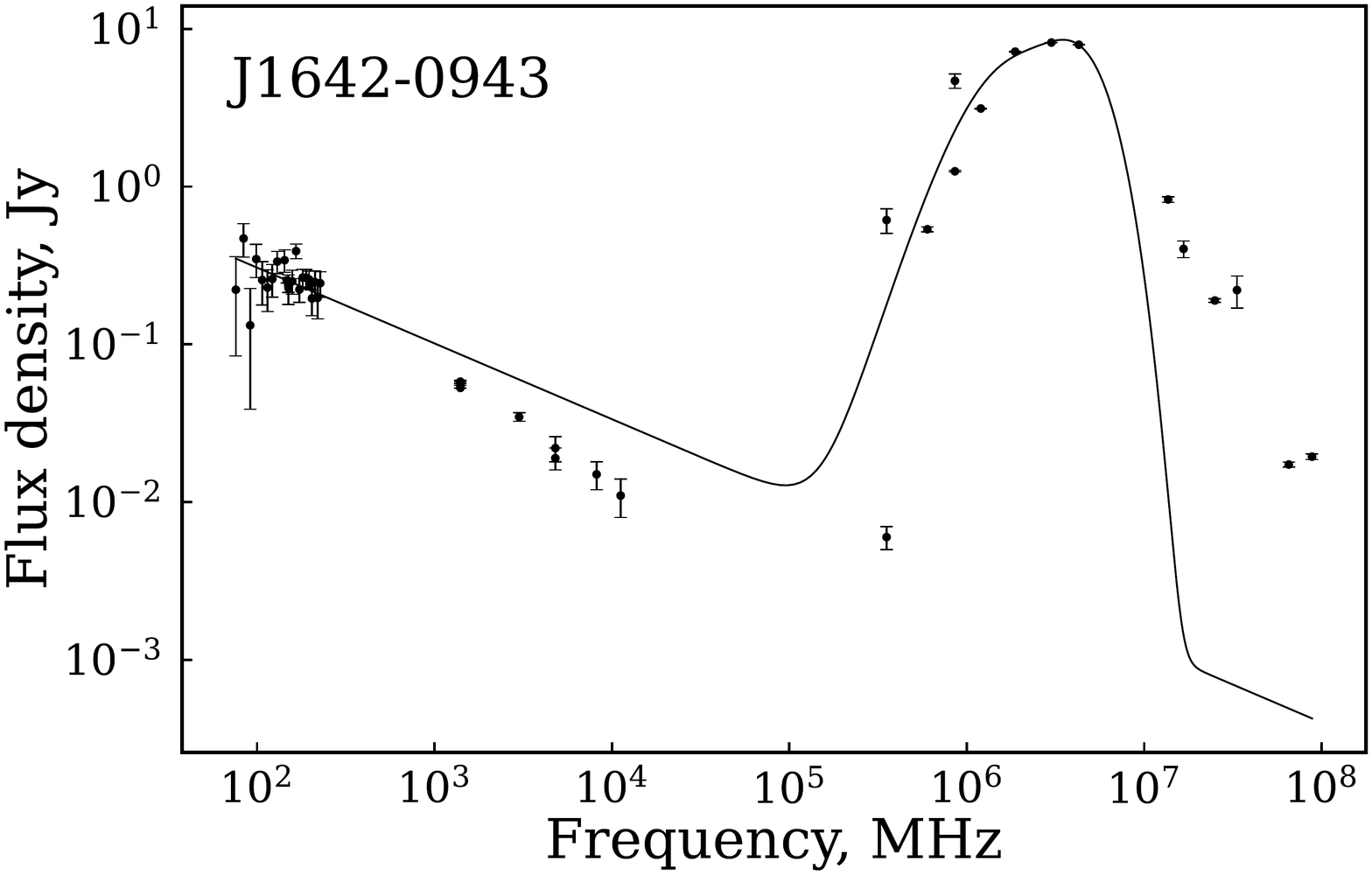}} \\
\caption{SED approximation for the OHMs having flux data in the sub-mm and IR regions (continued).}
\label{ris:SED_models3}
\end{figure*}

\begin{figure*}
    \centering
        \subfloat[]{\includegraphics[width=0.8\columnwidth]{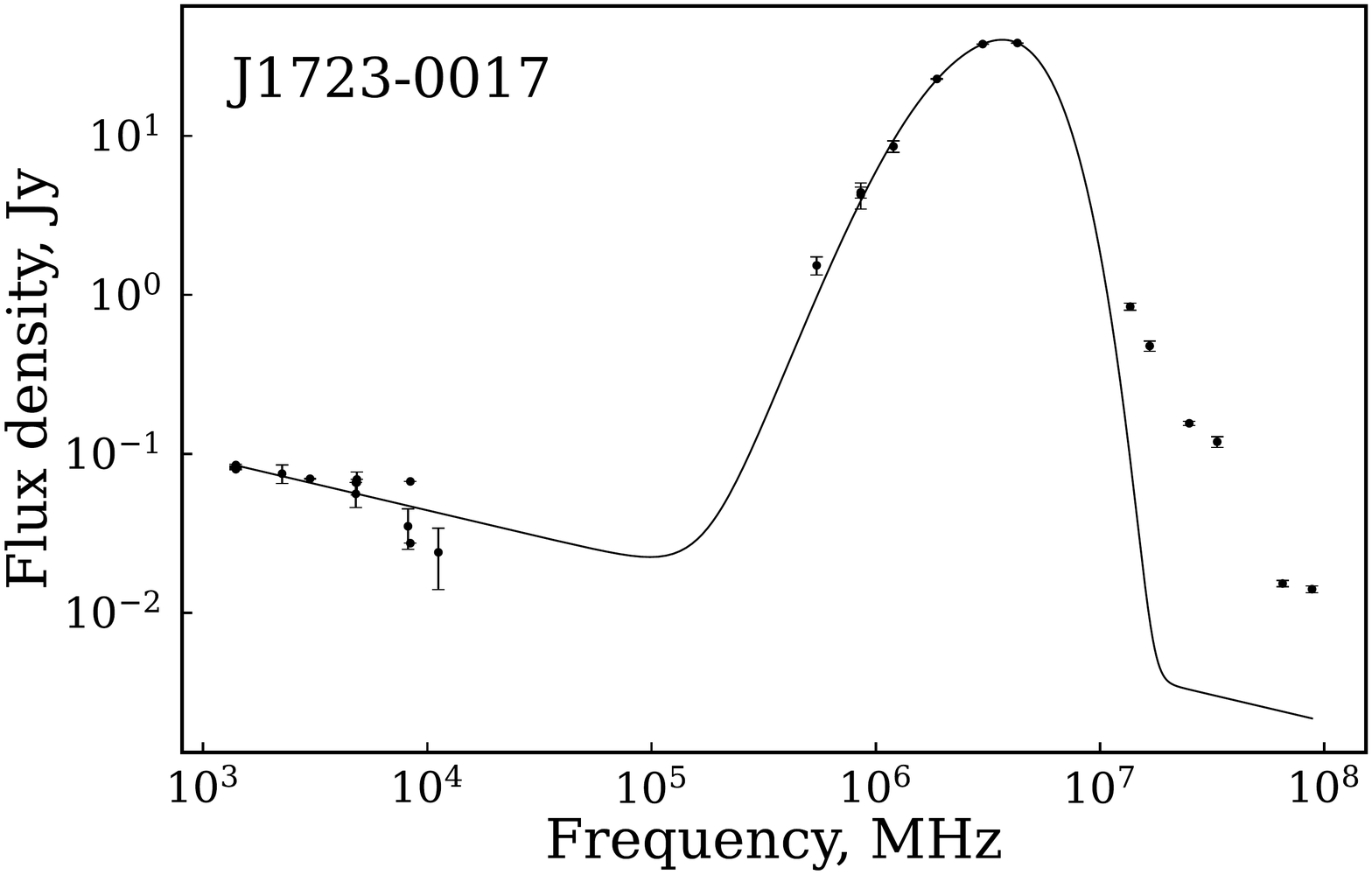}}
        \subfloat[]{\includegraphics[width=0.8\columnwidth]{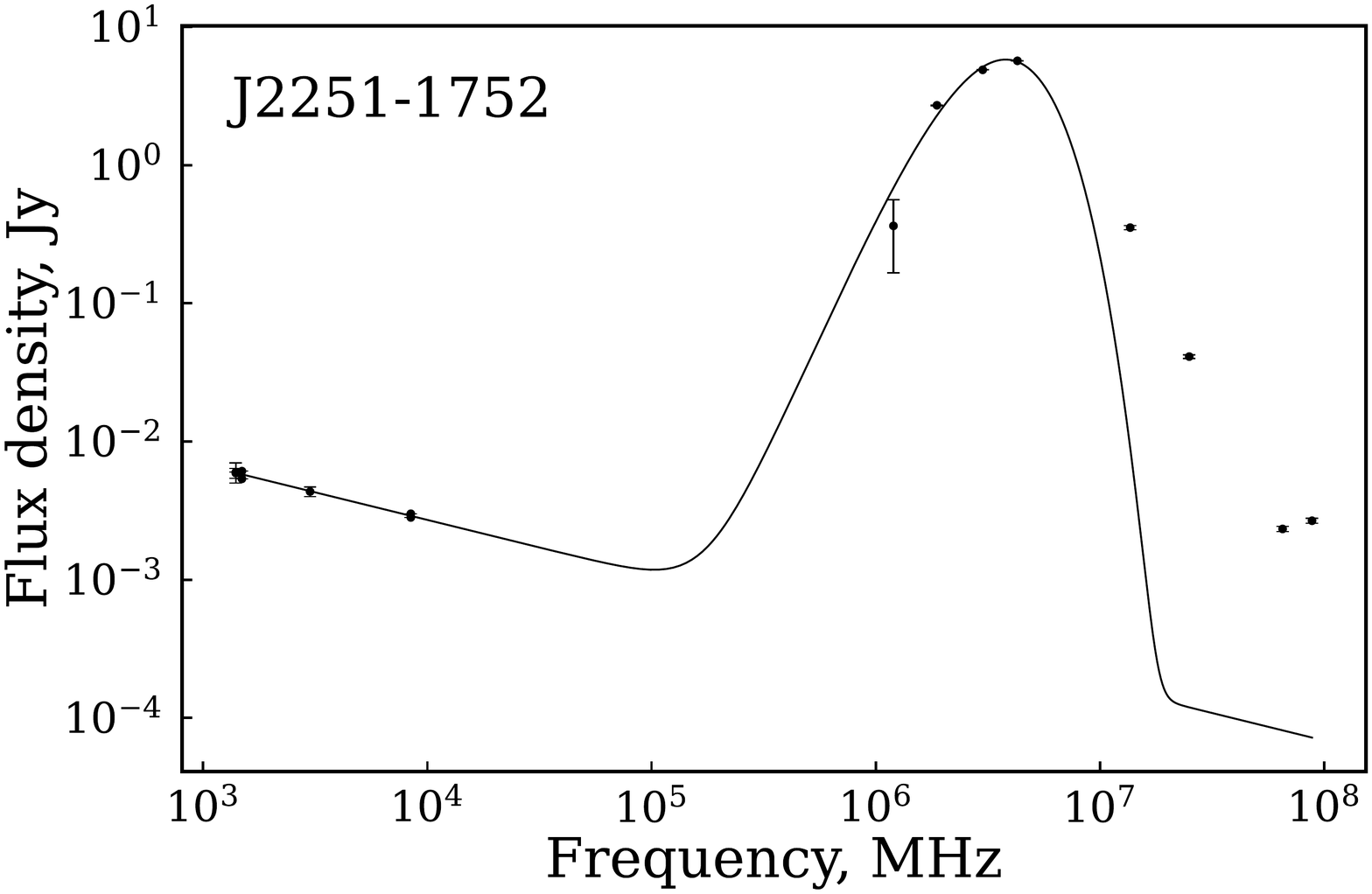}} \\ [-3ex]
        \subfloat[]{\includegraphics[width=0.8\columnwidth]{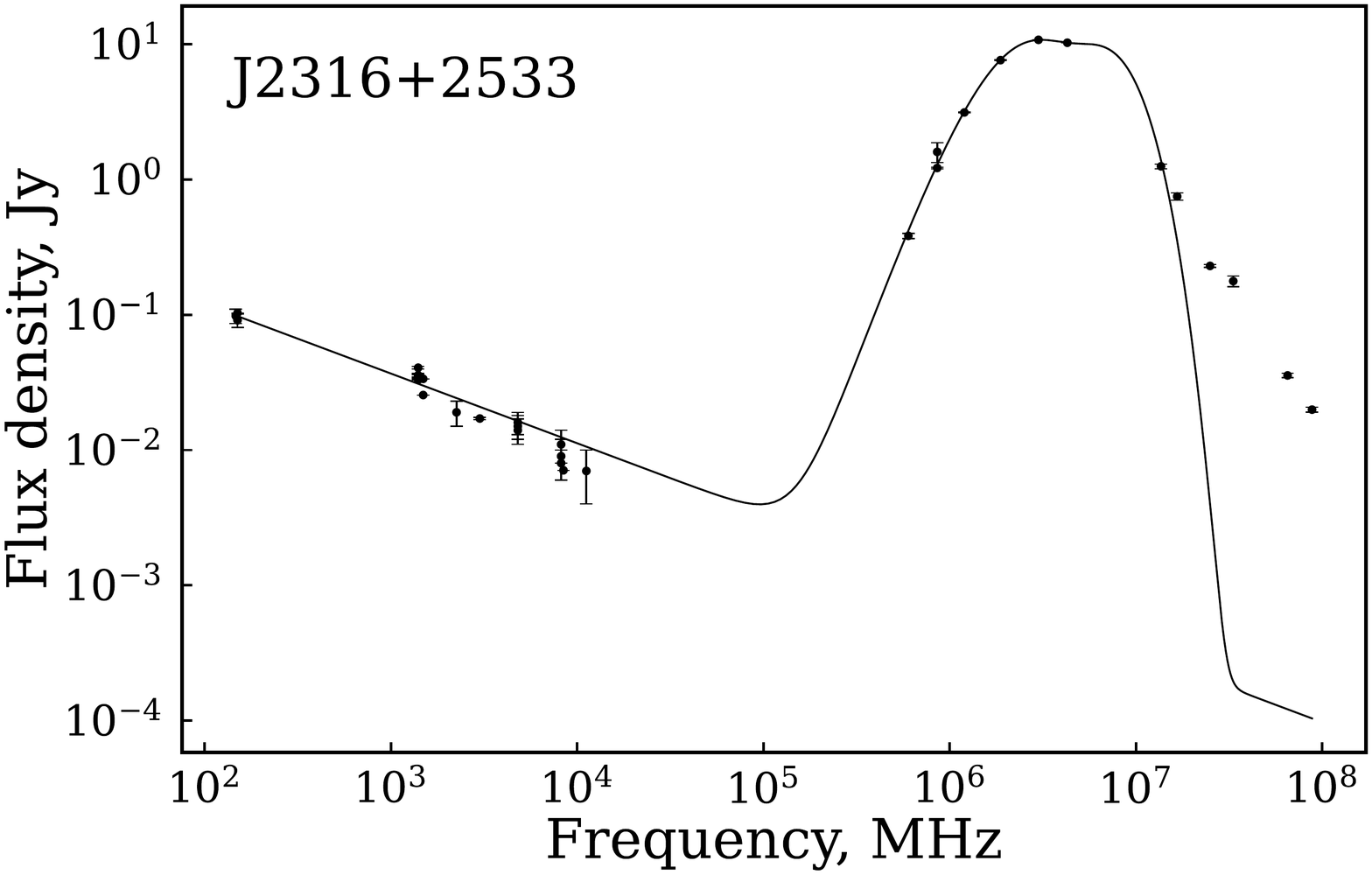}}
        \subfloat[]{\includegraphics[width=0.8\columnwidth]{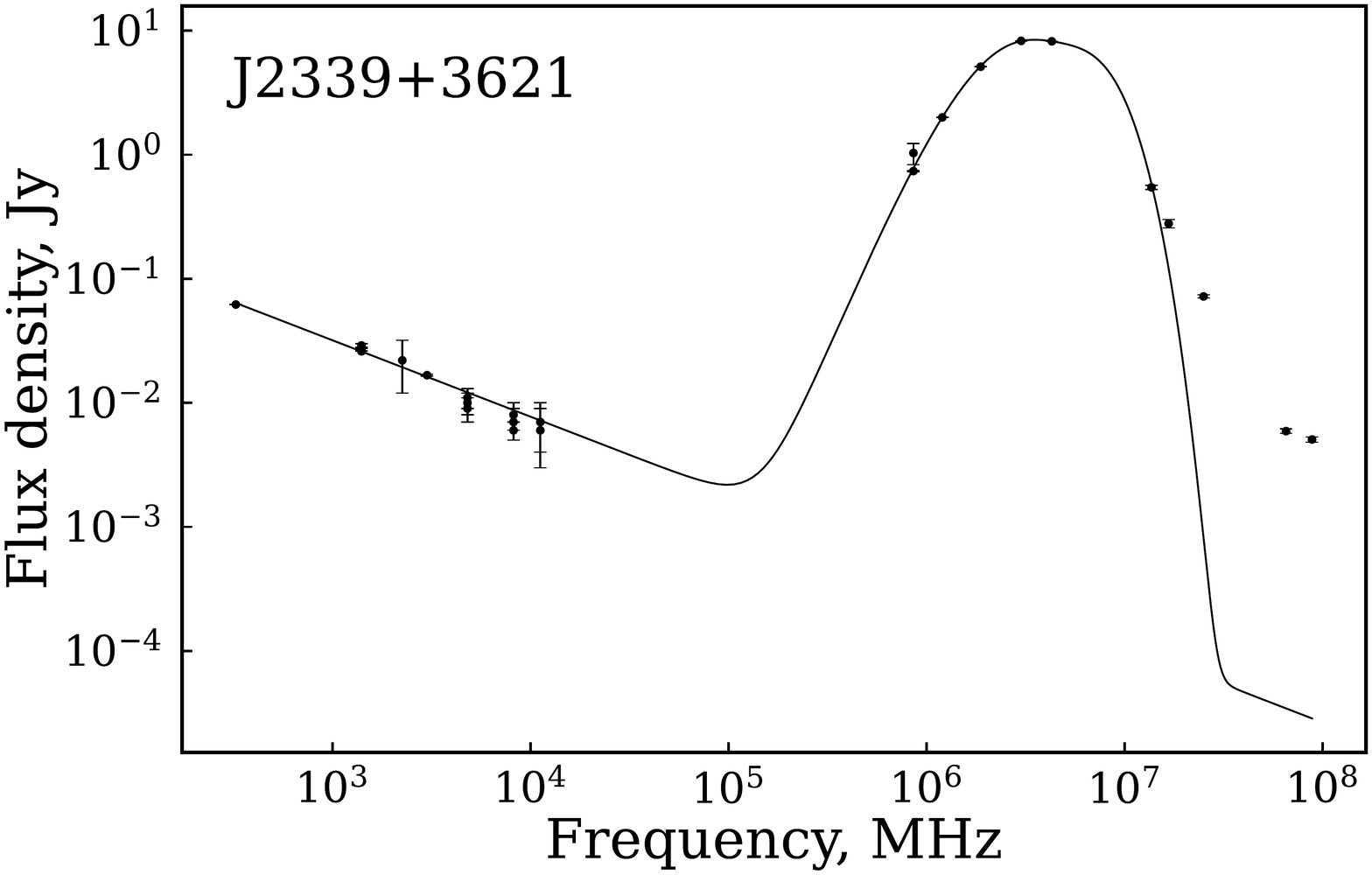}} \\
\caption{SED approximation for the OHMs having flux data in the sub-mm and IR regions (continued).}
\label{ris:SED_models4}
\end{figure*}

\begin{figure*}
    \centering
        \subfloat[]{\includegraphics[width=0.8\columnwidth]{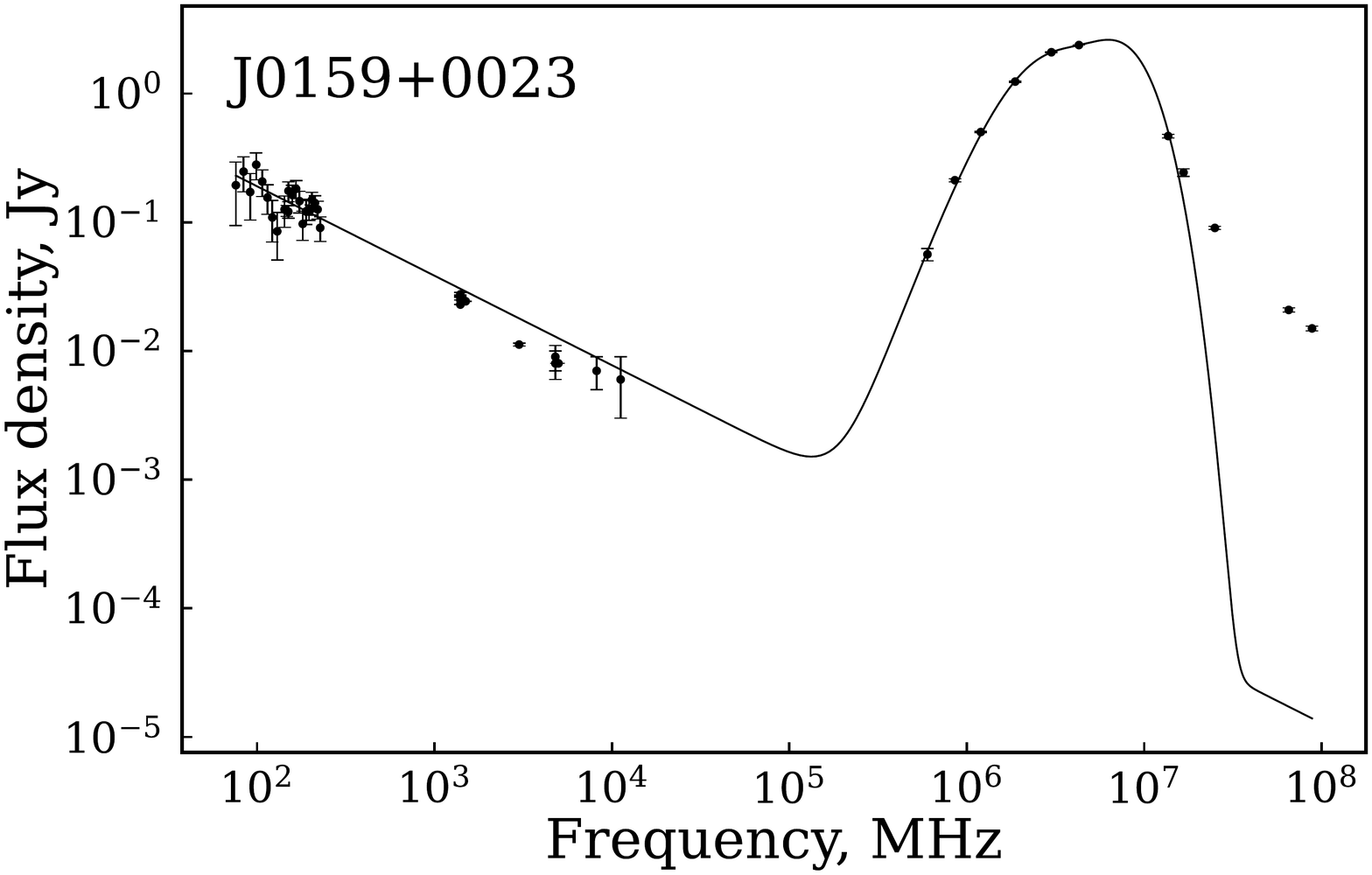}}
        \subfloat[]{\includegraphics[width=0.8\columnwidth]{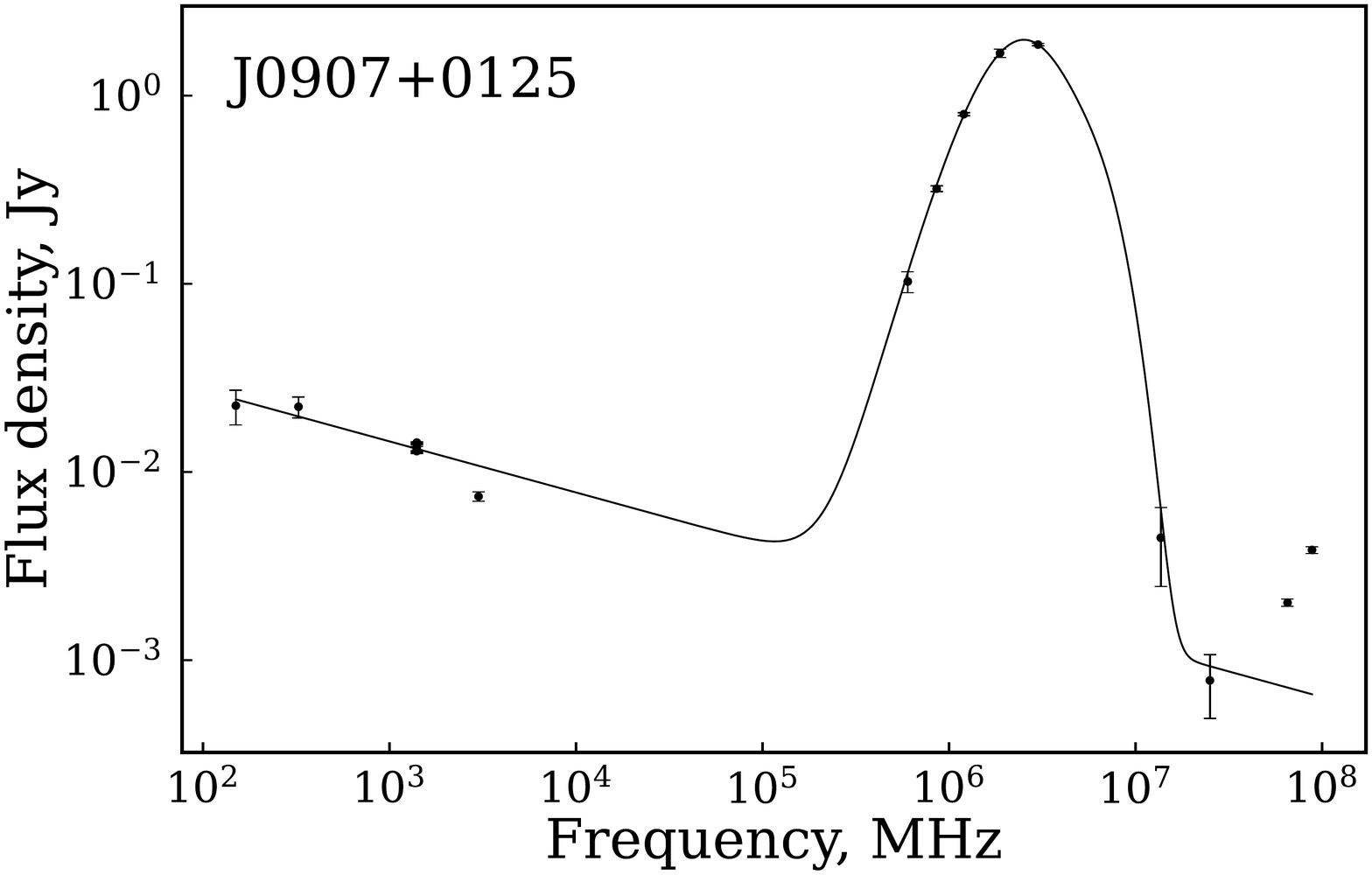}} \\ [-3ex]
        \subfloat[]{\includegraphics[width=0.8\columnwidth]{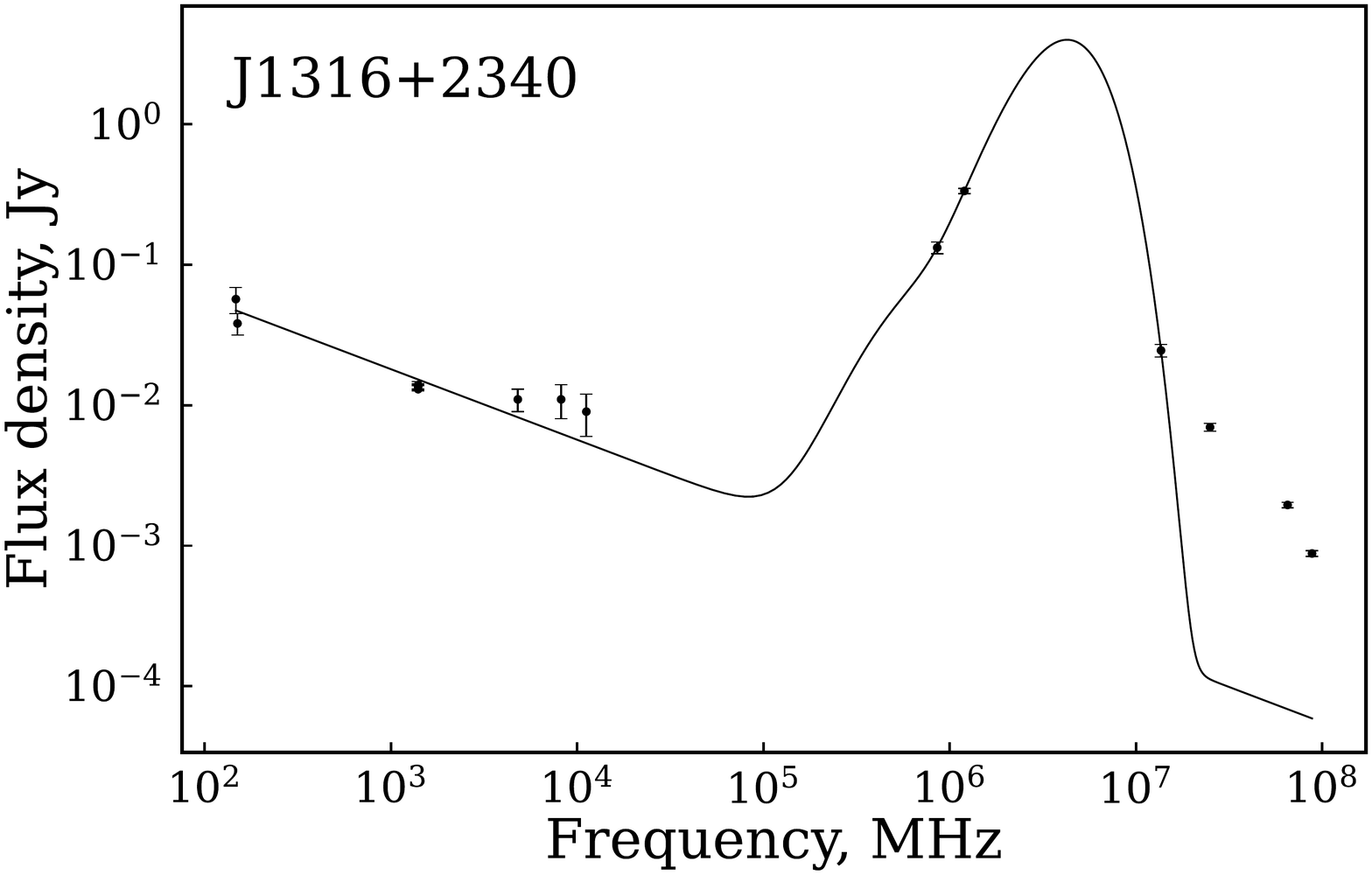}}
        \subfloat[]{\includegraphics[width=0.8\columnwidth]{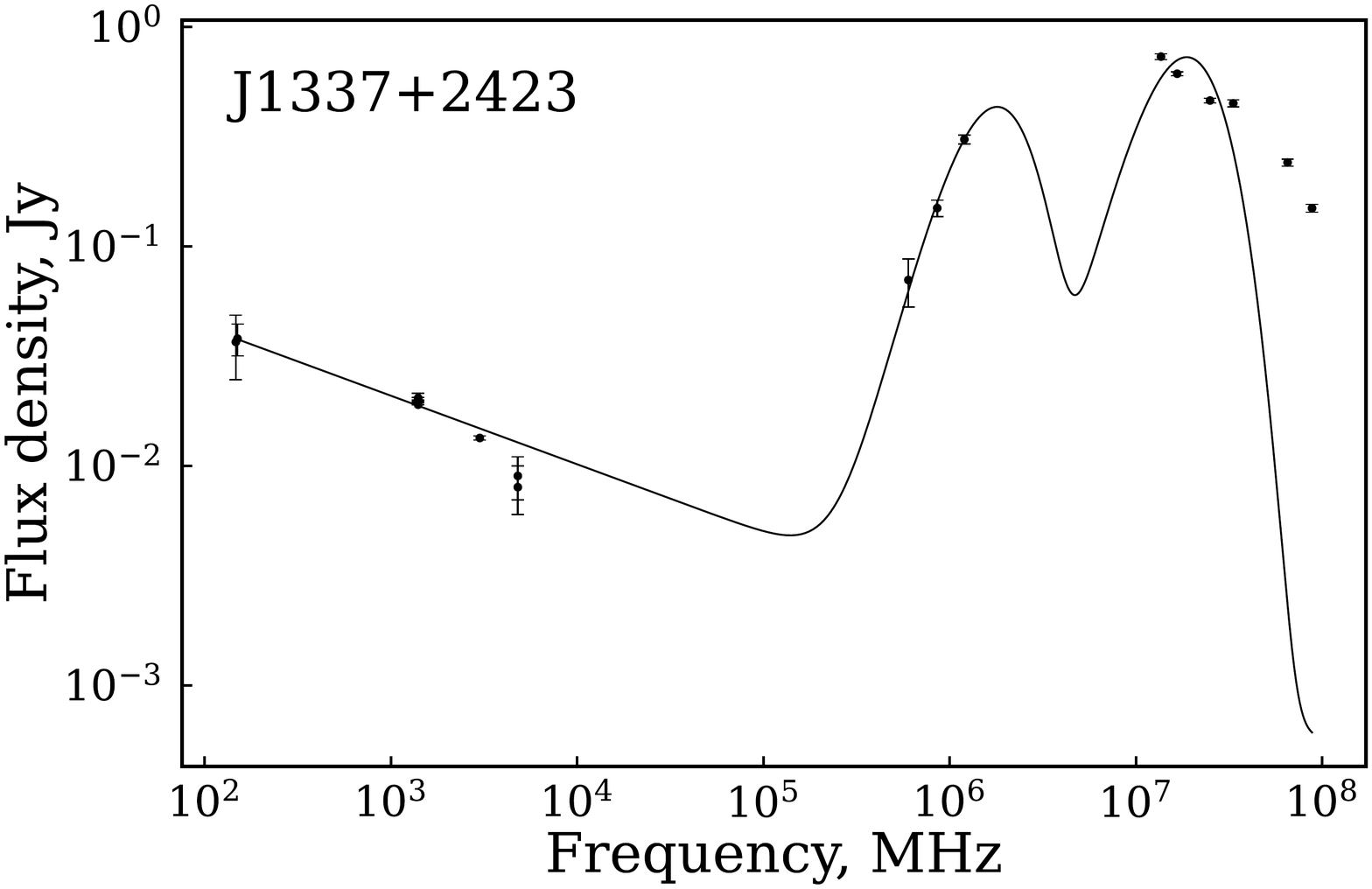}} \\ [-3ex]
        \subfloat[]{\includegraphics[width=0.8\columnwidth]{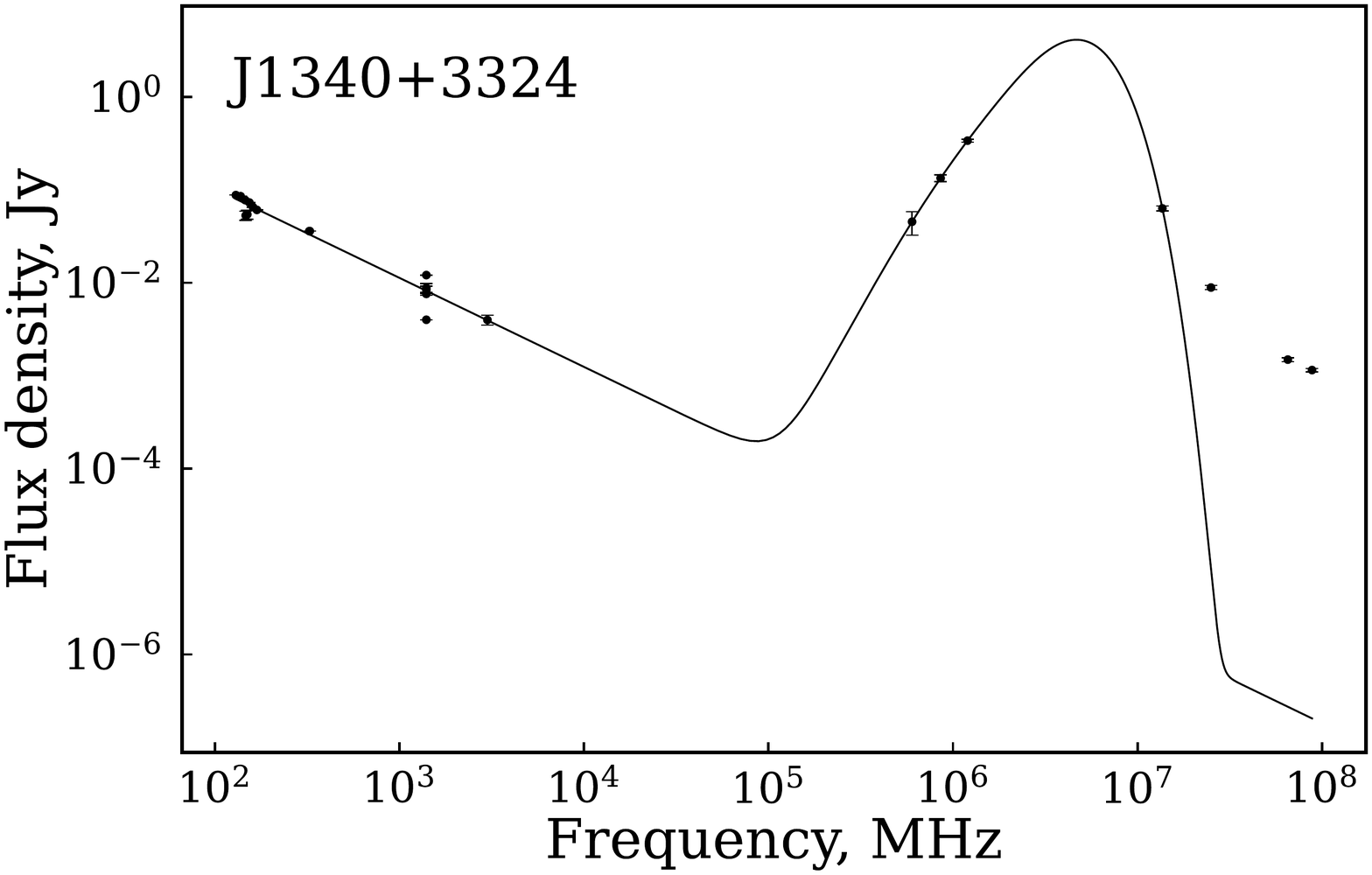}}
        \subfloat[]{\includegraphics[width=0.8\columnwidth]{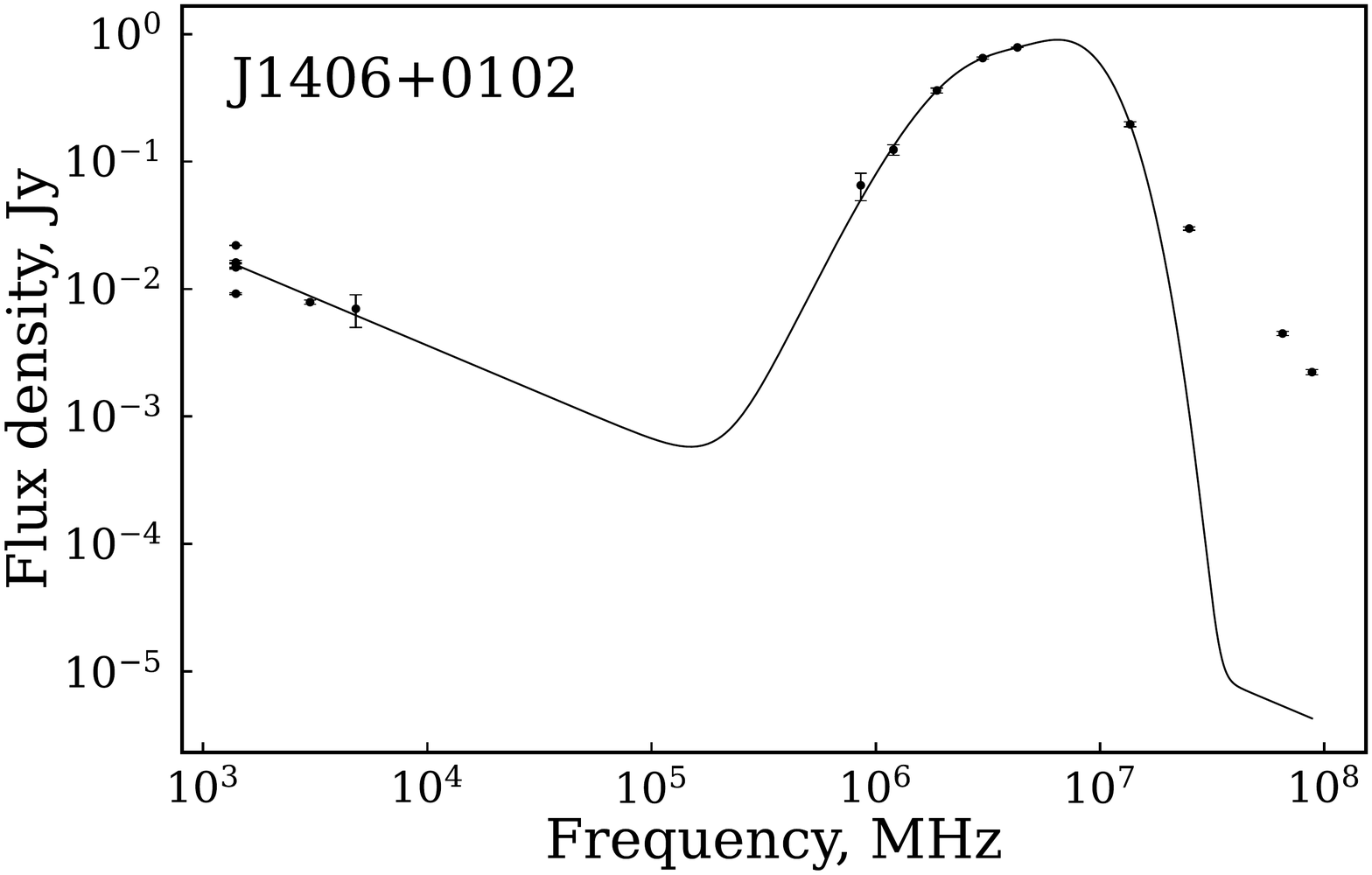}} \\ [-3ex]
        \subfloat[]{\includegraphics[width=0.8\columnwidth]{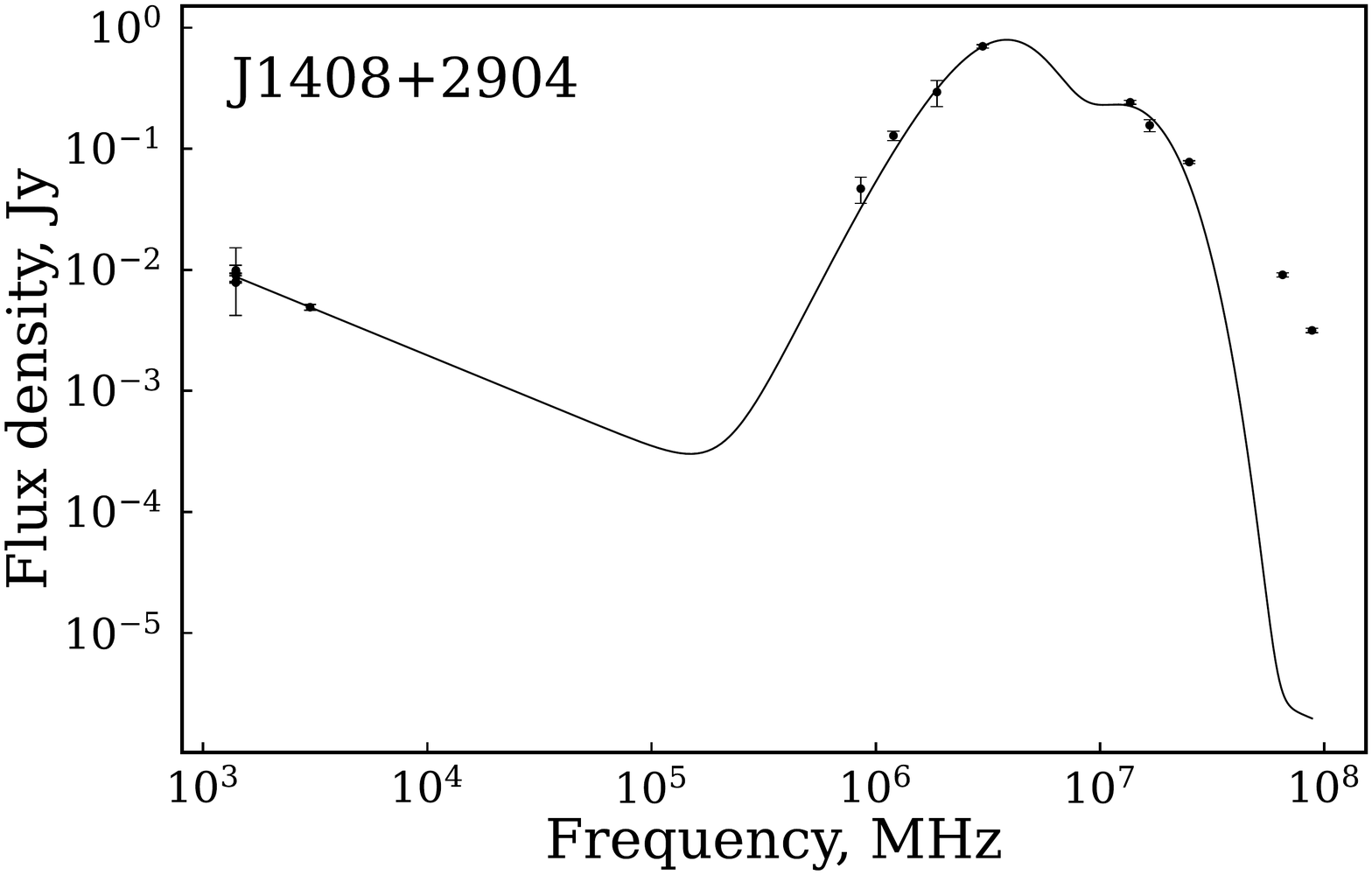}}
        \subfloat[]{\includegraphics[width=0.8\columnwidth]{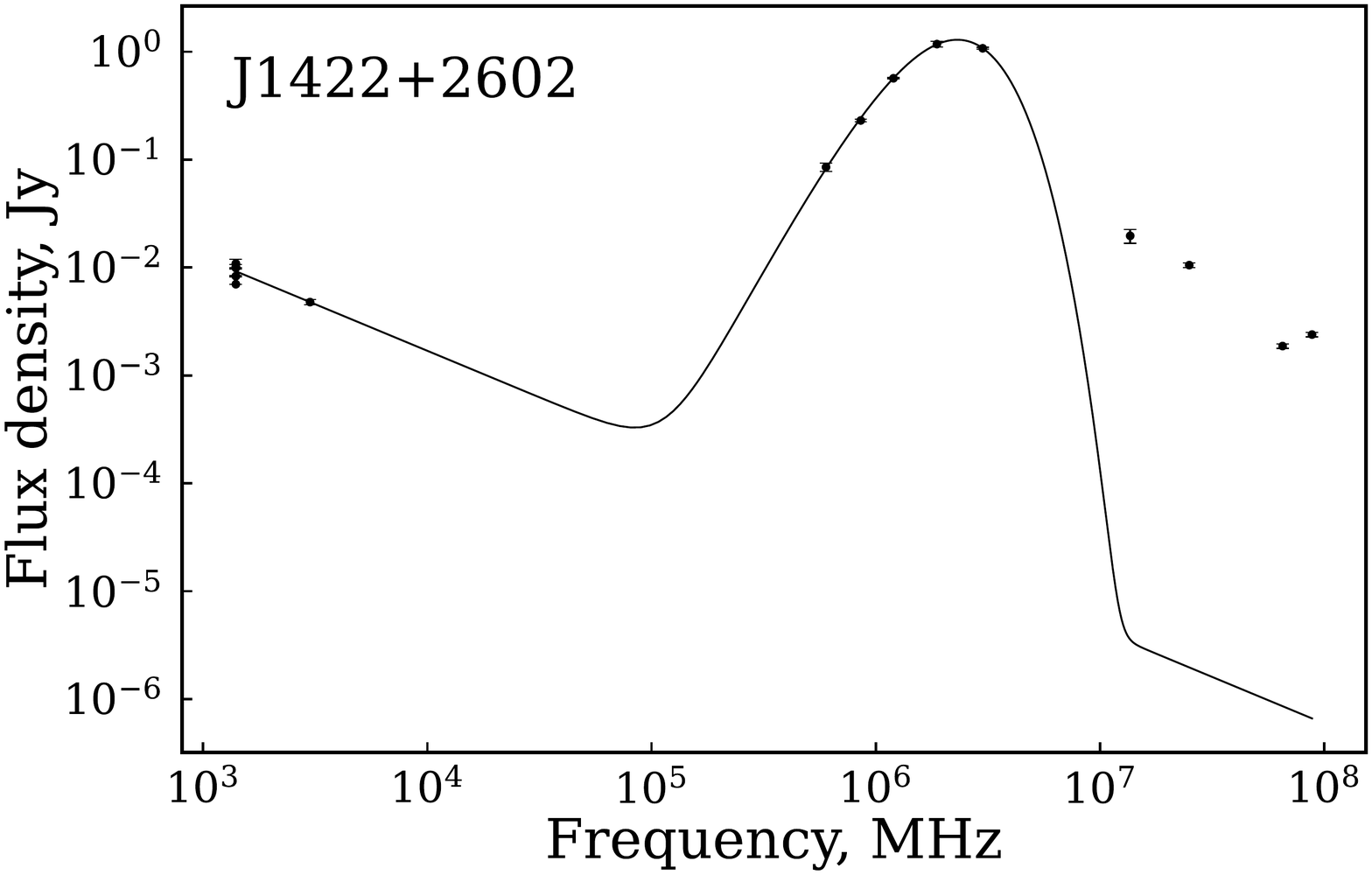}} \\
\caption{SED approximation for the control sample sources having flux data in the sub-mm and IR regions.}
\label{ris:SED_cntrl1}
\end{figure*}

\begin{figure*}
    \centering
        \subfloat[]{\includegraphics[width=0.8\columnwidth]{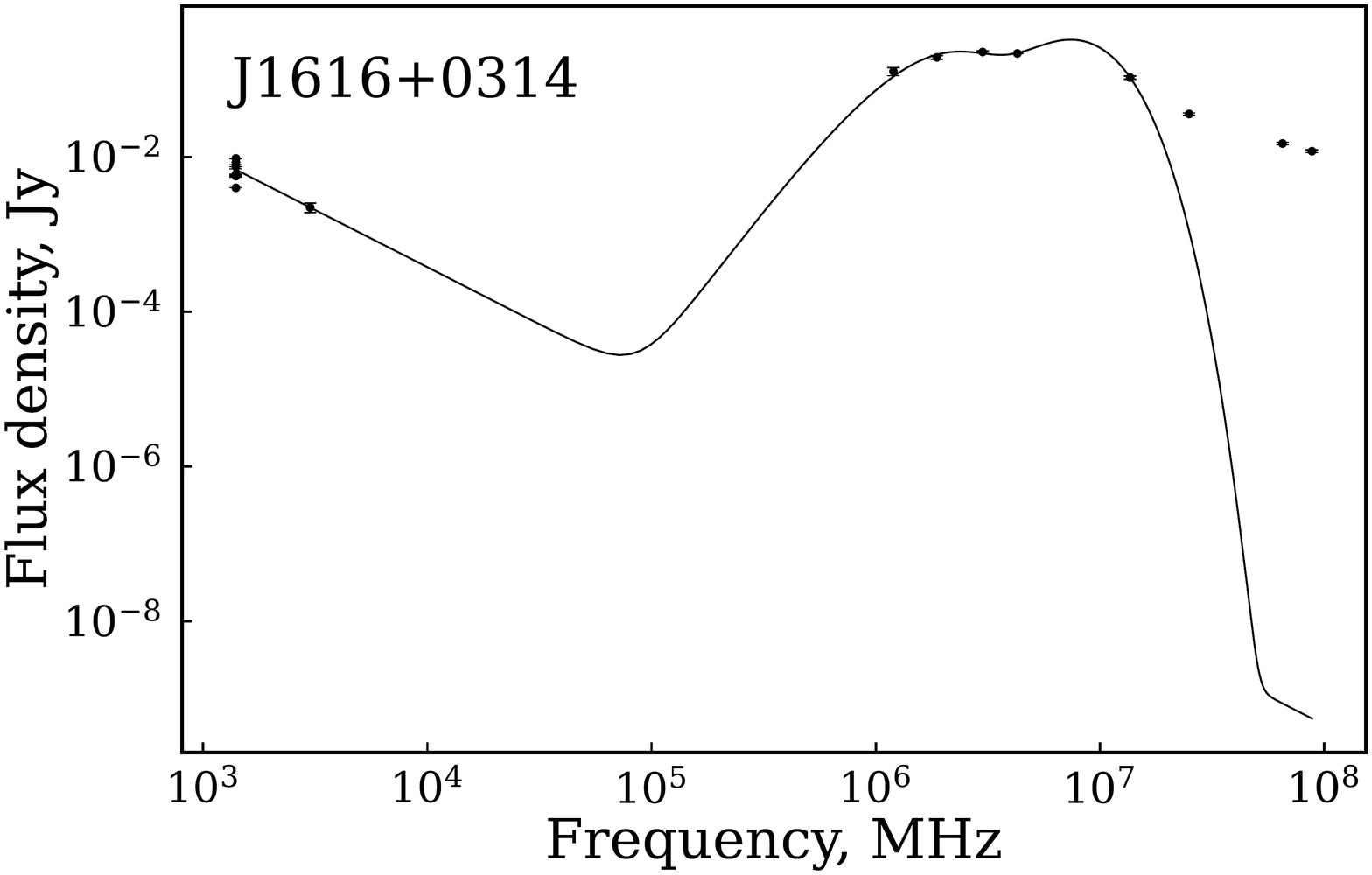}}
        \subfloat[]{\includegraphics[width=0.8\columnwidth]{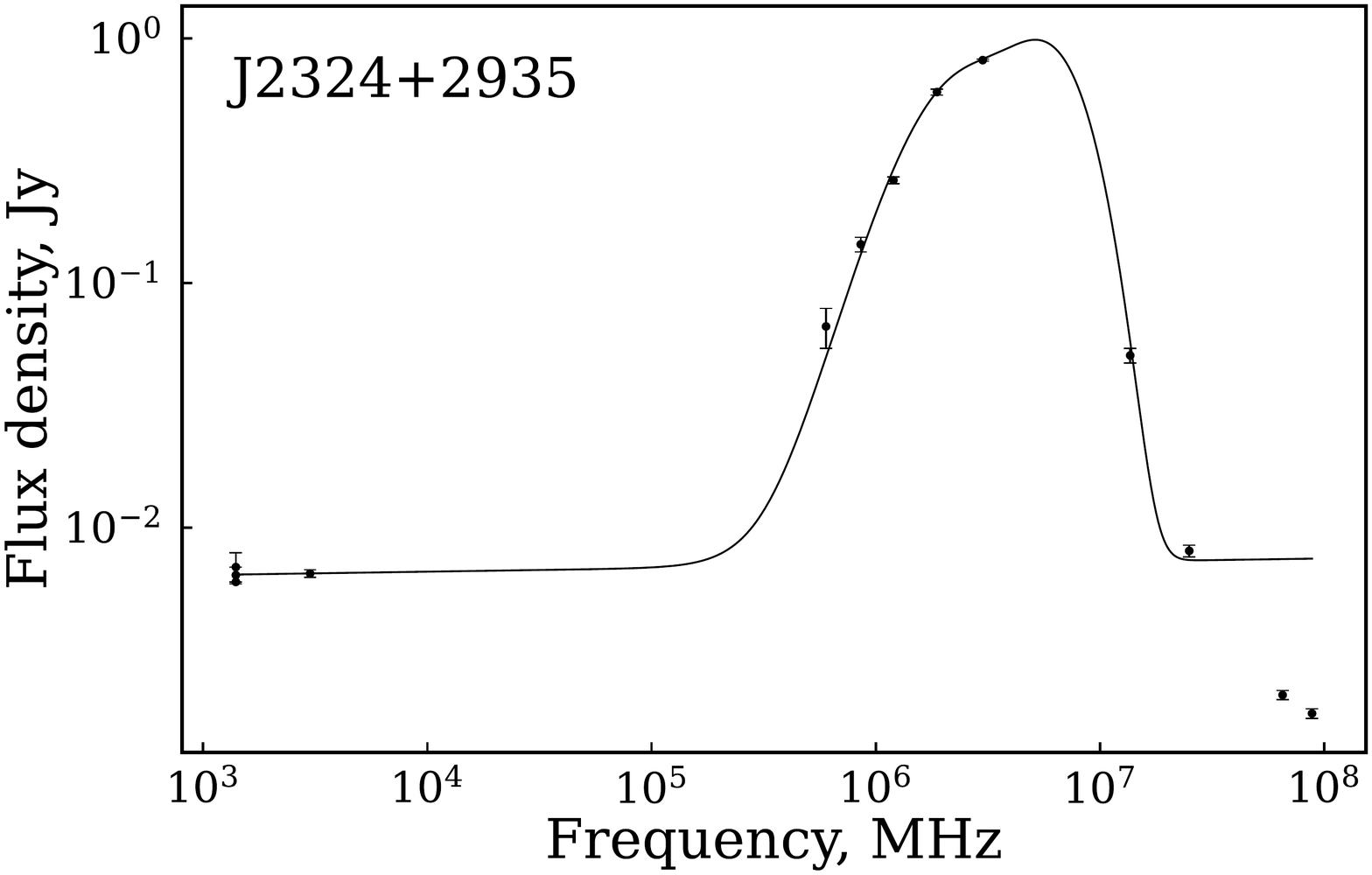}} 
\caption{SED approximation for the control sample sources having flux data in the sub-mm and IR regions (continued).}
\label{ris:SED_cntrl2}
\end{figure*}

The results of the SED approximation for megamasers are shown in Figs.~\ref{ris:SED_models1}--\ref{ris:SED_models4}, and for the control sample 
in Figs.~\ref{ris:SED_cntrl1}--\ref{ris:SED_cntrl2}. It should be noted that the fluxes in the middle and near-IR region with frequencies greater than 10~THz are formed due to the radiation of the PAH molecules and were not modeled in this work.

\begin{table}
\caption{Model parameters for the objects under study. The upper part of the table includes megamasers, the lower part includes objects from the control sample.}
\begin{center}
\begin{tabular}{ |l|l|l|l|l|l| }
\hline
NVSS name     & $\beta$ & $T_1$, & $T_2$, & $w_2$ & $T_{aver}$, \\
              &         &  K     & K      &       & K           \\
\hline
004733$-$251717 & -0.45 & 21.48 & 58.23 & 0.010 & 21.84 \\
005334$+$124133 & -0.89 & 24.46 & 78.97 & 0.004 & 24.70 \\ 
014430$+$170607 & -0.20 & 26.91 & 54.36 & 0.044 & 28.13 \\ 
024240$-$000047 & -0.57 & 23.00 & 159.70 & $2\cdot 10^{-5}$ & 23.01 \\ 
033336$-$360826 & -0.59 & 17.94 & 62.90 & 0.001 & 18.01 \\ 
052101$-$252145 & -0.60 & 29.66 & 69.02 & 0.024 & 30.61 \\ 
054548$+$584203 & -0.62 & 16.52 & 31.18 & 0.054 & 17.31 \\ 
093551$+$612112 & -0.44 & 24.60 & 56.94 & 0.010 & 24.92 \\ 
100605$-$335317 & -0.63 & 8.33 & 39.72 & 0.001 & 8.36 \\ 
102000$+$081335 & -0.30 & 26.44 & 52.23 & 0.058 & 27.94 \\ 
110353$+$405059 & -0.48 & 24.72 & 52.76 & 0.016 & 25.17 \\ 
112832$+$583343 & -0.66 & 21.71 & 61.12 & 0.013 & 22.24 \\ 
115311$-$390748 & -0.52 & 24.35 & 53.33 & 0.018 & 24.86 \\ 
121345$+$024840 & -0.21 & 27.05 & 54.78 & 0.036 & 28.04 \\ 
122654$-$005238 & -0.40 & 25.93 & 61.34 & 0.034 & 27.12 \\ 
125614$+$565222 & -0.31 & 29.32 & 68.79 & 0.022 & 30.18 \\ 
131226$-$154751 & -0.48 & 21.28 & 57.19 & 0.004 & 21.43 \\ 
131503$+$243707 & -0.26 & 24.99 & 51.53 & 0.044 & 26.18 \\ 
134442$+$555313 & -0.43 & 27.83 & 58.10 & 0.035 & 28.89 \\ 
134733$+$121724 & -0.52 & 32.90 & 117.39 & $4\cdot 10^{-4}$ & 32.93 \\ 
151313$+$071331 & -0.48 & 24.15 & 46.66 & 0.035 & 24.93 \\ 
152659$+$355839 & -0.24 & 27.43 & 67.53 & 0.029 & 28.59 \\ 
153457$+$233011 & -0.18 & 24.78 & 51.88 & 0.034 & 25.71 \\ 
164240$-$094315 & -0.48 & 15.32 & 37.04 & 0.026 & 15.88 \\ 
172321$-$001702 & -0.33 & 20.06 & 39.58 & 0.124 & 22.47 \\ 
225149$-$175225 & -0.40 & 5.39 & 39.56 & 0.200 & 12.22 \\ 
231600$+$253324 & -0.52 & 24.79 & 62.75 & 0.011 & 25.21 \\ 
233901$+$362109 & -0.62 & 27.19 & 59.91 & 0.022 & 27.91 \\ 
\hline
015950$+$002338 & -0.70 & 29.75 & 75.13 & 0.0178 & 30.56 \\ 
090734$+$012502 & -0.27 & 24.37 & 46.29 & 0.0175 & 24.75 \\ 
131653$+$234047 & -0.50 & 11.39 & 47.27 & 0.079 & 14.23 \\ 
133718$+$242302 & -0.31 & 19.23 & 200.71 & $1\cdot 10^{-5}$ & 19.23 \\ 
134015$+$332437 & -0.96 & 17.89 & 56.34 & 0.124 & 22.68 \\ 
140638$+$010255 & -0.74 & 32.75 & 81.72 & 0.022 & 33.83 \\ 
140819$+$290446 & -0.76 & 29.08 & 72.26 & 0.048 & 31.18 \\ 
142231$+$260203 & -0.86 & 25.89 & 34.76 & $1\cdot 10^{-6}$ & 25.89 \\ 
161644$+$031419 & -1.48 & 22.53 & 79.38 & 0.003 & 22.71 \\ 
232427$+$293541 & 0.01 & 18.49 & 37.20 & 0.148 & 21.28 \\ 
\hline
\end{tabular}
\end{center}
\label{tab:SED_models}
\end{table}

\section{Conclusion}

We have studied the integral flux density variability of infrared-bright galaxies with megamaser (OHM) sources and a control sample of galaxies without OH radiation. For most OHM galaxies the variability level in the radio continuum does not exceed 0.20, and the median values of the variability index for the entire OHM sample vary within 0.08--0.17 at the considered frequencies. A comparison with the control sample of non-OHM infrared galaxies has shown
that for this sample the variability level is 0.10--0.15, which does not
differ much in order of magnitude from the obtained variability of OHM galaxies. Individual bright representatives of OHM galaxies demonstrate variability of the order of \mbox{0.30--0.50} near the 5 GHz frequency 
on a time scale of 2--30 yrs. Such galaxies are usually associated with AGNs, for example J0047$-$2517, or active star formation is observed in them, like in J1509$-$1119.

The parameters of the spectral energy distribution for the galaxies with available literature measurements in the frequency range from MHz to THz have been determined. The low-frequency spectral indices in the range $<\!50$~GHz 
have been obtained, and the model parameters for thermal dust emission from 50~GHz to 10~THz have been estimated. A comparison of the dust component color temperature distributions for the two samples revealed no difference in their statistical properties. Also no significant correlations were found between the dust color temperature in  the OHM galaxies and such parameters as the variability index or luminosity in the OH line.

\section*{Acknowledgements}
We thank the referee, Dr. A.~M.~Sobolev, for providing valuable comments and recommendations. The observations have been made with the scientific equipment of the SAO~RAS radio telescope RATAN-600 and supported by the Ministry of Science and Higher Education of the Russian Federation.
%This paper has been supported by the Kazan Federal University Strategic Academic Leadership Program (``PRIORITY-2030'').
The study has been made with the financial support from the Russian Foundation for Basic Research (RFBR) and the National Natural Science Foundation of China (NSFC) within the framework of scientific project No.~\mbox{21-52-53035} ``The Radio Properties and Structure of OH Megamaser Galaxies.\!'' The study has been supported by NSFC projects No.~U1931203 and 12111530009. The research has made use of the CATS and VizieR databases and the NASA/IPAC Extragalactic Database (NED), which is funded by the National Aeronautics and Space Administration and operated by the California Institute of Technology.

\bibliographystyle{mnras}
\bibliography{OHM-var}

\newpage
\onecolumn 
\appendix
\section{Variability parameters}
\begin{longtable}{|l|c|c|c|c|c|c|c|c|c|}
\caption{\label{tab:OHM_var} Variability ($V_{S}$) and modulation indices ($M$) for OHM sample. N is the number of observations.}\\
\hline
NVSS name & N$_{2.3}$ &	$M_{2.3}$ &	$V_{S_{2.3}}$ &	N$_{4.7}$ &	$M_{4.7}$ &	$V_{S_{4.7}}$ &	N$_{8.2}$ &	$M_{8.2}$ &	$V_{S_{8.2}}$ \\
1 & 2 & 3 & 4 & 5 & 6 & 7 & 8 & 9 & 10 \\
\hline
\endfirsthead
\caption[]{continued.}\\
\hline
NVSS name & N$_{2.3}$ &	$M_{2.3}$ &	$V_{S_{2.3}}$ &	N$_{4.7}$ &	$M_{4.7}$ &	$V_{S_{4.7}}$ &	N$_{8.2}$ &	$M_{8.2}$ &	$V_{S_{8.2}}$ \\
1 & 2 & 3 & 4 & 5 & 6 & 7 & 8 & 9 & 10 \\
\hline
\endhead
\hline
\endfoot
000820$+$403755	&	0	&	-	&	-		    &	2	&	0.08	&	-0.17 (0.25)    &	0	&	-	 &	-		\\
%003600$-$271535	&	0	&	-	&	-		    &	1	&	-	    &	-		        &	0	&	-	 &	-		\\
004733$-$251717	&	15	& 0.20	&	0.34 (0.02)	&	27	&	0.24	&	0.29 (0.03)	    &	10	&	0.29 &	0.35 (0.05)	\\
005334$+$124133	&	1	&	-	&	-		    &	2	&	0.21	&	-0.13 (0.43)    &	0	&	-	 &	-		\\
%013852$-$102711	&	0	&	-	&	-		    &	1	&	-	   	&	-		        &	1	&	-	 &	-		\\
014430$+$170607	&	0	&	-	&	-	    	&	14	&	0.28	&	0.31 (0.07)    &	5	&	0.19 &	0.03	(0.17)	\\
%015903$+$254237	&	0	&	-	&	-	   		&	1	&	-	   	&	-	        	&	0	&	-	 &	-		\\
015913$-$292435	&	0	&	-	&	-		   	&	12	&	0.14	&	0.12	(0.07)	&	4	&	0.05 &	-0.13	(0.18)	\\
024240$-$000047	&	27	& 0.05  &	0.06 (0.05)	&	30	&	0.08	&	0.15	(0.03)	&	16	&	0.23 &	0.29	(0.07)	\\
025134$+$431514	&	16	& 0.16  &	0.16 (0.11)	&	53	&	0.21	&	0.45	(0.02)	&	22	&	0.17 &	0.33	(0.02)	\\
030830$+$204620	&	0	&	-	&	-	   		&	2	&	0.14	&	-0.05	(0.19)	&	0	&	-	 &	-		\\
032824$-$141207	&	0	&	-	&	-		   	&	2	&	0	   	&	0	         	&	0	&	-	 &	-		\\
033336$-$360826	&	7	& 0.24	&	0.31 (0.31)	&	4	&	0.17	&	0.19	(0.03)	&	2	&	0.38 &	0.18	(0.27)	\\
035441$+$003704	&	0	&	-	&	-		   	&	2	&	0.20	&	0		        &	1	&	-	 &	-		\\
051209$-$242156	&	0	&	-	&	-		   	&	11	&	0.35	&	0.35	(0.08)	&	1	&	-	 &	-		\\
052101$-$252145	&	0	&	-	&	-	   		&	3	&	0.46	&	0.33	(0.27)	&	2	&	0.47 &	-0.33	(0.55)	\\
054548$+$584203	&	0	&	-	&	-		   	&	5	&	0.13	&	0.11	(0.06)	&	0	&	-	&	-		\\
062222$-$364743	&	0	&	-	&	-		   	&	11	&	0.37	&	0.33	(0.08)	&	1	&	-	&	-		\\
065145$+$220427	&	0	&	-	&	-		   	&	4	&	0.24	&	0.05	(0.21)	&	1	&	-	&	-		\\
080947$+$050108	&	0	&	-	&	-		   	&	5	&	0.23	&	0.20	(0.07)	&	1	&	-	&	-		\\
%084750$+$232111	&	0	&	-	&	-		   	&	0	&	-	   	&	-		        &	0	&	-	&	-		\\
090634$+$045125	&	0	&	-	&	-		   	&	2	&	0.11	&	-0.11	(0.22)	&	0	&	-	&	-		\\
093551$+$612112	&	0	&	-	&	-		   	&	17	&	0.18	&	0.20	(0.12)	&	4	&	0.10&	-	\\
%095634$+$084303	&	0	&	-	&	-		   	&	1	&	-	   	&	-			    &	0	&	-	&	-		\\
100605$-$335317	&	0	&	-	&	-		   	&	4	&	0.18	&	-0.05	(0.26)	&	2	&	0	&	0		\\
%100626$+$272543	&	0	&	-	&	-		   	&	0	&	-	   	&	-		        &	0	&	-	&	-		\\
%102000$+$081335	&	0	&	-	&	-		   	&	1	&	-	   	&	-		        &	1	&	-	&	-		\\
%103638$+$153239	&	0	&	-	&	-		   	&	0	&	-	    &	-	         	&	0	&	-	&	-		\\
%104029$+$105320	&	0	&	-	&	-		   	&	1	&	-	   	&	-		        &	0	&	-	&	-		\\
110353$+$405059	&	0	&	-	&	-		   	&	3	&	0.22	&	0.17	(0.06)	&	3	&	0.18 &	0.08	(0.13)	\\
112832$+$583343	&	1	&	-	&	-	   		&	21	&	0.29	&	0.42	(0.02)	&	4	&	0.46 &	-		\\
115311$-$390748	&	0	&	-	&	-		   	&	12	&	0.10	&	0.06	(0.09)	&	2	&	0.09 &	-0.20	(0.25)	\\
%120424$+$192512	&	0	&	-	&	-		   	&	0	&	-	   	&	-		        &	0	&	-	&	-		\\
%120547$+$165108	&	0	&	-	&	-		   	&	0	&	-	   	&	-		        &	0	&	-	&	-		\\
%120944$-$050116	&	0	&	-	&	-		   	&	1	&	-	   	&	-		        &	0	&	-	&	-		\\
121345$+$024840	&	0	&	-	&	-		   	&	3	&	0.08	&	-0.13	(0.22)	&	4	&	0.26&	0.18	(0.17)	\\
122654$-$005238	&	1	&	-	&	-		   	&	5	&	0.06	&	-0.07	(0.16)	&	4	&	0.09&	-0.06	(0.18)	\\
125614$+$565222	&	0	&	-	&	-		   	&	23	&	0.32	&	0.16	(0.06)	&	6	&	0.15&	-	\\
131226$-$154751	&	0	&	-	&	-		   	&	2	&	0.02	&	-0.13	(0.15)	&	1	&	-	&	-		\\
131503$+$243707	&	1	&	-	&	-		   	&	2	&	0.02	&	-0.14	(0.16)	&	3	&	0.10&	-0.09	(0.20)	\\
%132732$+$473906	&	0	&	-	&	-		   	&	0	&	-	   	&	-		        &	0	&	-	&	-		\\
134442$+$555313	&	0	&	-	&	-		   	&	18	&	0.26	&	0.39	(0.04)	&	3	&	0.08&	-	\\
134733$+$121724	&	35	& 0.08	&	0.17 (0.02)	&	65	&	0.09	&	0.19	(0.01)	&	32	&	0.08&	0.14	(0.04)	\\
140649$+$061035	&	0	&	-	&	-		   	&	3	&	0.20	&	-0.08	(0.31)	&	0	&	-	&	-		\\
%140818$+$194622	&	0	&	-	&	-		   	&	0	&	-	   	&	-		        &	0	&	-	&	-		\\
150102$+$142001	&	0	&	-	&	-		   	&	3	&	0.09	&	-0.11	(0.22)	&	0	&	-	&	-		\\
150916$-$111925	&	0	&	-	&	-		   	&	3	&	0.58	&	0.49	(0.06)	&	0	&	-	&	-		\\
151313$+$071331	&	4	& 0.08	&	-0.15 (0.22)&	17	&	0.14	&	0.13	(0.12)	&	6	&	0.08&	-0.10	(0.19)	\\
%152549$+$052249	&	0	&	-	&	-		   	&	1	&	-	   	&	-		        &	0	&	-	&	-		\\
152659$+$355839	&	0	&	-	&	-		   	&	4	&	0.09	&	-0.08	(0.21)	&	4	&	0.18&	0		\\
%152727$-$095550	&	0	&	-	&	-		   	&	0	&	-	   	&	-	         	&	0	&	-	&	-		\\
153457$+$233011	&	8	& 0.19	&	0.10 (0.06)	&	19	&	0.06	&	0.06	(0.04)	&	12	&	0.04&	-0.04	(0.07)	\\
161140$-$014707	&	0	&	-	&	-		   	&	3	&	0.12	&	-0.05	(0.19)	&	2	&	0.08&	-0.17	(0.25)	\\
%161611$+$422401	&	0	&	-	&	-		   	&	1	&	-	   	&	-	         	&	0	&	-	&	-		\\
163221$+$155146	&	0	&	-	&	-		   	&	2	&	0.09	&	-0.09	(0.18)	&	0	&	-	&	-		\\
164240$-$094315	&	0	&	-	&	-		   	&	12	&	0.19	&	0.14	(0.11)	&	3	&	0.13 &	-0.11	(0.20)	\\
172321$-$001702	&	2	& 0.02  &	-0.12 (0.13)&	13	&	0.12	&	0.09	(0.08)	&	5	&	0.41 &	-0.05	(0.11)	\\
175429$+$325312	&	1	&	-	&	-		   	&	13	&	0.23	&	0.15	(0.23)	&	2	&	0.18 &	-0.05	(0.24)	\\
%183835$+$355219	&	0	&	-	&	-		   	&	1	&	-	   	&	-	         	&	1	&	-	&	-		\\
%190041$+$352125	&	0	&	-	&	-		   	&	0	&	-	   	&	-	         	&	0	&	-	&	-		\\
205125$+$185804	&	0	&	-	&	-		   	&	3	&	0.10	&	-0.13	(0.25)	&	1	&	-	&	-		\\
%205724$+$170741	&	0	&	-	&	-		   	&	0	&	-	   	&	-	         	&	0	&	-	&	-		\\
220436$+$421940	&	0	&	-	&	-		   	&	2	&	0.07	&	-0.13	(0.20)	&	2	&	0.07&	-0.14	(0.21)	\\
%220749$+$303942	&	0	&	-	&	-		   	&	0	&	-	   	&	-	         	&	0	&	-	&	-		\\
%221134$-$181704	&	0	&	-	&	-		   	&	0	&	-	   	&	-	         	&	0	&	-	&	-		\\
%221410$+$045226	&	0	&	-	&	-		   	&	1	&	-	   	&	-	         	&	0	&	-	&	-		\\
225149$-$175225	&	0	&	-	&	-		   	&	0	&	-	   	&	-	         	&	2	&	0.03&	0.03	(0.01)	\\
%230421$+$342149	&	0	&	-	&	-		   	&	0	&	-	   	&	-       		&	0	&	-	&	-		\\
%230520$+$074145	&	0	&	-	&	-		   	&	0	&	-	   	&	-	         	&	0	&	-	&	-		\\
230735$+$041559	&	0	&	-	&	-		   	&	2	&	0.05	&	-0.16	(0.21)	&	0	&	-	&	-		\\
231600$+$253324	&	1	&	-	&	-		   	&	5	&	0.08	&	-0.10	(0.21)	&	5	& 0.15  &	0.06	(0.20)	\\
%232556$+$100248	&	0	&	-	&	-		   	&	1	&	-	   	&	-	         	&	1	&	-	&	-		\\
233511$+$293000	&	0	&	-	&	-	    	&	2	&	0.11	&	-0.11	(0.22)	&	0	&	-	&	-		\\
233901$+$362109	&	2	& 0.04	&  -0.23 (0.27) &	4	&	0.11	&	-0.05	(0.19)	&	4	& 0.15  &	0		\\
\hline
\end{longtable}
\newpage
\begin{longtable}{|l|c|c|c|c|c|c|c|c|c|}
\caption{\label{tab:control_var} Variability ($V_{S}$) and modulation indices ($M$) for control sample. N is the number of observations.}\\
\hline
NVSS name & N$_{2.3}$ &	$M_{2.3}$ &	$V_{S_{2.3}}$ &	N$_{4.7}$ &	$M_{4.7}$ &	$V_{S_{4.7}}$ &	N$_{8.2}$ &	$M_{8.2}$ &	$V_{S_{8.2}}$ \\
1 & 2 & 3 & 4 & 5 & 6 & 7 & 8 & 9 & 10 \\
\hline
\endfirsthead
\caption[]{continued.}\\
\hline
NVSS name & N$_{2.3}$ &	$M_{2.3}$ &	$V_{S_{2.3}}$ &	N$_{4.7}$ &	$M_{4.7}$ &	$V_{S_{4.7}}$ &	N$_{8.2}$ &	$M_{8.2}$ &	$V_{S_{8.2}}$ \\
1 & 2 & 3 & 4 & 5 & 6 & 7 & 8 & 9 & 10 \\
\hline
\endhead
\hline
\endfoot
002651$+$340122	&	0	&	-	&	-		    &	3	&	0.07&	-0.17 (0.25)	&	1	&	-	&	- \\
013741$+$330935	&	710	&	0.09& 0.24 (0.01)	&	126	&	0.04&	0.18 (0.01)	&	552	&	0.07	&	0.28 (0.05)	\\
015328$+$260939	&	0	&	-	&	-		    &	4	&	0.08&	-0.10 (0.21)	&	1	&	-	&	-		\\
015950$+$002338	&	0	&	-	&	-		    &	4	&	0.05&	-0.18 (0.24)	&	1	&	-	&	-		\\
021008$+$235049	&	0	&	-	&	-		    &	2	&	0.08&	-0.23 (0.31)	&	0	&	-	&	-		\\
041634$+$122457	&	0	&	-	&	-		    &	3	&	0.31&	0.08 (0.31)	&	1	&	-	&	-		\\
063956$+$280956	&	0	&	-	&	-		    &	2	&	0.29&	0.08 (0.23)	&	1	&	-	&	-		\\
072128$+$040146	&	0	&	-	&	-		    &	3	&	0.04&	-0.18 (0.23)	&	0	&	-	&	-		\\
081755$+$312827	&	0	&	-	&	-		    &	18	&	0.13&	0.07 (0.10)	&	6	&	0.37	&	0.23 (0.09)	\\
085448$+$200630	&	1573&	0.32&	0.68 (0.02)	&	776	&	0.37&	0.81 (0.03)	&	2830	&	0.44	&	0.82 (0.02)	\\
094521$+$173753	&	0	&	-	&	-		    &	2	&	0.04&	-0.19 (0.22)	&	1	&	-	&	-		\\
111438$+$324133	&	2 	&	0	&	0		    &	8	&	0.06&	-0.09 (0.17)	&	2	&	0.09	&	-0.12 (0.21)	\\
115314$+$131427	&	0	&	-	&	-		    &	3	&	0.44&	0.25 (0.25)	&	1	&	-	&	-		\\
121320$+$314053	&	0	&	-	&	-		    &	5	&	0.26&	0.24 (0.12)	&	3	&	0.09	&	-0.15 (0.27)	\\
122906$+$020308	&	109	&	0.08&	0.22 (0.02)	&	739	&	0.12&	0.48 (0.07)	&	1424	&	0.19	&	0.57 (0.01)	\\
131653$+$234047	&	0	&	-	&	-		    &	2	&	0.22&	0		&	2	&	0.29	&	0		\\
133718$+$242302	&	0	&	-	&	-		    &	3	&	0.09&	-0.11 (0.22)	&	0	&	-	&	-		\\
135646$+$102609	&	0	&	-	&	-		    &	5	&	0.30&	0.17 (0.22)	&	2	&	0.08	&	-0.17 (0.25)	\\
142056$-$000429	&	0	&	-	&	-		    &	2	&	0.22&	0		&	1	&	-	&	-		\\
161413$+$260415	&	0	&	-	&	-		    &	5	&	0.19&	0		&	1	&	-	&	-		\\
175105$+$265903	&	1	&	-	&	-		    &	8	&	0.10&	0.07 (0.06)	&	3	&	0.20	&	-0.03 (0.26)	\\
181700$+$155449	&	1	&	-	&	-		    &	2	&	0.07&	-0.14 (0.21)	&	1	&	-	&	-		\\
183336$+$225201	&	0	&	-	&	-		    &	3	&	0.14&	-0.09 (0.27)	&	0	&	-	&	-		\\
204817$+$193655	&	1	&	-	&	-		    &	3	&	0.17&	0		&	1	&	-	&	-		\\
214633$+$354835	&	0	&	-	&	-		    &	2	&	0.27&	0.09 (0.18)	&	1	&	-	&	-		\\
224509$+$323128	&	0	&	-	&	-		    &	2	&	0.09&	-0.09 (0.18)	&	1	&	-	&	-		\\
231354$+$033055	&	0	&	-	&	-		    &	2	&	0.10&	-0.10 (0.19)	&	1	&	-	&	-		\\
231635$+$040518	&	23	&	0.06&	0.08 (0.07)	&	46	&	0.09&	0.23 (0.01)	    &	4	&	0.05	&	-0.02 (0.09)	\\
232549$+$283421	&	1	&	-	&	-		    &	4	&	0.08&	-0.13 (0.22)	&	4	&	0.14	&	-0.05 (0.26)	\\
235630$+$233849	&	0	&	-	&	-		    &	2	&	0.13&	-0.13 (0.25)	&	1	&	-	&	-		\\
\hline
\end{longtable}

\bsp
\label{lastpage}

\end{document}